\documentclass[
twocolumn,
aps,
prb,
reprint,
superscriptaddress,
amsmath,amssymb
]{revtex4-1}
\usepackage[pdftex]{graphicx} 
\usepackage[hidelinks]{hyperref} 
\hypersetup{
 setpagesize=false,
 bookmarksnumbered=true,%
 bookmarksopen=true,%
 colorlinks=true,%
 linkcolor=blue,
 citecolor=blue,
}

\usepackage{braket}
\usepackage{times,multirow,amsfonts,bm,xspace,pifont}
\usepackage{soul}
\usepackage{ulem}
\usepackage{mathrsfs}
\begin{document}
\newcommand{\rr}{{\bm r}}
\newcommand{\q}{{\bm q}}
\renewcommand{\k}{{\bm k}}
\newcommand*\wien    {\textsc{wien}2k\xspace}
\newcommand*\textred[1]{\textcolor{red}{#1}}
\newcommand*\textblue[1]{\textcolor{blue}{#1}}
\newcommand*\YY[1]{\textcolor{blue}{#1}}
\newcommand*\YYS[1]{\textcolor{blue}{\sout{#1}}}
\newcommand*\AD[1]{\textcolor{magenta}{#1}}
\newcommand*\ADS[1]{\textcolor{magenta}{\sout{#1}}}
\newcommand*\ADC[1]{\textcolor{green}{#1}}
\newcommand*\TK[1]{{\color{red}{#1}}}
\newcommand*\TKS[1]{{\color{red}{\sout{#1}}}}
\newcommand*\BdG{{\rm BdG}}

\title{
Quantum geometric effect on Fulde–Ferrell–Larkin–Ovchinnikov superconductivity}

\author{Taisei Kitamura}
 \email[]{kitamura.taisei.67m@st.kyoto-u.ac.jp}
\affiliation{Department of Physics, Graduate School of Science, Kyoto University, Kyoto 606-8502, Japan}

\author{Akito Daido}
\affiliation{Department of Physics, Graduate School of Science, Kyoto University, Kyoto 606-8502, Japan}

\author{Youichi Yanase}
\affiliation{Department of Physics, Graduate School of Science, Kyoto University, Kyoto 606-8502, Japan}
\date{\today}

\begin{abstract}
Quantum geometry characterizes the geometric properties of Bloch electrons in the wave space, represented by the quantum metric and the Berry curvature. 
Recent studies have revealed that the quantum geometry plays a major role in various physical phenomena, from multipole to non-Hermitian physics.
For superconductors, the quantum geometry is clarified to appear in the superfluid weight, an essential quantity of superconductivity.
Although the superfluid weight was considered to be determined by the Fermi-liquid contribution for a long time, the geometric contribution is not negligible in some superconductors such as artificial flat-band systems and monolayer FeSe. 
While the superfluid weight is essential for many superconducting phenomena related to the center of mass momenta of Cooper pairs (CMMCP), the full scope of the quantum geometric effect on superconductivity remains unresolved.
In this paper, we study the quantum geometric effect on the Fulde–Ferrell–Larkin–Ovchinnikov (FFLO) state acquiring a finite CMMCP in equilibrium. 
As a benchmark, the phase diagrams of effective models for monolayer FeSe in an in-plane magnetic field are calculated.
In the case of the isotropic $s$-wave pairing, the quantum geometry stabilizes the BCS state, and a metastable BCS state appears in the high magnetic field region.
In addition, the quantum geometry induces the phase transition from the FFLO state to the BCS state with increasing temperature.
On the other hand, for the inter-sublattice pairing, the quantum geometry gives a negative contribution to the superfluid weight; this can induce the FFLO superconductivity in particular parameter sets.

\end{abstract}

\maketitle

\section{Introduction}
For the past decades, the geometric properties of Bloch electrons have been intensively studied from the viewpoint of topological phenomena~\cite{hasan2010topological,qi2011topological}.
The Berry curvature~\cite{berry1984quantal}, defined as the imaginary part of the quantum geometric tensor~\cite{provost1980riemannian,resta2011the}, plays an essential role in various topological phenomena as it determines the anomalous quantum/non-quantized Hall effects~\cite{thouless1982quantized,xiao2010berry,Nagaosa2010}.
On the other hand, while the real part of the quantum geometric tensor defines the quantum metric~\cite{resta2011the,provost1980riemannian}, which represents the distance between two adjacent states in the wave space, the quantum metric has not been intensively studied until these days.

Since the quantum metric is closely related to the Berry curvature, it is expected that the quantum metric may affect various physical phenomena.
Indeed, recent studies have revealed essential roles of the quantum metric in properties of solids~\cite{piechon2016geometric,marzari1997maximally,Neupert2013,Srivastava2015,gao2014field,gao2019nonreciprocal,lapa2019semiclassical,daido2020thermodynamic,kitamura2021thermodynamic,mitscherling2022bound,ahn2021reimannian,rhim2020quantum,wang2021exact,hwang2021geometric,mera2022nontrivial}, artificial quantum systems~\cite{julku2021excitations,julku2021quantum,topp2021light}, and non-Hermitian systems~\cite{solnyshkov2021quantum,liao2021experimental}.
Furthermore, the quantum metric can be divided into the contribution from each band, which is especially called the band-resolved quantum metric.
It has been shown that the band-resolved quantum metric is an essential ingredient of linear and nonlinear optical responses such as photocurrent generation~\cite{ahn2020low-frequency,watanabe2021chiral} and spectral weight transfer~\cite{ahn2021superconductivity-induced}.
In addition, the relationship between the quantum metric and other geometric quantities such as the Berry phase, Berry curvature, and Chern number is recently studied\cite{ozawa2021relations,mera2021kahler,mera2021engineering,mera2021relating}.
Thus, the quantum geometry containing the quantum metric is becoming a fundamental property for understanding the physical phenomena of quantum materials.

The quantum geometry is known to be essential for superconductors 
after theoretical works showed that the quantum metric appears in the superfluid weight $D^{\rm s}$~\cite{peotta2015superfluidity,liang2017band}.
Based on the Fermi-liquid theory, it has been traditionally believed that the superfluid weight is determined by the effective mass $m^*$ and density $n^*$ of Bloch electrons 
i.e. $n^*/m^*$~\cite{tinkham2004introduction,jujo2001fermi}.
However, the geometric contribution beyond the Fermi-liquid theory was recently pointed out, and it is particularly important in the flat-band system~\cite{peotta2015superfluidity}, in which the Fermi-liquid contribution vanishes as $m^{*} \rightarrow\infty$, leading to $n^{*}/m^{*} \rightarrow0$.
This idea has been applied to some artificial fermion systems, such as cold atoms on the optical Lieb lattice~\cite{taie2015coherent,ozawa2017interaction-driven,julku2016geometric,he2021geometry,huhtinen2022revisiting} 
and superconducting twisted bilayer graphene (TBG)~\cite{cao2018unconventional,hu2019geometric,julku2020superfluid,xie2020topology-bonded,peri2021fragile,rossi2021quantum,torma2022superconductivity,hu2022quantum}. 
In addition, a perfectly flat-band model with strictly local obstructed Wannier functions was studied~\cite{herzogarbeitman2021superfluid}.
The theoretical prediction has been verified in the TBG, as the geometric origin of the superfluid weight has been reported in the recent experiment~\cite{tian2021evidence}.
In these systems, the quantum geometry determines the magnetic penetration depth by $\lambda(T) = 1/\sqrt{4\pi D^{\rm s}(T)}$.
For the two-dimensional systems, the quantum geometry also determines the zero-resistance transition temperature, since Berezinskii-Kosterlitz-Thouless (BKT) transition temperature $T_{\rm BKT}$ is given by the superfluid weight according to the formula $D^{\rm s}(T_{\rm BKT}) = 8T_{\rm BKT}/\pi$.

While the intensive studies introduced above pay attention to the flat-band systems, a significant enhancement of superconductivity by the quantum geometry is possible in other systems as well.
Indeed, it has been shown that the quantum geometry enhances the BKT transition temperature in monolayer FeSe~\cite{kitamura2021superconductivity}, which exhibits a high superconducting transition temperature of more than $65$ K~\cite{wang2012interface-induced,he2013phase,xu2020spectroscopic}.
The origin of the $T_c$ enhancement in FeSe without flat band can be attributed to a small carrier density $n^*$, which suppresses the Fermi-liquid contribution to the superfluid weight $n^*/m^*$. 
A small superfluid weight may also be related to the experimental implication of the Bardeen-Cooper-Schrieffer to Bose-Einstein-Condensation (BCS-BEC) crossover~\cite{nozieres1985bose,kasahara2014field-induced,kasahara2016giant,hanaguri2019quantum,kasahara2020evidence}.
Furthermore, because FeSe is a mother compound of a topological superconductor candidate FeSe$_{1-x}$Te$_x$~\cite{wang2015topological,xu2016topological,wang2018evidence,zhang2018obsevation,machida2019zero-energy}, the geometric properties of Bloch electrons should be nontrivial, causing a sizable geometric contribution to the superfluid weight.
Thus, by focusing on quantum geometry, we will obtain a better understanding of superconducting phenomena in monolayer FeSe.


Considering the superfluid weight as the second-order coefficient of the superconducting free energy with respect to the center of mass momenta of Cooper pairs (CMMCP), we expect that the quantum geometry is closely related to a wide range of phenomena related to the CMMCP.
However, the quantum geometric effect on superconductivity has been mostly unexplored.
An example which we study in this paper is the Fulde–Ferrell–Larkin–Ovchinnikov (FFLO) superconductivity~\cite{flude1964superconductivity,larkin1964nonuniform}, in which Cooper pairs spontaneously have a finite center of mass momentum.
As the bulk FeSe is a candidate FFLO superconductor~\cite{kasahara2020evidence}, the quantum geometric effect on the FFLO superconductivity in monolayer FeSe is an intriguing issue. 

In this paper, calculating the temperature-magnetic field phase diagram of superconductivity in a model for monolayer FeSe, we clarify the quantum geometric effect on the FFLO superconductivity.
Considering that the monolayer FeSe is a typical system for sizeable quantum geometry without artificial structure, we expect that our work is a milestone for the relation of quantum geometry and FFLO superconductivity.
We find that the role of quantum geometry qualitatively depends on the structure of Cooper pair wave functions. 
In the case of the intra-sublattice pairing, the geometric superfluid weight is always positive and stabilizes the Bardeen-Cooper-Schrieffer (BCS) superconductivity rather than the FFLO superconductivity. 
On the other hand, in the case of the inter-sublattice pairing, quantum geometry has a negative contribution to the superfluid weight, and it can induce the FFLO superconductivity in particular parameter sets.
We discuss an essential role of glide-mirror symmetry breaking in monolayer FeSe due to the substrate.

The rest of this paper is organized as follows.
In Sec.~\ref{sec:gl_and_fflo}, we review the superfluid weight, whose sign determines the thermodynamic stability of the FFLO and BCS states.
In Sec.~\ref{sec:sfw_mf}, we formulate the superfluid weight in the magnetic field based on the properties of Bloch electrons. 
Using the obtained formula and considering isotropic $s$-wave superconductivity, we reproduce the conventional Fermi-liquid mechanism of FFLO superconductivity, which is induced by the negative Fermi-liquid contribution to the superfluid weight.
In Sec.~\ref{sec:geomtric_sfw_two_band}, analyzing a simple two-band model, we show that the geometric contribution to the superfluid weight can be negative, although it has been believed to be positive in literature. 
A main result of this paper is discussed in Sec.~\ref{sec:spd_fese}, where we show the superconducting phase diagrams of models for monolayer FeSe in an in-plane magnetic field. 
It is shown that the quantum geometry leads to an unusual superconducting phase diagram. 
For instance, quantum-geometry-induced FFLO superconductivity is revealed.
Finally, we give a summary of this paper in Sec. \ref{sec:summary}.

\section {Superfluid weight and FFLO superconductivity\label{sec:gl_and_fflo}}

\begin{figure}[tbp]
    \centering
    \includegraphics[width=0.45\textwidth]{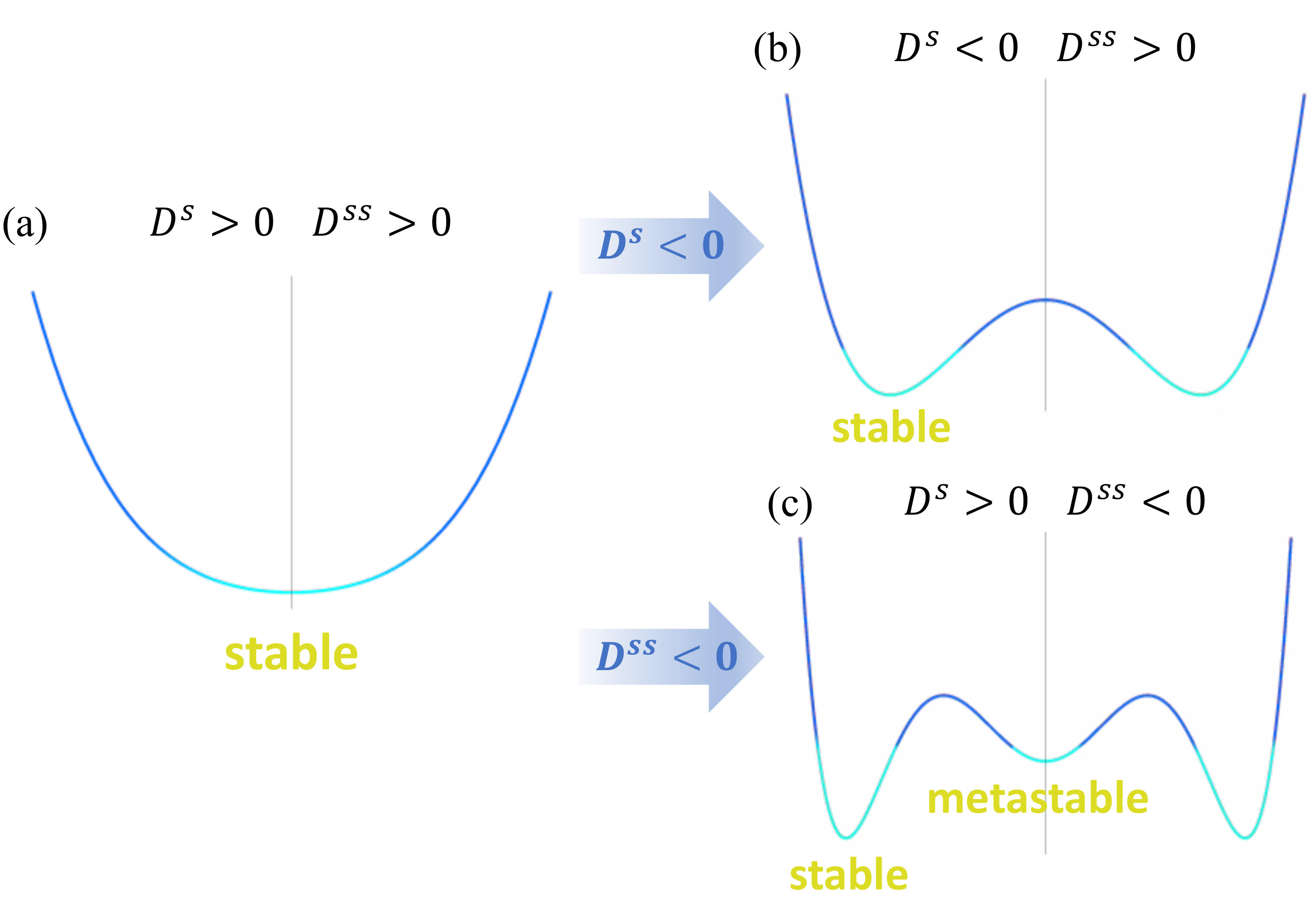}
    \caption{Schematic figure of the CMMCP dependence of the superconducting free energy in the case of (a) $D^{\rm s} > 0$ and $D^{\rm ss} > 0$, (b) $D^{\rm s} < 0$ and $D^{ss} > 0$, and (c) $D^{\rm s} > 0$, $D^{\rm ss} < 0$. (a) The BCS state is most stable.  (b) Superconducting free energy takes the minimum value for a finite CMMCP, and the FFLO state is most stable. (c) Since the superfluid weight is positive, $\bm q = 0$ realizes a local minimum. However, negative $D^{\rm ss}$ makes the FFLO state stable, and the BCS state becomes a metastable state. }
    \label{fig:fe_schematic}
\end{figure}

For discussing the stability between the BCS and FFLO states, it is useful to consider the CMMCP dependence of the superconducting free energy.
For simplicity, we assume the space inversion symmetry.
In this case, the odd-order expansion coefficients with respect to the CMMCP vanish.
Thus, the superconducting free energy can be expanded by the CMMCP as, 
\begin{eqnarray}
    F(\bm q ) = \dfrac{1}{2}\left.\partial^2_{q_\mu}F(\bm q )\right\vert_{\bm q = 0}q^2_{\mu} + \dfrac{1}{24}\left.\partial^4_{q_\mu}F(\bm q )\right\vert_{\bm q = 0}q_{\mu}^4+\ldots, \notag\\
\end{eqnarray}
We also assume that the minimization with respect to a single momentum direction $q_\mu$ is sufficient owing to the point group symmetry of the system.
Here, $F(\bm q)$ is the free energy with a finite CMMCP $\bm q$, and we take the sum of repeated indices $\mu = x,y,z$.
The first term $\partial^2_{q_\mu}F(\bm q )\vert_{\bm q = 0}=D_{\mu\mu}^{\rm s}$ is the superfluid weight.
Hereafter, we write the second term $\partial^4_{q_\mu}F(\bm q )\vert_{\bm q = 0}$ as $D^{\rm ss}$ for the simplicity of notation.

Here, we assume that the higher-order coefficients above the sixth order are positive for simplicity, 
and illustrate the CMMCP dependence of the superconducting free energy in Fig.~\ref{fig:fe_schematic}.
First, we consider the simplest situation of $D^{\rm ss} >0$.
When the superfluid weight is positive, $F(0)$ realizes the minimum free energy, and the BCS state with zero CMMCP is most stable (Fig.~\ref{fig:fe_schematic}(a)).
On the other hand, when the superfluid weight is negative, a finite CMMCP satisfying $F(0) > F(\bm q)$ exists; this means that the FFLO state is more stable than the BCS state (Fig.~\ref{fig:fe_schematic}(b)).
Thus, from the sign of the superfluid weight, we can determine whether the FFLO or the BCS state is stable. 

Next, we discuss a bit complicated case of $D^{\rm ss} < 0$.
When $D^{\rm s}$ is negative, the FFLO state becomes most stable like in the case of Fig.~\ref{fig:fe_schematic}(b).
On the other hand, a positive superfluid weight $D^{\rm s} > 0$ does not ensure the globally stable BCS state. 
In other words, $F(0)$ is the local minimum of free energy, and $F(\bm q)$ with $\q \ne 0$ may be another local minimum due to the negative $D^{\rm ss}$. 
Thus, the FFLO state may be more stable than the metastable BCS state (Fig.~\ref{fig:fe_schematic}(c)).
Indeed, $D^{\rm ss}$ can be negative in the high magnetic field region, and thus, it is not satisfactory to calculate only the superfluid weight. 
Nevertheless, calculation of the superfluid weight $D^s$ for a given BCS state offers a convenient criterion of the FFLO state without requiring the knowledge of finite-$q$ information.
In Sec.~\ref{sec:spd_fese}, we calculate both $D^s$ and the CMMCP dependence of superconducting free energy to obtain the phase diagram of the BCS and FFLO states. 
In Sec.~\ref{sec:pd_s}, we show that the metastable BCS state as in Fig.~\ref{fig:fe_schematic}(c) can be induced by the geometric contribution to the superfluid weight.

\section {Superfluid weight in magnetic field\label{sec:sfw_mf}}
In this section, we derive the formula of the superfluid weight in an in-plane magnetic field based on Refs.~\onlinecite{peotta2015superfluidity,liang2017band,kitamura2021superconductivity}.
Then, using the obtained formulas, we show the conventional Fermi-liquid mechanism of FFLO superconductivity for the isotropic $s$-wave pairing state. The negative Fermi-liquid contribution to the superfluid weight leads to the FFLO state when $\vert\Delta\vert\approx \vert h\vert$.
We note that only the magnetic field along the in-plane direction of two-dimensional superconductors is considered so that the effect of the vortex can be ignored.
The detailed calculations in this section are shown in Appendix~\ref{sec:sfw_dirivation}.

\subsection{Superfluid weight formula based on Bloch electrons}
We start from the Bogoliubov-de Gennes (BdG) Hamiltonian for a finite CMMCP $\bm q$, written by,
\begin{eqnarray}
	\hat{H}_{\BdG}(\bm q) = \sum_{\bm k}\hat{\psi}^\dagger(\bm k,\bm q)H_{\BdG}(\bm k,\bm q)\hat{\psi}(\bm k,\bm q).
\end{eqnarray}
We adopt the matrix representation of the BdG Hamiltonian
\begin{align}
	H_{\BdG}(\bm k,\bm q) =& \left(
		\begin{array}{cc}
			H_0(\bm k+\bm q)&\bm \Delta(\bm k)\\
			\bm \Delta^\dagger(\bm k)&-H^T_0(-\bm k+\bm q)
		\end{array}
	\right)\notag\\
	&+h\gamma_0\otimes \bm 1_{2f},
	\label{eq:BdG}
\end{align}
and the Nambu spinor
\begin{eqnarray}
    \hat{\psi}^\dagger(\bm k,\bm q) = \left(
    \begin{array}{cc}
        \hat{\bm c}_\uparrow^\dagger(\bm k+\bm q) &  \hat{\bm c}_\downarrow^T(-\bm k+\bm q)\\
    \end{array}
    \right),
\end{eqnarray}
where $\hat{\bm c}_{\sigma}^\dagger(\bm k) = \left(
\begin{array}{ccc}
    \hat{c}^\dagger_{1\sigma}(\bm k) & \cdots &\hat{c}^\dagger_{f\sigma}(\bm k) \\
\end{array}
\right)$, $\hat{c}_{l\sigma}(\bm k)$ ($\hat{c}_{l\sigma}^\dagger(\bm k)$) is the annihilation (creation) operator with wave vector $\bm k$ and spin $\sigma = \uparrow(\downarrow)$, and other degrees of freedom in the normal-state Hamiltonian such as orbitals and sublattices are represented by $l=1 ... f$.
Here, $H_0(\bm k)$ is the matrix representation of the Fourier transform of the hopping integral, which satisfies the relationship $H_0(\bm k) = H_0^T(-\bm k)$ since we assume the time-reversal symmetry at zero magnetic field $h=0$.
For simplicity, we ignore the spin-orbit coupling.
Thus, because of the spin rotation symmetry, we can choose an arbitrary spin quantization axis. Here, it is chosen to be in the direction of the magnetic field, and
as a result, the in-plane magnetic field is introduced by the Zeeman energy term, $h \gamma_0\otimes\bm 1_{2f}$, where $\bm 1_{2f}$ and $\gamma_0$ are the identity matrices of dimension $2f$ and Nambu space, respectively, and  $\otimes$ denotes the tensor product.
The gap function $\bm \Delta(\bm k)$ introduced above is self-consistently determined by solving the gap equations for the matrix elements,
\begin{eqnarray}
\Delta_{ls}(\bm k) =  \sum_{\bm k^\prime}V_{ls}(\bm k,\bm k^\prime)\braket{\hat{c}_{s\downarrow}(-\bm k^\prime+\bm q)\hat{c}_{l\uparrow}(\bm k^\prime+\bm q)}\label{eq:gap_equation},
\end{eqnarray}
in the later calculations in Sec.~\ref{sec:spd_fese}. 

The superfluid weight is given by the second-order derivative of the superconducting free energy with respect to the CMMCP,
\begin{eqnarray}
    D_{\mu \nu}^{\rm s} = \lim_{\bm q\rightarrow0}\partial_{q_\mu}\partial_{q_\nu}F(\bm q).
\end{eqnarray}
The free energy of the Bogoliubov-de Gennes Hamiltonian Eq. \eqref{eq:BdG} is written by up to a constant shown in Appendix~\ref{sec:sfw_dirivation},
\begin{eqnarray}
    F(\bm q) =  -k_{\rm B}T\sum_{\bm k}\sum_{a}\ln\left[1+e^{-\beta\left(E_a(\bm k,\bm q)+h\right)}\right],
\end{eqnarray}
with the inverse temperature $\beta = 1/k_{\rm B}T$. Hereafter, we set $k_{\rm B} = 1$ in the natural unit. 
$E_{a}(\bm k)$ is obtained by the eigenvalue equation of the BdG Hamiltonian,
\begin{eqnarray}
    H_{\BdG}(\bm k,\bm q)\ket{\psi_a(\bm k,\bm q)} = (E_{a}(\bm k,\bm q)+h)\ket{\psi_a(\bm k,\bm q)}.
\end{eqnarray}
We find that the magnetic field only shifts the energy level.
This change decreases the condensation energy and destabilizes the superconductivity.
Differentiating the free energy, we obtain the superfluid weight as 
\begin{align}
    D_{\mu \nu}^{\rm s} =& D_{\mu\nu}^{\rm para} + D_{\mu\nu}^{\rm diag},\label{eq:D_s}\\
    D_{\mu\nu}^{\rm diag} =& -\sum_{\bm k}\sum_{ab}\dfrac{f_h(E_a(\bm k))-f_h(E_b(\bm k))}{E_a(\bm k)-E_b(\bm k)}\notag\\
    &\times J_{ab}^{-\mu}(\bm k)\left(J_{ba}^{-\mu}(\bm k)+ d \bm \Delta_{ba}^{\nu}(\bm k)\right)\label{eq:sfw_dia},\\
    D_{\mu\nu}^{\rm para} =& \sum_{\bm k}\sum_{ab}\dfrac{f_h(E_a(\bm k))-f_h(E_b(\bm k))}{E_a(\bm k)-E_b(\bm k)}J_{ab}^{+\mu}(\bm k)J_{ba}^{+\nu}(\bm k)\label{eq:sfw_para},\notag\\
\end{align}
with
\begin{eqnarray}
    &J_{ab}^{\pm\mu}(\bm k) = \bra{\psi_a(\bm k)}\partial_{k_\mu}H_{\pm}(\bm k)\ket{\psi_b(\bm k)},\\
    &d\bm \Delta_{ab}^{\mu}(\bm k)= \bra{\psi_a(\bm k)}\partial_{k_\mu}\left(
        \begin{array}{cc}
         0& \bm \Delta(\bm k) \\
         \bm \Delta^\dagger(\bm k)& 0
        \end{array}
    \right)\ket{\psi_b(\bm k)}. \label{eq:d_Delta} \notag\\
\end{eqnarray}
Here, $f_h(E) = \{\exp(\beta(E+h))+1\}^{-1}$ is the Fermi distribution function with the magnetic field, and we define the Block diagonal Hamiltonian,
\begin{eqnarray}
   	H_{\pm}(\bm k) =& \left(
		\begin{array}{cc}
			H_0(\bm k)&0\\
			0&\pm H_0(\bm k)
		\end{array}
	\right).
\end{eqnarray}
We also notice that the magnetic field appears only in the Fermi distribution function.

Next, we divide the superfluid weight into the total Fermi-liquid contribution $D^{\rm s:conv}$ and the total geometric contribution $D^{\rm s:geom}$, as $D^{\rm s}=D^{\rm s:conv}+D^{\rm s:geom}$.
For this purpose, based on the previous studies~\cite{liang2017band,kitamura2021superconductivity}, we express the superfluid weight using the normal state Bloch wave functions defined by
\begin{eqnarray}
    H_0(\bm k)\ket{u_n(\bm k)} = \epsilon_{n}(\bm k)\ket{u_n(\bm k)}.
\end{eqnarray}
Eigenstates of the BdG Hamiltonian are represented as,
\begin{eqnarray}
    \ket{\psi_a(\bm k)} = \sum_n\left(
        \begin{array}{cc}
             \phi_n^{a\uparrow}(\bm k)\ket{u_n(\bm k)},  &
             \phi_n^{a\downarrow}(\bm k)\ket{u_n(\bm k)} 
        \end{array}
    \right)^T\label{eq:expansion}.\notag\\
\end{eqnarray}
Here, $\phi_n^{a\uparrow(\downarrow)}(\bm k)$ are the matrix elements of the unitary matrix which diagonalizes the band representation of the BdG Hamiltonian,
\begin{eqnarray}
   &\tilde{H}_\BdG(\bm k) = [\gamma_0 \otimes U(\bm k)] H_\BdG(\bm k) [\gamma_0 \otimes U^\dagger(\bm k)],
\end{eqnarray}
and $U^\dagger(\bm k) = \left(
    \begin{array}{ccc}
     \ket{u_1(\bm k)}& \cdots&\ket{u_f(\bm k)} \\
    \end{array}
    \right)$. 
Inserting Eq.~\eqref{eq:expansion} into Eqs.~\eqref{eq:D_s}-\eqref{eq:d_Delta}, we get
\begin{widetext}
\begin{eqnarray}
  &&D^{\rm s}_{\mu\nu} = D^{\rm conv}_{\mu\nu} + D^{\rm geom}_{\mu\nu} + D^{\rm multi}_{\mu\nu} + D^{\rm gap}_{\mu\nu},\label{sfw_all}\\
  &&D^{\rm conv}_{\mu\nu} = 2\sum_{\bm k}\sum_{nm}C_{nnmm}^{\uparrow\uparrow\downarrow\downarrow}(\bm k) \left\{j_{nn}^{\mu}(\bm k)j_{mm}^{\nu}(\bm k)
  +j_{nn}^{\nu}(\bm k)j_{mm}^{\mu}(\bm k) \right\},\label{conv}\\
  &&D^{\rm geom}_{\mu\nu} = 2\sum_{\bm k}\sum_{n\neq m, p\neq q}C_{nmpq}^{\uparrow\uparrow\downarrow\downarrow}(\bm k) \{j_{nm}^{\mu}(\bm k)j_{pq}^{\nu}(\bm k)
  +j_{nm}^{\nu}(\bm k)j_{pq}^{\mu}(\bm k)\},\label{geom}\\
  &&D^{\rm multi}_{\mu\nu} = 2\sum_{\bm k}\sum_{n,p\neq q}\left[C_{nnpq}^{\uparrow\uparrow\downarrow\downarrow}(\bm k)
  \{j_{nn}^{\mu}j_{pq}^{\nu}(\bm k)
  +j_{nn}^{\nu}(\bm k)j_{pq}^{\mu}(\bm k)\}
  +C_{pqnn}^{\uparrow\uparrow\downarrow\downarrow}(\bm k)
  \{j_{pq}^{\mu}(\bm k)j_{nn}^{\nu}(\bm k)
  +j_{pq}^{\nu}(\bm k)j_{nn}^{\mu}(\bm k)\}\right],\\
  &&D^{\rm gap}_{\mu\nu} =-
  \sum_{\bm k}\sum_{nmpq\sigma}
  S_{\sigma}\left[C_{nmpq}^{\uparrow\downarrow\sigma\sigma}(\bm k)\delta\Delta_{nm}^{\mu}(\bm k)+C_{nmpq}^{\downarrow\uparrow\sigma\sigma}(\bm k)\delta\Delta_{nm}^{\dagger\mu}(\bm k)\right]
  j_{pq}^{\nu}(\bm k),\label{gap}
\end{eqnarray}
\end{widetext}
where $S_{\sigma}$ takes $+(-)$ when $\sigma = \uparrow(\downarrow)$.
Here, 
$J_{nm}^{\mu}(\bm k)$, $\delta\Delta_{nm}^{\mu}(\bm k)$, $\delta\Delta_{nm}^{\dagger\mu}(\bm k)$, and
$C_{nmpq}^{\sigma_1\sigma_2\sigma_3\sigma_4}(\bm k)$
are given by
\begin{align}
  j_{nm}^{\mu}(\bm k) =& \bra{u_{n}(\bm k)}\partial_{k_\mu} H_{0}(\bm{k})\ket{u_{m}(\bm k)},\\
  \delta\Delta_{nm}^{(\dagger)\mu}(\bm k) =& \bra{u_{n}(\bm k)}\partial_{k_\mu} \bm\Delta^{(\dagger)}(\bm k)\ket{u_{m}(\bm k)},\\
  C_{nmpq}^{\sigma_1\sigma_2\sigma_3\sigma_4}(\bm k) =& \sum_{\bm k}\sum_{ab}\dfrac{f_h(E_{a}(\bm k))-f_h(E_{b}(\bm k))}{E_{a}(\bm k)-E_{b}(\bm k)}\notag\\
  \times & \phi_{n}^{a\sigma_1*}(\bm k)\phi_{m}^{b\sigma_2}(\bm k)\phi_{p}^{b\sigma_3*}(\bm k)\phi_{q}^{a\sigma_4}(\bm k)\label{eq:c_sfw}.
\end{align}

In the above formula, the superfluid weight is expressed by summation of the four terms. Now let us briefly discuss each term. First, the conventional term, $D^{\rm conv}_{\mu\nu}$, represents the Fermi-liquid contribution, and this term contains the group velocity of normal state quasiparticles $\partial_{k_\mu} \epsilon(\bm k)$.
In other words, the conventional term is purely the intra-band contribution.
In the case of the isotropic $s$-wave superconductivity, the well-known formula $D^{\rm conv} = n^*/m^*$ is reproduced at $T = 0$~\cite{liang2017band}.

Second, $D^{\rm geom}_{\mu\nu}$, the geometric term, arises from geometrically nontrivial properties of the Bloch electrons because this term contains off-diagonal components of the Berry connection, 
$A_{nm}^{\mu}(\bm k) = i\braket{u_n(\bm k)\vert\partial_{k_\mu} u_m(\bm k)}$.
The geometric term is divided into two parts; one is
\begin{eqnarray}
    D^{\rm geom1}_{\mu\nu} = 2\sum_{\bm k}\sum_{n\neq m}C_{nmmn}^{\uparrow\uparrow\downarrow\downarrow}(\bm k)\{j_{nm}^{\mu}(\bm k)j_{mn}^{\nu}(\bm k)
  +{\rm c.c.} \},\notag\\\label{eq:geom1}
\end{eqnarray}
which reduces to the band-resolved quantum metric, $g_{nm}^{\mu\mu}(\bm k)=\frac{1}{2}\{A_{nm}^{\mu}(\bm k)A_{mn}^{\nu}(\bm k)+{\rm c.c.}\}$, and the other is
\begin{eqnarray}
    &&D^{\rm geom2}_{\mu\nu} =\notag\\
    &&2\sum_{\bm k}\sideset{}{'}\sum_{n\neq m, p\neq q} C_{nmpq}^{\uparrow\uparrow\downarrow\downarrow}(\bm k) \{j_{nm}^{\mu}(\bm k)j_{pq}^{\nu}(\bm k)
  +j_{nm}^{\nu}(\bm k)j_{pq}^{\mu}(\bm k)\},\label{eq:geom2}\notag\\
\end{eqnarray}
which is induced by the inter-band pairing.
Here, $\sum_{n\neq m,p\neq q}^\prime$ takes the sum over $n,m,q,p$ satisfying $n\neq q$ and/or $m\neq p$.
It has been shown that $D^{\rm geom1}$ exactly reduces to the quantum metric in the isolated band limit~\cite{peotta2015superfluidity,liang2017band}.

Third, $D^{\rm multi}_{\mu\nu}$ comes from the multi-gap properties and vanishes in the case of the purely intra-band gap function.
We call this term the multi-gap term.
This term requires both the Berry connection and the group velocity, and in that sense, the multi-gap term also reflects the quantum geometry of Bloch electrons.

Fourth, the gap term, $D^{\rm gap }_{\mu\nu}$, arises owing to the $\bm k$-dependence of the gap function, since this term contains the $\bm k$-derivative of the gap function.
From the viewpoint of the quantum geometry, the gap term can be divided into two terms as $D^{\rm gap}_{\mu\nu}=D^{\rm gap1}_{\mu\nu}+D^{\rm gap2}_{\mu\nu}$ with
\begin{align}
    D^{\rm gap1}_{\mu\nu}=\sum_{\bm k}\sum_{nmp\sigma}
  &S\left[C_{nmpp}^{\uparrow\downarrow\sigma\sigma}(\bm k)\delta\Delta_{nm}^{\mu}(\bm k)\right.\notag\\
  &+\left.C_{nmpp}^{\downarrow\uparrow\sigma\sigma}(\bm k)\delta\Delta_{nm}^{\dagger\mu}(\bm k)\right]
  j_{pp}^{\nu}(\bm k),\label{gap1}\\
    D^{\rm gap2}_{\mu\nu}=\sum_{\bm k}\sum_{nmp\neq q\sigma}
  &S\left[C_{nmpq}^{\uparrow\downarrow\sigma\sigma}(\bm k)\delta\Delta_{nm}^{\mu}(\bm k)\right.\notag\\
  &+\left.C_{nmpq}^{\downarrow\uparrow\sigma\sigma}(\bm k)\delta\Delta_{nm}^{\dagger\mu}(\bm k)\right]
  j_{pq}^{\nu}(\bm k).\label{gap2}
\end{align}
The first term contains the group velocity, while it does not contain the quantum geometry. 
Furthermore, for the band-independent gap function, $\Delta(\bm k) = \bm 1\Delta_0(\bm k)$, this term combined with the conventional term reproduces the Fermi-liquid formula $D^{\rm conv}+D^{\rm gap1} = n^*/m^*$~\cite{kitamura2021superconductivity}. Thus, $D^{\rm gap1}$ can be considered a part of the Fermi-liquid contribution.
On the other hand, the second term $D^{\rm gap2}_{\mu\nu}$ contains the Berry connection 
as the geometric term and multi-gap term are, 
which means that this term has a geometric origin.
Thus, we consider $D^{\rm s:geom}=D^{\rm geom} + D^{\rm mulit} + D^{\rm gap2}$ to be the total geometric contribution, while $D^{\rm s:conv}=D^{\rm conv} + D^{\rm gap1}$ is considered the total Fermi-liquid contribution.

An essential point of the formulas is that effects of the magnetic field are reflected only in $C_{nmpq}^{\sigma_1\sigma_2\sigma_3\sigma_4}(\bm k)$ through the Fermi distribution function.
The magnetic field causes the Zeeman shift of the energy level, that changes each term.
Importantly, the function $C_{nmpq}^{\sigma_1\sigma_2\sigma_3\sigma_4}(\bm k)$ depends on the superconducting symmetry, making a variety in behaviors of each term.
Consequently, we will obtain various superconducting phase diagrams, phase transitions, and in particular, mechanisms of FFLO superconductivity.

\subsection{Conventional Fermi-liquid mechanism of FFLO superconductivity\label{sec:conv_fl_mecha_fflo}}
In this subsection, we review the conventional Fermi-liquid mechanism of FFLO superconductivity.
Since this mechanism is understood based on the Fermi-liquid contribution,  we mainly focus on the conventional term, while we also show the formula of the geometric term.
Here, we consider the isotropic $s$-wave pairing,
\begin{eqnarray}
   \bm \Delta(\bm k) = \Delta\bm 1,
\end{eqnarray}
for simplicity. In this case, because the multi-gap term and the gap term vanish, we can rewrite the superfluid weight by
\begin{widetext}
\begin{align}
    &D_{\mu\nu}^{\rm conv} = \sum_{\bm k}\sum_{\sigma}\sum_{n}
    \left(
    \dfrac{\vert\Delta\vert^2}{E^{\rm s}_{n}(\bm k)^2}f_h^\prime(S_\sigma E^{\rm s}_{n}(\bm k))
    -S_\sigma\dfrac{\vert\Delta\vert^2}{E_{n}^{\rm s}(\bm k)^3}f_h(S_\sigma E^{\rm s}_{n}(\bm k))
    \right)
    \partial_{k_\mu}\epsilon_{n}(\bm k)\partial_{k_\nu}\epsilon_{n}(\bm k)
    ,\label{eq:Ds_conv_ts}\\
    &D_{\mu\nu}^{\rm geom} = \dfrac{1}{2}\sum_{\bm k}\sum_{\sigma\sigma^\prime}\sum_{n\neq m}
    \dfrac{f_h(S_\sigma E^{\rm s}_{n}(\bm k))-f_h(S_{\sigma^\prime} E^{\rm s}_{m}(\bm k))}{S_\sigma E^{\rm s}_{n}(\bm k)-S_{\sigma^\prime} E^{\rm s}_{m}(\bm k)}
    \left(
    S_\sigma S_{\sigma^\prime}\dfrac{\vert\Delta\vert^2}{E^{\rm s}_{n}(\bm k)E^{\rm s}_{m}(\bm k)}
    \right)
    (\epsilon_n(\bm k)-\epsilon_m(\bm k))^2g_{nm}^{\mu\nu}(\bm k)
    ,\label{eq:Ds_geom_ts}
\end{align}
\end{widetext}
with $E_{n}^{\rm s}(\bm k) = \sqrt{\epsilon_n(\bm k)^2+\vert\Delta\vert^2}$ and $S_\sigma=\pm1$ for $\sigma=\uparrow\downarrow$ (see Appendix~\ref{sec:sfw_dirivation} for more details).

To see the conventional Fermi-liquid mechanism of FFLO superconductivity, we focus on the low temperature region, i.e. $\vert \Delta \vert \gg T$, and we assume $h > 0$.
In this region, since $f_h(E_n^{\rm s}(\bm k)),f_h^\prime(E_n^{\rm s}(\bm k)) \simeq 0$, the superfluid weight can be written as, 
\begin{widetext}
\begin{align}
\label{eq:sfw_conv_iso_s}
    &D_{\mu\nu}^{\rm conv} = \sum_{\bm k}\sum_{n}
    \left(
    \dfrac{\vert\Delta\vert^2}{E^{\rm s}_{n}(\bm k)^2}f_h^\prime(- E^{\rm s}_{n}(\bm k))
    +\dfrac{\vert\Delta\vert^2}{E_{n}^{\rm s}(\bm k)^3}f_h(- E^{\rm s}_{n}(\bm k))
    \right)
    \partial_{k_\mu}\epsilon_{n}(\bm k)\partial_{k_\nu}\epsilon_{n}(\bm k)
    ,\\
    &D_{\mu\nu}^{\rm geom} = \dfrac{\vert\Delta\vert^2}{2}\sum_{\bm k}\sum_{n\neq m}
    \left(
    \dfrac{f_h(- E^{\rm s}_{m}(\bm k))}{E^{\rm s}_{m}(\bm k)}
    - \dfrac{f_h(- E^{\rm s}_{n}(\bm k))}{E^{\rm s}_{n}(\bm k)}
    \right)
    \dfrac{\epsilon_n(\bm k)-\epsilon_m(\bm k)}{\epsilon_n(\bm k)+\epsilon_m(\bm k)}g_{nm}^{\mu\nu}(\bm k).\label{eq:sfw_geom_iso_s}
\end{align}
\end{widetext}
In the low magnetic field region  $|\Delta|\gg h$, $f_h^\prime(-E_n^{\rm s}(\bm k))$ is nearly zero since $-E_n^{\rm s}(\bm{k})+h\lesssim-|\Delta|$ in the whole Brillouin zone, while the Fermi-sea term proportional to $f_h(-E)$ is always a positive-definite tensor. Thus, the superfluid weight $D_{\mu\nu}^{\rm conv}$ becomes positive definite.
This leads to a well-known conclusion; the BCS state is stable in the low magnetic field region.
However, in the high magnetic field region where $h \approx \vert\Delta\vert$ is satisfied, $f_h^\prime(-E_n^{\rm s}(\bm k))\simeq -\delta(-E_n^{\rm s}(\bm{k})+h)$ in Eq.~\eqref{eq:sfw_conv_iso_s} may contribute with a negative value.
When this contribution dominates over the other, the superfluid weight can be negative, indicating that the BCS state is unstable. This in turn means that the FFLO superconductivity should be realized.
This is an explanation of the conventional Fermi-liquid mechanism of FFLO superconductivity from the perspective of the superfluid weight.

However, as shown in Eq.~\eqref{eq:sfw_geom_iso_s}, the superfluid weight of multi-band superconductors has the contribution from the geometric term.
For instance, in the monolayer FeSe, because the geometric term may be comparable to the conventional term~\cite{kitamura2021superconductivity}, the conventional understanding should be refined.
In the case of the isotropic $s$-wave superconductivity, the geometric term is always positive and may be disadvantageous for the FFLO superconductivity. However, the geometric term is not always positive for the other pairing states, as we will see in the next section. 
Thus, superconducting phase diagrams should be modified by taking into account the geometric contribution.
In Sec.~\ref{sec:spd_fese}, we show various roles of the geometric term in the models of monolayer FeSe with various superconducting symmetries.

\section{
Negative geometric contribution to the superfluid weight
\label{sec:geomtric_sfw_two_band}}
Before going to the analysis of models, we show a general mechanism of negative geometric contribution to the superfluid weight.
The geometric contribution to the superfluid weight is positive in situations discussed in the literature~\cite{peotta2015superfluidity,liang2017band}.
However, the geometric term can be negative in general, reducing the stability of the BCS state.
To illustrate the negative geometric contribution, we consider the minimal model with a two-fold degree of freedom in this section.
Although the geometric term is negative in the following model at zero magnetic field, we derive analytical formulas applicable in the presence of the magnetic field for generality.

When the system has a two-fold degree of freedom, the normal state Hamiltonian and the gap function can be written as, 
\begin{eqnarray}
    H_0(\bm k) &=& \xi(\bm k)\rho_0+ \bm f(\bm k)\cdot\bm \rho
    \label{eq:two},\\
    \bm \Delta(\bm k) &=& \phi(\bm k)\rho_0+\bm d(\bm k)\cdot\bm \rho,\label{eq:twog}
\end{eqnarray}
where $(\rho_0, {\bm \rho})$ are the Pauli matrices for the normal-state degree of freedom such as orbital and sublattice. 
When $\bm{d}(\bm{k})=0$, the situation is similar to that in Sec.~\ref{sec:conv_fl_mecha_fflo}, and therefore, both the geometric and conventional terms are positive in the absence of the magnetic field.

On the other hand, a different situation is realized when the component $\bm d(\bm k)$ is finite and $\phi(\bm k)=0$.
In this case, the intra-band pairing component in the band representation of the gap function, $
    \tilde{\bm \Delta}(\bm k) = U(\bm k)\bm \Delta(\bm k)U^\dagger(\bm k)
$, is proportional to $\rho_z$ because the $\rho_0$ component vanishes.
For simplicity, neglecting the inter-band pairing, we assume the gap function in the band representation, 
$\tilde{\bm \Delta}(\bm k) = \tilde{d}_z(\bm k)\rho_z$.
In contrast to the plain $s$-wave state $\tilde{\bm{\Delta}}(\bm{k})\propto\rho_0$, the gap function has opposite signs on the two bands, whose intriguing feature is captured by the geometric contribution to the superfluid weight.
Indeed, the conventional term and the geometric term are obtained as,
\begin{widetext}
\begin{align}
    &D_{\mu\nu}^{\rm conv} = \sum_{\bm k}\sum_{\sigma}\sum_{n}
    \{(\rho_z)_{nn}\}^2\left(
    \dfrac{\vert\tilde{d}_z(\bm k)\vert^2}{E^{\rm s}_{n}(\bm k)^2}f_h^\prime(S_\sigma E^{\rm s}_{n}(\bm k))
    -S_\sigma\dfrac{\vert\tilde{d}_z(\bm k)\vert^2}{E_{n}^{\rm s}(\bm k)^3}f_h(S_\sigma E^{\rm s}_{n}(\bm k))
    \right)
    \partial_{k_\mu}\epsilon_{n}(\bm k)\partial_{k_\nu}\epsilon_{n}(\bm k)
    ,\label{eq:conv_dz}\\
    &D_{\mu\nu}^{\rm geom} = \dfrac{1}{2}\sum_{\bm k}\sum_{\sigma\sigma^\prime}\sum_{n\neq m}
    (\rho_z)_{nn}(\rho_z)_{mm}\dfrac{f_h(S_\sigma E^{\rm s}_{n}(\bm k))-f_h(S_{\sigma^\prime} E^{\rm s}_{m}(\bm k))}{S_\sigma E^{\rm s}_{n}(\bm k)-S_{\sigma^\prime} E^{\rm s}_{m}(\bm k)}
    \left(
    \dfrac{S_\sigma S_{\sigma^\prime}\vert\tilde{d}_z(\bm k)\vert^2}{E^{\rm s}_{n}(\bm k)E^{\rm s}_{m}(\bm k)}
    \right)
    (\epsilon_n(\bm k)-\epsilon_m(\bm k))^2g_{nm}^{\mu\nu}(\bm k)\label{eq:geom_dz}.
\end{align}
\end{widetext}
The detailed derivation is shown in Appendix~\ref{sec:two_degree}.
Here, only $D^{\rm geom1}$ is finite, and $D^{\rm geom2}$ vanishes since we ignore the inter-band pairing.

The essential point is that the formula for the geometric term $D^{\rm{geom}}$ contains the factor $(\rho_z)_{nn}(\rho_z)_{mm}$, namely, the relative sign of the gap function for the bands $n$ and $m$.
The conventional term is insensitive to the phase of the order parameter, as $\{(\rho_z)_{nn}\}^2 = 1$ and the formula is the same as that shown in Sec.~\ref{sec:conv_fl_mecha_fflo}.
On the other hand, since $(\rho_z)_{nn}(\rho_z)_{mm} = -1$ for $n\neq m$, the sign of the geometric term is opposite to the plain $s$-wave case. 
This consideration reveals the negative geometric contribution even in the absence of the magnetic field, although it is positive in the plain $s$-wave superconducting state.
The negative geometric term in the superfluid weight may be advantageous to realize the FFLO state.
Possibility of such a quantum-geometry-induced FFLO superconductivity is discussed in Sec.~\ref{sec:inter_sublattice}.

\section{Superconducting phase diagram of monolayer FeSe\label{sec:spd_fese}}
Now, we move on to the main result of this paper, the superconducting phase diagram of monolayer FeSe with an in-plane magnetic field.
This section is divided into three parts.
In Sec.~\ref{sec:gao_model}, we explain a minimal model for monolayer FeSe which takes account of glide-mirror symmetry breaking \cite{gao20216hidden}.
In the following two subsections, we show various superconducting phase diagrams for isotropic $s$-wave, extended $s$-wave, and nodeless $d$-wave pairing states, which have been proposed as the symmetry of superconductivity in monolayer FeSe~\cite{huang2017monolayer,yamakawa2017superconductivity,gao20216hidden,khodas2012interpocket,chen2015electron,agterberg2017resilient,ge2019evidence,schrodi2020multichannel,kang2016superconductivity,zhang2016distinctive,bang2019phonon,rademaker2021enchnanced,fan2015plain}. 
The quantum geometry of Bloch electrons strongly impacts the phase diagrams and may open a route to realizing the FFLO superconductivity. Furthermore, the variety in the phase diagram can verify the superconducting symmetry.

\subsection{Gao's model\label{sec:gao_model}}
To discuss the superconducting phase diagram of monolayer FeSe, we adopt the Gao's model~\cite{gao20216hidden}, 
\begin{eqnarray}
	 H_{0}(\bm k) &&=
	\dfrac{h_{A}(\bm k)+h_{B}(\bm k)}{2}\tau_{0}\otimes\rho_{0}
	+h_{xy}(\bm k)\tau_0\otimes\rho_{x}\notag\\
	&&+\dfrac{h_{A}(\bm k)-h_{B}(\bm k)}{2}\tau_{z}\otimes\rho_{0}
	+h_{T}(\bm k)\tau_{x}\otimes\rho_{0},\label{eq:hamiltonian_gao}
\end{eqnarray}
where,
\begin{align}
	&h_{A}(\bm k) = -2(t_2\cos k_x + t_3\cos k_y)-\mu ,\\
	&h_{B}(\bm k) = -2(t_3\cos k_x + t_2\cos k_y)-\mu,\\
	&h_{xy}(\bm k) = -2t_4(\cos k_x + \cos k_y), \label{eq:hxy}\\
	&h_{T}(\bm k) = -2t_1\cos k_x/2 \cos k_y/2. \label{eq:ht}
\end{align}
In Eq.~\eqref{eq:hamiltonian_gao}, $\tau_{\mu}$ and $\rho_{\mu}$ are the Pauli matrices for the sublattice and orbital space. Here, we consider two sublattices and $(d_{xz}$, $d_{yz})$ orbitals of Fe ions, and the total degree of freedom is  $f = 4$.

The bare hopping integrals are $(t_1, t_2, t_3, t_4) = (0.16, 0.04, -0.2, 0.004)$, and the energies are in the unit of $1 {\rm eV}$.
$t_2$ and $t_3$ are intra-orbital hopping integrals in the same sublattice. 
In the bulk iron-based superconductors, the hopping integrals along the $x$ and $y$ directions are equivalent, and $t_2 = t_3$ is satisfied. 
However, in the monolayer FeSe on substrate, the position of Se ions along the $z$-direction 
is different between the two subalattices, which induces the glide-mirror symmetry breaking; this leads to $t_2 \neq t_3$ and makes $\tau_z$ component finite.
Thus, we can not diagonalize the sublattice space by $\bm k$-independent unitary matrix, making the Berry connection between in the sublattice space finite.

Equation~\eqref{eq:ht} with 
$t_1$ is the intra-orbital hopping between the different sublattices.
Although the hopping parameters are different between the $(x+y)/2$ and $(x-y)/2$ directions in iron-based superconductors, the Gao's model ignores the difference for simplicity, and $\rho_z$ components vanish in $H_0(\bm{k})$.
On the other hand, Equation~\eqref{eq:hxy} with $t_4$ represents the inter-orbital hopping in the same sublattice.
Because the $\rho_y$ and $\rho_z$ components are absent in the Hamiltonian, the orbital space is diagonalized by the $\bm k$-independent unitary matrix $(\rho_z+\rho_x)/\sqrt{2}$.
Thus, it turns out that the Berry connection for the orbital space vanishes.

\begin{figure}[tbp]
    \centering
    \includegraphics[width=0.48\textwidth]{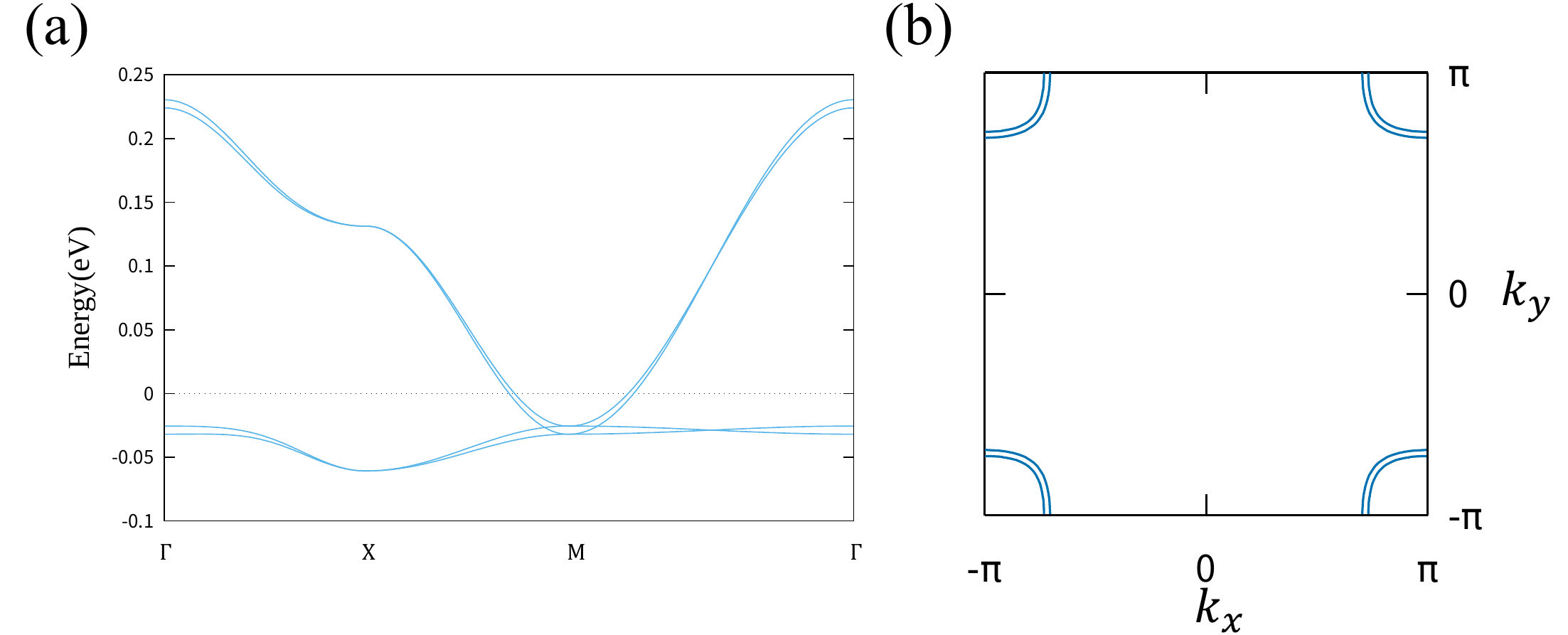}
    \caption{(a) The energy dispersion and (b) the Fermi surface of Gao's model for monolayer FeSe. Here we set the particle number $n=2.08$.}
    \label{fig:gao_model}
\end{figure}

Eigenvalue equation is written as $H_0(\bm k)\ket{u_{n_{\tau}n_{\rho}}(\bm k)}=\epsilon_{n_{\tau}n_{\rho}}(\bm k)\ket{u_{n_{\tau}n_{\rho}}(\bm k)}$ with the single particle's energy $\epsilon_{n_{\tau}n_{\rho}}(\bm k)$ and Bloch wave function $\ket{u_{n_{\tau}n_{\rho}}(\bm k)}$. 
The energy dispersion and the Fermi surface are shown in Fig.~\ref{fig:gao_model}(a) and \ref{fig:gao_model}(b), respectively.
We determine the chemical potential $\mu$ so that the particle number is $n = 2.08$, unless we mention otherwise. 
We find that the Gao's model reproduces the Fermi surface of monolayer FeSe which has only the electron-like Fermi surfaces due to the excessive electron doping~\cite{miyata2015high-temperature,hanzawa2016electric,shiogai2016electric}.

Let us show a simplified form of the Bloch wave functions.
As mentioned before, the orbital space can be diagonalized as,
\begin{eqnarray}
	&&\dfrac{1}{2} \left[\tau_{0}\otimes(\rho_z+\rho_x)\right] H_{0}(\bm k) \left[\tau_{0}\otimes(\rho_z+\rho_x)\right]= \notag\\
	&&\dfrac{h_{A}(\bm k)+h_{B}(\bm k)}{2}\tau_{0}\otimes\rho_{0}+\dfrac{h_{A}(\bm k)-h_{B}(\bm k)}{2}\tau_{z}\otimes\rho_{0}\notag\\
	&&+h_{xy}(\bm k)\tau_0\otimes\rho_{z}
	+h_{T}(\bm k)\tau_{x}\otimes\rho_{0}.
\end{eqnarray}
Since this matrix is block-diagonalized and does not commute with $\tau_x\otimes\rho_0$ and $\tau_z\otimes\rho_0$, we also diagonalize the sublattice space by the unitary matrix which depends on $\bm k$ (see Appendix~\ref{appendix:negative_geom_gao}).
As a result, we can write the Bloch wave function by the tensor product,
\begin{eqnarray}
    \ket{u_{n_\tau n_\rho}(\bm k)}=\ket{\tau_{n_\tau}(\bm k)}\otimes\ket{\rho_{n_\rho}}.
\end{eqnarray}
Therefore, the Berry connection is given by, 
\begin{eqnarray}
    \braket{u_{n_\tau n_\rho}\vert\partial_{k_\mu}u_{n_\tau^\prime n_\rho^\prime}(\bm k)}=\braket{\tau_{n_\tau}(\bm k)\vert\partial_{k_\mu}\tau_{n^\prime_\tau}(\bm k)}\braket{\rho_{n_\rho}\vert\rho_{n^\prime_\rho}}.\notag\\
\end{eqnarray}
We notice that the Berry connection is finite only when $n_{\rho} = n_{\rho}^\prime$; the quantum geometry appears only in the sublattice space.
In Sec.~\ref{sec:inter_sublattice}, we show that this property of the Berry connection can make the geometric contribution to the superfluid weight negative.

To take the mass renormalization effect\cite{maletz2014unusual,aichhorn2010theoretical,yin2011kinetic} into account, we introduce a renormalization factor $z=1/5$ or $1/10$ 
for the normal state Hamiltonian as $z H_0(\bm k)$, which enhances the quantum geometric effect on the FFLO superconductivity.
For the mass renormalization factor $z = 1/5$, we can reproduce the geometric contribution to the superfluid weight in the realistic 10-orbital model of monolayer FeSe derived from the first-principles calculation~\cite{kitamura2021superconductivity} (see Appendix~\ref{sec:geom_gao}).

In the following subsections, solving the gap equation, we set the superconducting transition temperature as $T_{\rm c} = 83$~K at the zero magnetic field\cite{wang2012interface-induced,he2013phase,xu2020spectroscopic}.
The high superconducting transition temperature in monolayer FeSe is considered to be closely related to the film thickness\cite{huang2017monolayer} and the effect of the substrate\cite{lee2014interfacial,song2011molecular,song2019evidence}. 
In particular, an electron doping and a substrate-induced  electron-phonon coupling are expected to be essential for the high transition temperature.
These effects making the monolayer FeSe different from bulk FeSe are phenomenologically contained as the high mean-field transition temperature.

\subsection{Intra-sublattice pairing: isotropic $s$-wave superconductivity\label{sec:pd_s}}

\begin{figure*}[tbp]
    \centering
    \includegraphics[width=1.0\textwidth]{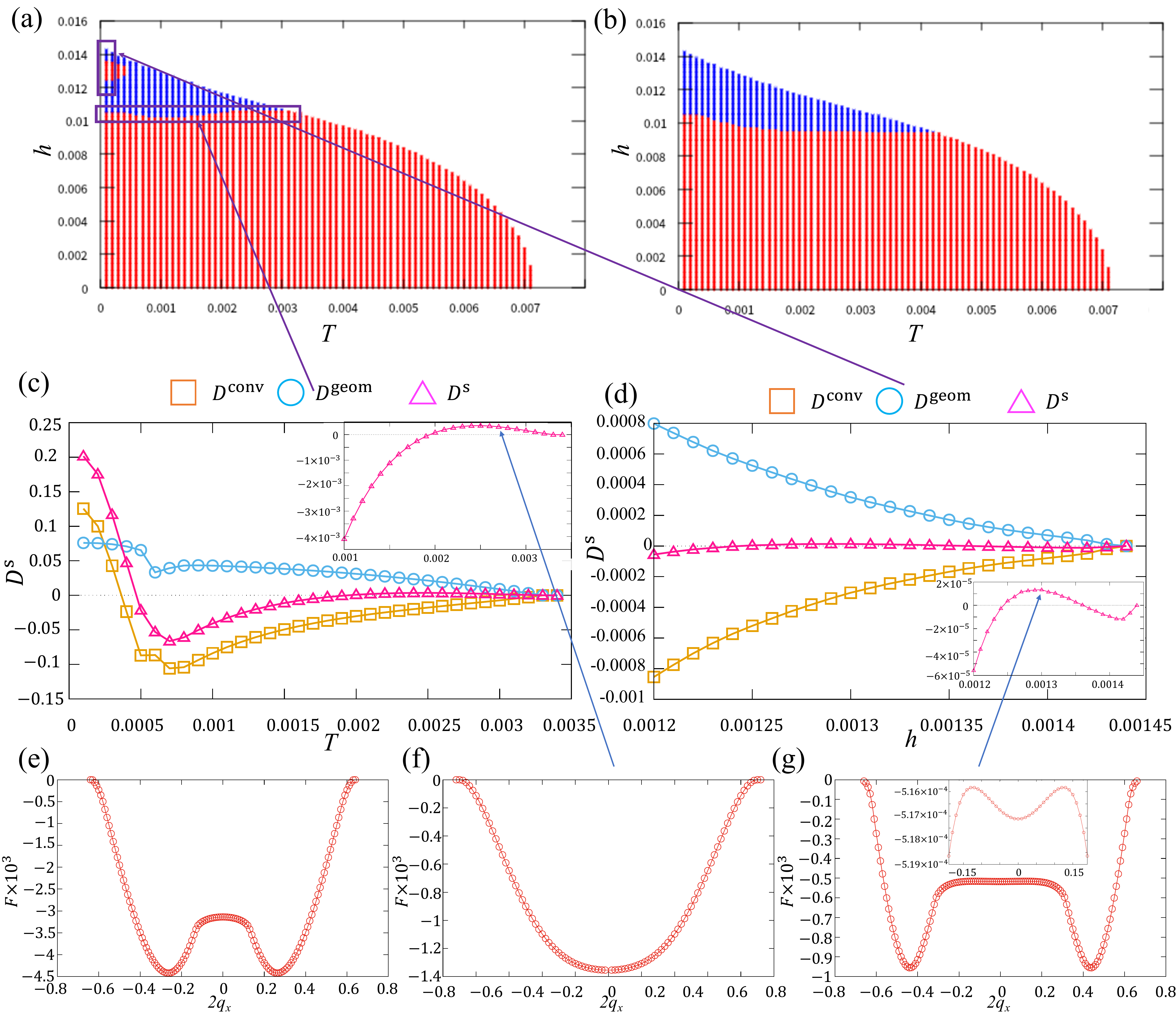}
    \caption{(a) and (b) The temperature-magnetic field superconducting phase diagram of monolayer FeSe with a model for isotropic $s$-wave superconductivity. We set the mass renormalization factor $z=1/5$. The red and blue color show a positive and negative superfluid weight, indicating the (meta-)stability and instability of the BCS state, respectively.
    The latter also indicates that the FFLO state is stable.
    We show the sign of (a) the total superfluid weight, $D^{\rm conv} + D^{\rm geom}$, 
    and (b) the conventional term, $D^{\rm conv}$. By comparing panel~(a) with panel~(b), effects of quantum geometry on the superconducting phase diagram are clarified. 
    The purple boxes in panel~(a) show the region in which quantum geometry induces unusual superconducting phase transitions. (c) The temperature dependence of superfluid weight at $h=0.0105$, corresponding to the lower purple box in panel (a). The orange, blue, and pink lines show the conventional term $D^{\rm conv}$, the geometric term $D^{\rm geom}$, and the total superfluid weight $D^{\rm conv} + D^{\rm geom}$, respectively. The inset shows the region $T>0.001$ in which the total superfluid weight changes the sign. 
    (d) The magnetic field dependence of the superfluid weight at $T=0.0001$, corresponding to the upper purple box in panel (a). The colors indicate the same terms as in panel~(c). The inset is an enlarged figure which shows the sign change of the total superfluid weight. 
    Panels (e), (f), and (g) show the $\bm q$ dependence of the condensation energy, i.e. $F_{\rm s}(\bm q)-F_{\rm n}$. 
    $F_{\rm s}(\bm q)$ and $F_{\rm n}$ are the free energy of the superconducting state and the normal state, respectively.
    (e) We set $T = 0.0001$ and $h=0.011$, in which the FFLO state is stable, consistent with the negative superfluid weight. (f) We set $T = 0.0027$ and $h=0.0105$ in the lower purple box of panel~(a). Although the conventional term is negative $D^{\rm conv} < 0$, the BCS state is stable because the geometric term is positive and $\vert D^{\rm geom}\vert > \vert D^{\rm conv}\vert$. (g) $T = 0.0001$ and $h = 0.013$ corresponding to the red region in the upper purple box of panel~(a). The superfluid weight is positive due to the geometric term as in the case of panel (f). However, the FFLO state is stable and the BCS state is metastable because of higher-order derivative terms such as $D^{\rm ss}$. The inset shows that $F(0)$ is a local minimum.} 
    \label{fig:pd_gao_s}
\end{figure*}

First, we consider the isotropic $s$-wave superconductivity, in which only the intra-sublattice pairing is finite, i.e.
\begin{eqnarray}
    \bm \Delta(\bm k) = \Delta_0\tau_0\otimes\rho_0.
    \label{eq:s-wave}
\end{eqnarray}
This state is obtained when we solve the self-consistent gap equation for an isotropic interaction $V_{ls}({\bm k},{\bm k'}) = -V \delta_{ls}$.
Since all orbitals and sublattices are equivalent in the sense that they are related to each other through four-fold rotational symmetry and mirror symmetry, the 
${\bm k}$-independent gap function is independent of the orbital and sublattice. 
As a result, the gap term and the multi-gap term vanish, and therefore, $D^{\rm conv}=D^{\rm s:conv}$ and $D^{\rm geom}=D^{\rm s:geom}$ in this subsection. 
Consistent with the pairing state in Eq.~\eqref{eq:s-wave}, a weakly $\bm k$-dependent $s$-wave pairing state has been theoretically predicted~\cite{yamakawa2017superconductivity} and supported by an experiment~\cite{fan2015plain}.

In Figs.~\ref{fig:pd_gao_s}(a) and \ref{fig:pd_gao_s}(b), we show the superconducting phase diagram as a function of the temperature and magnetic field. 
Figure~\ref{fig:pd_gao_s}(a)  
shows the sign of the total superfluid weight, i.e. $D^{\rm conv} + D^{\rm geom}$, indicating the (meta-)stability or instability of the BCS state. On the other hand, Fig.~\ref{fig:pd_gao_s}(b) shows the sign of the Fermi-liquid contribution $D^{\rm conv}$. 
Thus, we understand the effect of quantum geometry on the phase diagram by comparing Fig.~\ref{fig:pd_gao_s}(a) with Fig.~\ref{fig:pd_gao_s}(b). 
Because the geometric term has a sizable contribution to the superfluid weight in monolayer FeSe~\cite{kitamura2021superconductivity} and is positive in the isotropic $s$-wave pairing state, 
the BCS state is stabilized by the geometric term.

Here we discuss two features induced by the geometric term.
One is the reentrant BCS phase transition highlighted by the horizontal purple box in Fig.~\ref{fig:pd_gao_s}(a); as increasing the temperature, the phase transition from the BCS state to the FFLO state occurs, and further increase of temperature stabilizes the BCS state again.
The origin of this reentrant behavior is understood by Fig.~\ref{fig:pd_gao_s}(c), which shows the temperature dependence of the superfluid weight.
While the conventional term is negative except for in the low temperature region, the geometric term is positive. Thus, the competition of the two terms,
$D^{\rm geom}$ and  $D^{\rm conv}$, leads to the multiple sign changes of the superfluid weight. 
In particular, the positive superfluid weight near the transition temperature is due to the geometric contribution beyond the Fermi-liquid theory, and it is 
consistent with the CMMCP dependence of the condensation energy, which is shown in Fig.~\ref{fig:pd_gao_s}(f).
In this figure, we see the minimum of free energy at ${\bm q}=0$, consistent with the positive superfluid weight.
We have also confirmed that the FFLO state is stable when $D^{\rm s} < 0$, as shown in Fig.~\ref{fig:pd_gao_s}(e). 

Another intriguing feature is the metastable BCS state, which appears inside the FFLO phase as highlighted by the vertical purple box in Fig.~\ref{fig:pd_gao_s}(a).
To understand this phase, we show the magnetic field dependence of the superfluid weight in Fig.~\ref{fig:pd_gao_s}(d). 
In this parameter range, the conventional term is negative, while the geometric term is positive. As a result of the competition of the two contributions, the total superfluid weight is tiny and changes the sign two times. 
In the intermediate field region, the superfluid weight is positive, suggesting the stable BCS state. 
However, 
the BCS state is metastable in this phase, corresponding to the case of Fig.~\ref{fig:fe_schematic}(c).
Indeed, the CMMCP dependence of the superconducting free energy (Fig.~\ref{fig:pd_gao_s}(g)) is  qualitatively the same as Fig.~\ref{fig:fe_schematic}(c). This means that although the positive superfluid weight ensures the metastable BCS state, higher-order derivative such as $D^{ss}$ makes the FFLO state more stable. 

From the result in Fig.~\ref{fig:pd_gao_s}, we conclude the geometric contribution can affect the superconducting phase diagram.
Especially, the geometric term is essential for the metastability of the BCS state.
We obtain a conventional phase diagram for the FFLO state as in Fig.~\ref{fig:pd_gao_s}(b), when only the Fermi-liquid contribution is taken into account. Thus, the unusual behaviors result from the sizable quantum geometric effect on superconductivity.

\subsection{Inter-sublattice pairing\label{sec:inter_sublattice}}
Next, we show the results for the inter-sublattice pairing states.
The gap function is written as,
\begin{eqnarray}
    &\bm \Delta(\bm k) = \Delta(\bm k)\tau_x\otimes\rho_0,
    \label{eq:D_inter_sublattice}
\end{eqnarray}
where $\bm{\Delta}(\bm{k})$ belongs to an irreducible representation, $A_{1g}$ or $B_{1g}$, corresponding to an extended $s$-wave superconductivity and nodeless $d$-wave superconductivity, respectively. 
These states are stable as solutions of the gap equation for the pairing interaction $V_{ls}({\bm k}, {\bm k'}) = V(\bm k,\bm k^\prime)\left(\tau_x\otimes\rho_0\right)_{ls}$.

Before showing the numerical results of the model calculation, we discuss the possibility of the negative geometric contribution to the superfluid weight in the inter-sublattice pairing state.
For the analogy with the discussion in Sec.~\ref{sec:geomtric_sfw_two_band}, we consider the contribution from the lines on $\vert k_x\vert =\vert k_y\vert$,
where the band representation of the gap function is given by $\tilde{\bm \Delta}(\bm k) = \tilde{\Delta}(\bm k) \tau_z\otimes\rho_0$.
Since it is proportional to $\tau_z$ in the sublattice space, the finite Berry connection arising from the sublattice space gives a negative contribution, similarly to Eq.~\eqref{eq:geom_dz}.
On the other hand, contribution due to the orbital degree of freedom, which may be positive because of $\rho_0$ in the normal part Hamiltonian, vanishes since the Berry connection in the orbital space is absent in the Gao's model.

Indeed, the geometric term is shown to be negative in the following part of this subsection.
We would like to emphasize that the negative geometric contribution is attributed to the Berry connection in the sublattice space $\braket{\tau_{n_\tau}(\bm k)\vert\partial_{k_\mu}\tau_{n^\prime_\tau}(\bm k)}$, which becomes finite owing to the glide-mirror symmetry breaking.
We have confirmed that the geometric term is positive in the model for bulk iron-based superconductors, which preserves glide-mirror symmetry, since the Berry connection due to the sublattice space vanishes (see Appendix~\ref{sec:geom_bulk_iron}). 
Thus, the quantum geometry arising from the glide-mirror symmetry breaking is essential for the negative geometric term of the superfluid weight in the inter-sublattice pairing state.

\subsubsection{Extended $s$-wave superconductivity\label{sec:ex_s_pd}}
\begin{figure}[htbp]
    \centering
    \includegraphics[width=0.48\textwidth]{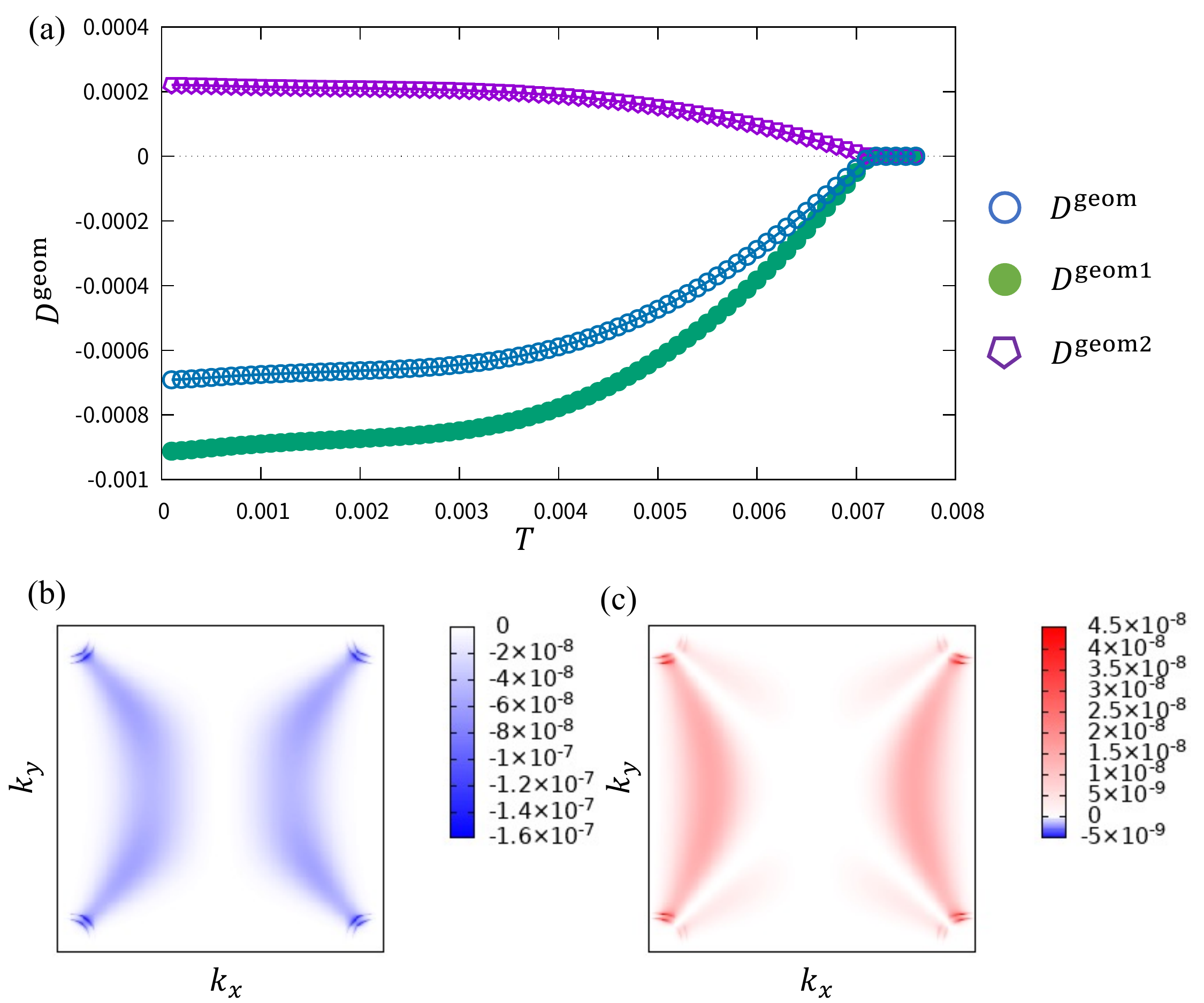}
    \caption{The geometric term of the superfluid weight in the extended $s$-wave pairing state for $h = 0$ and $z = 1/5$.
    (a) The blue, green, and purple lines show the total geometric term $D^{\rm geom}$, $D^{\rm geom1}$, and $D^{\rm geom2}$, respectively.
    Panels (b) and (c) show the $\bm k$-resolved $D^{\rm geom1}$ term and $D^{\rm geom2}$ term, respectively.}
    \label{fig:geom_u-37_mf0}
\end{figure}

\begin{figure*}[tbp]
    \centering
    \includegraphics[width=1.0\textwidth]{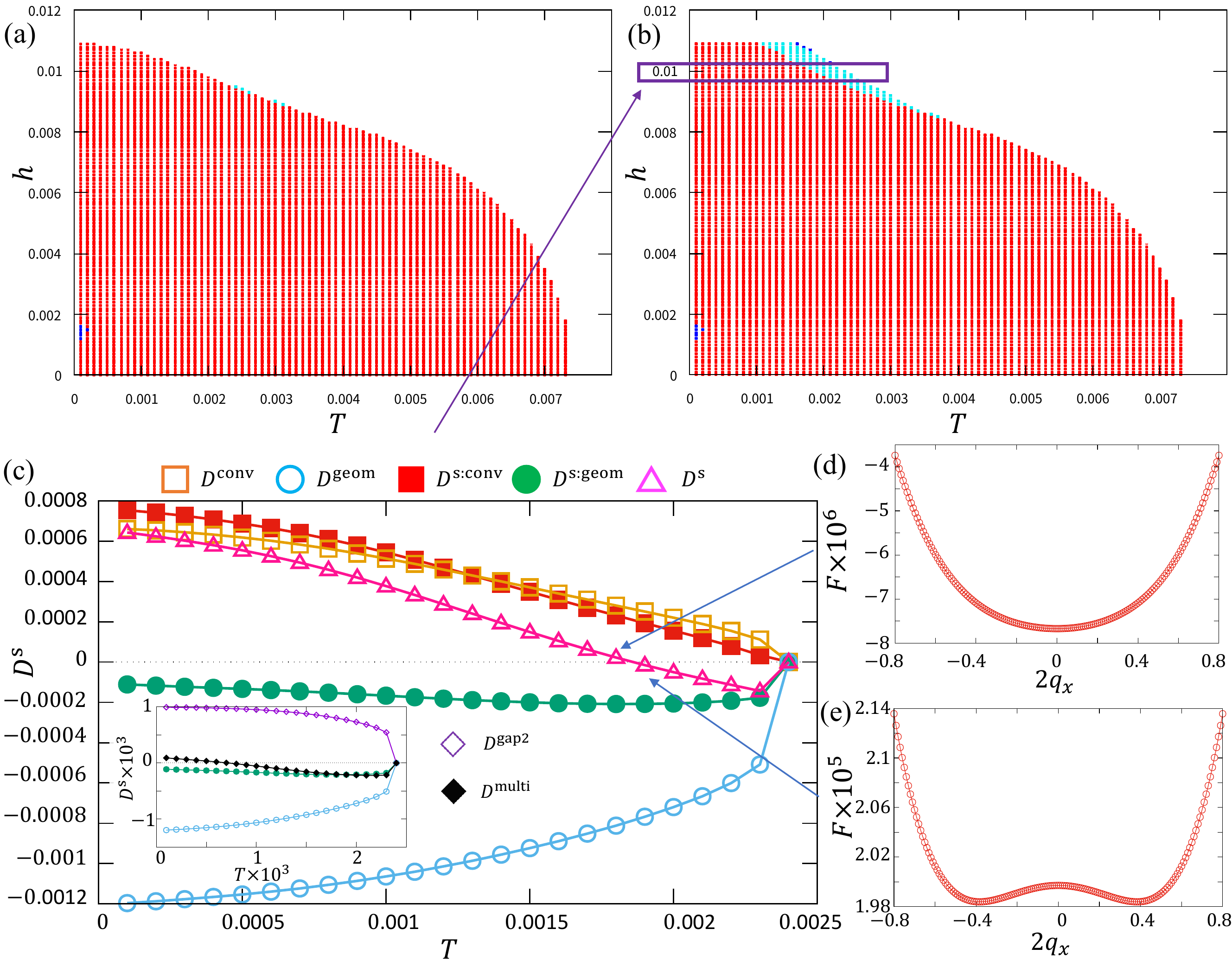}
    \caption{The superconducting phase diagram of the extended $s$-wave pairing state for $z=1/10$.
    We show the sign of the superfluid weight only for thermodynamically stable superconducting states in panel~(a), while we also show metastable superconducting states in panel~(b).
    In the cyan region, the FFLO state with $D^{\rm s} < 0$ is induced by the negative total geometric contribution $D^{\rm s:geom}<0$, although the total Fermi-liquid contribution is positive, $D^{\rm s:conv}>0$.
    (c) Temperature dependence of the superfluid weight at $h=0.01$.
    The green and red lines show $D^{\rm s:geom}$ and $D^{\rm s:conv}$, respectively.
    The other lines show the same terms as in Figs.~\ref{fig:pd_gao_s}(c) and \ref{fig:pd_gao_s}(d).
    The inset shows each term of the total geometric contribution ($D^{\rm s:geom}$, green line).
    The purple, black, and blue lines show $D^{\rm gap2}$, $D^{\rm multi}$, and $D^{\rm geom}$, respectively.
    Panels (d) and (e) show the $\bm q$ dependence of the condensation energy at $T=0.0018$ and $T=0.0019$, respectively.}
    \label{fig:pd_gao_m10_exs}
\end{figure*}

Here, we show the superfluid weight and phase diagram of the extended $s$-wave pairing state, in which Eq.~\eqref{eq:D_inter_sublattice} with $\Delta(\bm k)\propto \cos(\bm k_x/2)\cos(\bm k_y/2)$ belongs to the $A_{1g}$ irreducible representation.
Note that the gap node appears on the $k_x=\pi$ and $k_y=\pi$ lines since we neglect the intra-sublattice pairing.
We conduct numerical calculations for the two mass renormalization factors, $z=1/5$ and $1/10$.

First, we demonstrate the negative geometric term of the superfluid weight. 
Figure~\ref{fig:geom_u-37_mf0}(a) for $h=0$ and $z=1/5$ shows the temperature dependence of the geometric term; the blue, green, and purple lines show $D^{\rm geom}$, $D^{\rm geom1}$, and $D^{\rm geom2}$, respectively.
We see that the total geometric term $D^{\rm geom}$ is indeed negative in contrast to the case of isotropic $s$-wave superconductivity.
This is due to the negative $D^{\rm geom1}$ term as we expected from the discussion in Appendix~\ref{appendix:negative_geom_gao} (see Eq.~\eqref{eq:gao_geom_x=y}).
A positive finite $D^{\rm geom2}$ is induced by the inter-band pairing, indicating that the inter-band pairing stabilizes the superconductivity. However, the magnitude of this term is smaller than the intra-band pairing term $|D^{\rm geom1}|$, and the total geometric term is negative. 
As shown in Figs.~\ref{fig:geom_u-37_mf0}(b) and ~\ref{fig:geom_u-37_mf0}(c), contribution to $D^{\rm geom1}$ ($D^{\rm geom2}$) from each momentum ${\bm k}$ is negative (positive).

The negative geometric term is expected to change the superconducting phase diagram in a different way from the isotropic $s$-wave superconductivity. 
Here, to show the effect of the negative geometric term on the superconducting phase diagram,
we assume large mass enhancement $z=1/10$; the geometric contribution becomes essential as increasing $m^*/m = z^{-1}$ since the Fermi-liquid contribution is suppressed.
We find two FFLO phases in Fig.~\ref{fig:pd_gao_m10_exs}(b); one is in the low magnetic field and low temperature region (blue region), and the other is in the high magnetic field region (cyan region). 
Although the low-field FFLO phase looks unusual, it is stabilized owing to a characteristic feature of monolayer FeSe, as is explained later. 

Here, we discuss the high-field FFLO phase which is shown by the cyan region.
In this phase, the negative superfluid weight, $D^{\rm s}<0$, is induced by the negative total geometric contribution, $D^{\rm s:geom} < 0$, while the total Fermi-liquid contribution is positive, $D^{\rm s:conv} > 0$.
To see this, we show the temperature dependence of the superfluid weight for $h = 0.01$ in Fig.~\ref{fig:pd_gao_m10_exs}(c).
Thus, quantum geometry induces the FFLO superconductivity in the cyan region.
To obtain further insights, we show each term of the total geometric contribution in the inset of Fig.~\ref{fig:pd_gao_m10_exs}(c). 
The geometric term $D^{\rm geom }$ is dominant to the negative contribution.
Furthermore, $D^{\rm geom1}$ is the main origin of the negative geometric contribution, as shown in  Fig.~\ref{fig:geom_u-37_mf0}.
Therefore, the band-resolved quantum metric plays the main role on the quantum-geometry-induced FFLO superconductivity.

The condensation energy $F(\bm q)$ of the high-field FFLO state is shown in Fig.~\ref{fig:pd_gao_m10_exs}(e), and it actually takes the minimum at finite CMMCP.
However, the condensation energy is positive, meaning that the normal state is more stable than the superconducting state.
In this way, the high-field FFLO states obtained in this model are mostly the metastable states.
Actually, most part of the cyan region vanishes in Fig.~\ref{fig:pd_gao_m10_exs}(a), where only the superconducting state with negative condensation energy is illustrated.
We expect that the metastable FFLO state can be verified through the hysteresis measurement.

\begin{figure}[htbp]
    \centering
    \includegraphics[width=0.45\textwidth]{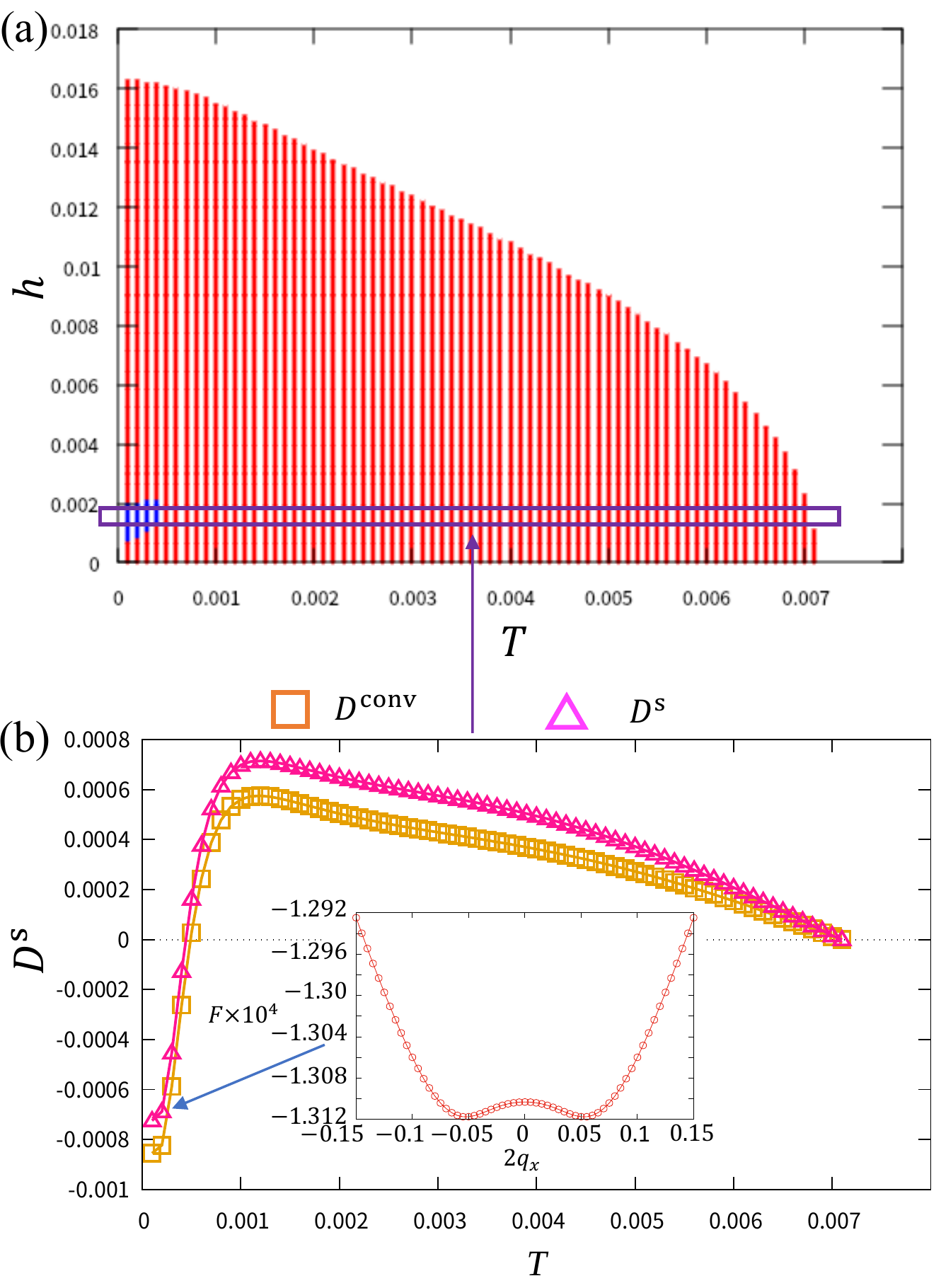}
    \caption{(a) The 
    superconducting phase diagram of the extended $s$-wave pairing state for $z=1/5$.
    We plot only thermodynamically stable states.
    (b) Temperature dependence of the superfluid weight at $h=0.0015$.
    The orange and pink lines show the conventional term $D^{\rm conv}$ and the total superfluid weight $D^{\rm s}$, respectively.
    The inset shows the CMMCP dependence of the superconducting free energy at $T=0.0002$.}
    \label{fig:pd_gao_m5_exs}
\end{figure}

Now we discuss an unusual feature of the superconducting phase diagram, namely, the low-field FFLO phase. 
This phase is induced by the negative conventional term, and therefore, it is enhanced in the phase diagram for $z=1/5$ (Fig.~\ref{fig:pd_gao_m5_exs}(a)).
We find an FFLO phase in the low magnetic field region around $h=0.0015$.
In this region, the FFLO state is more stable than the BCS state because of the negative conventional term, i.e $D^{\rm conv} < 0$ leading to $D^{\rm s} < 0$,
as shown in Fig.~\ref{fig:pd_gao_m5_exs}(b). 
The geometric term is negligible as it is almost canceled by the gap term and the multi-gap term.
The stable FFLO state is also confirmed from the inset of Fig.~\ref{fig:pd_gao_m5_exs}(b), as the superconducting free energy takes the minimum at finite CMMCP. 


The negative conventional term in such a low magnetic field region below the paramagnetic limiting field may originate from unusual properties of the gap function and band structure.
In the monolayer FeSe, the Fermi surfaces exist only near the $M$ point due to electron doping. In the extended $s$-wave pairing state, the gap function is small near the $M$ point because the factor $\cos(k_x/2)\cos(k_y/2)$ vanishes at the $M$ point.
Therefore, the gap size near the $M$ point is comparable to the Zeeman field $h$ even when it is much smaller than the paramagnetic limiting field, leading to a negative superfluid weight and the low-field FFLO superconducting phase. 
On the other hand, the gap function has the maximum at the $\Gamma$ point, around which the incipient bands exist below the Fermi level. Therefore, a sizable contribution to the superconducting condensation energy comes from the bands near the $\Gamma$ point. 
Even when the magnetic field is larger than the gap size near the $M$ point, the superconducting phase is stable owing to the contribution from the incipient bands. 
Thus, we consider that the incipient bands below the Fermi level play a major role if the extended $s$-wave superconductivity occurs in the monolayer FeSe.
Important roles of the incipient bands for 
iron-based superconductors 
were also pointed out in previous studies~\cite{gao20216hidden,bang2014a,bang2019phonon,chen2015electron,linscheid2016high,maier2019effective,rademaker2021enchnanced,mishra2016s}.
We confirmed that the low-field FFLO state is stable even in the presence of a finite intra-sublattice pairing (see Appendix~\ref{sec:fflo_low_magnetir}).

\subsubsection{Nodeless $d$-wave superconductivity\label{sec:d_pd}}

\begin{figure*}[t]
    \centering
    \includegraphics[width=1.0\textwidth]{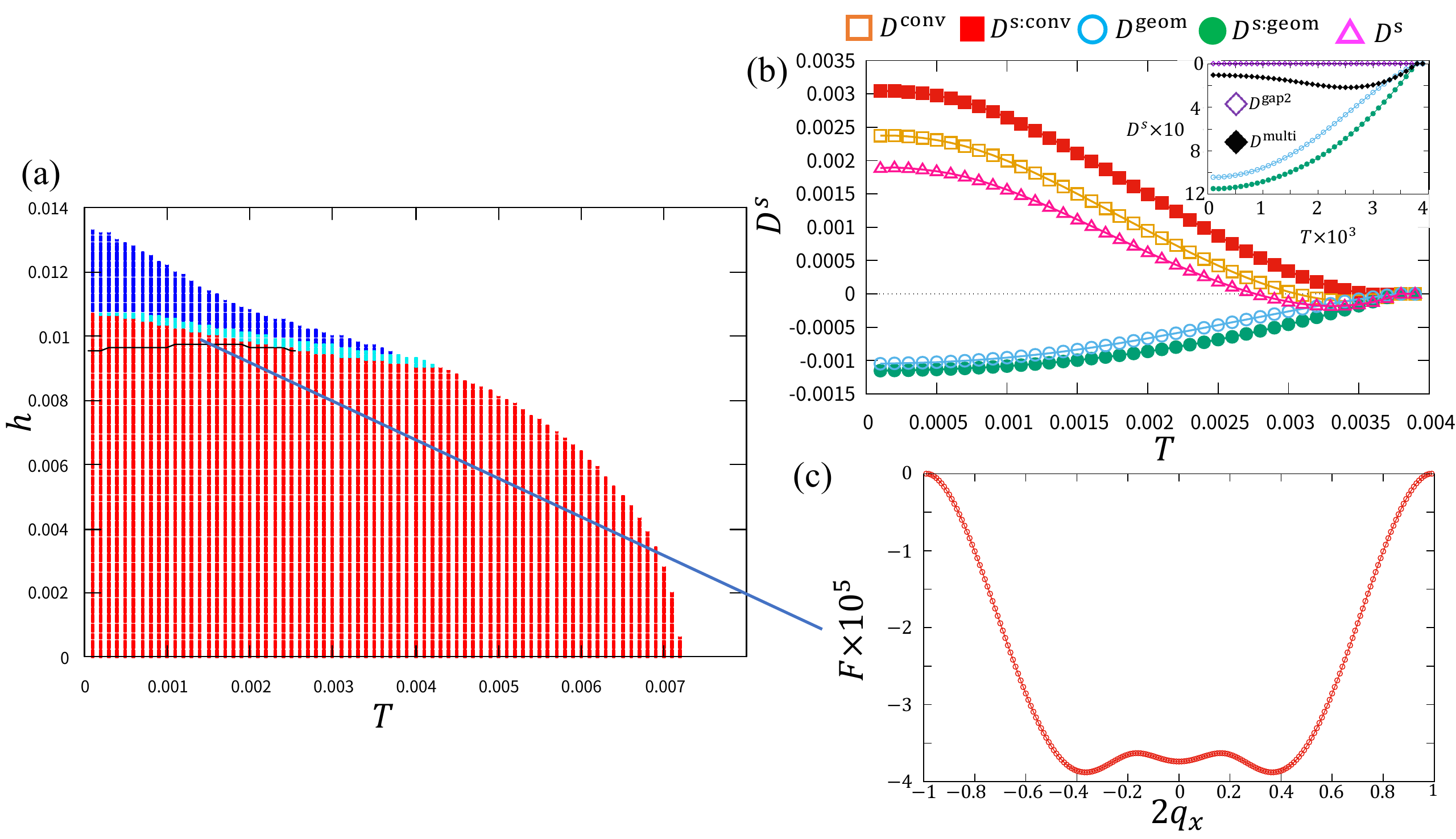}
    \caption{(a) The superconducting phase diagram of the nodeless $d$-wave pairing state for $z=1/5$ and $n=2.07$. Red and blue regions indicate the positive and negative superfluid weight, respectively. The cyan region shows the quantum-geometry-induced FFLO state. 
    In the upper side of the black line, although $D^{\rm s} > 0$, the FFLO state is most stable. 
    Only thermodynamically stable states are plotted. 
    (b) The temperature dependence of the superfluid weight at $h=0.0095$.
    All colors show the same quantities as in Fig.~\ref{fig:pd_gao_m10_exs}(c).
    (c) The CMMCP dependence of the superconducting free energy at $h=0.0095$ and $T=0.0014$, which is in the upper side of the black line. We confirm that the FFLO state is most stable, and the BCS state is metastable.}
    \label{fig:pd_gao_d}
\end{figure*}

Finally, we show the results of the nodeless $d$-wave pairing state, in which Eq.~\eqref{eq:D_inter_sublattice} with $\Delta(\bm k)\propto\sin(k_x/2)\sin(k_y/2)$ belongs to the $B_{1g}$ irreducible representation.
In this case, the geometric term is negative due to the same reason as the extended $s$-wave pairing state, and quantum-geometry-induced FFLO superconductivity is realized. 

In Fig.~\ref{fig:pd_gao_d}(a), we show the superconducting phase diagram 
and find two unique features.
One is shown in the cyan region of Fig.~\ref{fig:pd_gao_d}(a), in which
the FFLO state is induced by the  negative total geometric contribution.
This is confirmed by Fig.~\ref{fig:pd_gao_d}(b), which shows the temperature dependence of the superfluid weight at $h = 0.0095$.
Thus, the quantum-geometry-induced FFLO superconductivity occurs in this region.
As shown in the inset of Fig.~\ref{fig:pd_gao_d}(b), the geometric term ($D^{\rm geom}$, blue line) is dominant in the total geometric contribution ($D^{\rm s:geom}$, green line), which means that the negative $D^{\rm geom}$ plays the main role on the quantum-geometry-induced FFLO superconductivity as in the extended $s$-wave pairing state.
Note that we plot only the thermodynamically stable state while a metastable FFLO state appears as in the case of Fig.~\ref{fig:pd_gao_m10_exs}.

Another feature is illustrated by the black line in Fig.~\ref{fig:pd_gao_d}(a).
In the red region above the black line, 
although the BCS state is metastable as indicated by the positive superfluid weight, the FFLO state is the most stable state as a consequence of negative  higher-order derivatives of the free energy.
Figure~\ref{fig:pd_gao_d}(c) shows the $\bm q$-dependence of the superconducting free energy at $h=0.0095$ and $T = 0.0014$, which illustrates the stable FFLO state and the metastable BCS state.

We note that the low-field FFLO state, which appears in Figs.~\ref{fig:pd_gao_m10_exs} and \ref{fig:pd_gao_m5_exs} for the extended $s$-wave state, does not appear in the nodeless $d$-wave pairing state.
The difference comes from the momentum dependence of the gap magnitude.
In contrast to the extended $s$-wave pairing state, the magnitude of the $d$-wave gap function takes the maximum value at the $M$ point, while it disappears at the $\Gamma$ point. Thus, superconductivity is mainly caused by the Fermi surfaces near the $M$ point, and the incipient bands near the $\Gamma$ point do not play an essential role. 
Therefore, the phase diagram is similar to the conventional one, and the conventional term of the superfluid weight is negative only in the high magnetic field region. We conclude that the negative total geometric contribution enhances the FFLO state in the nodeless $d$-wave pairing state.

\section{summary\label{sec:summary}}
In this paper, we have studied the quantum geometric effect on the superconducting phase diagram.
The quantum geometry appears in the superfluid weight in superconductors, and the superfluid weight determines the thermodynamic stability between the BCS and FFLO states.
Thus, the quantum geometry affects the superconducting phase diagram through the superfluid weight.

To investigate the quantum geometric effect on the FFLO superconductivity, we formulated the superfluid weight in a magnetic field based on the properties of Bloch electrons.
Similarly to the previous studies, the superfluid weight is divided into the Fermi-liquid contribution and the geometric contribution.
Using the obtained formula, we reproduced the conventional FFLO superconductivity induced by the negative Fermi-liquid contribution when the gap size is almost equivalent to the magnitude of the Zeeman field.
Then, based on a simple two-band model, we showed that the sign of the geometric term depends on the superconducting symmetry; this can cause various superconducting phase diagrams.
In contrast to the belief that the geometric contribution is positive, we clarified a ubiquitous mechanism of the negative geometric contribution.

As a benchmark of the quantum geometric effect on the FFLO superconductivity, we investigated models of monolayer FeSe, in which the quantum geometry plays an essential role. We assumed three different superconducting states, isotropic $s$-wave, extended $s$-wave, and nodeless $d$-wave pairing states.

In the isotropic $s$-wave pairing state, the geometric contribution is always positive, and a large positive geometric contribution stabilizes the BCS state.
As a result, two unconventional behaviors are obtained; one is the superconducting phase transition from the FFLO to BCS state as the temperature is increased, and the other is the metastable BCS state in the high magnetic field region.

On the other hand, the geometric contribution to the superfluid weight is negative in the inter-sublattice pairing state, due to the glide-mirror symmetry breaking.
In the extended $s$-wave pairing state, the FFLO state is stable in the low and high magnetic field regions.
In the high magnetic field region, the quantum-geometry-induced FFLO superconductivity occurs because of the negative total geometric contribution.
Although the FFLO state is metastable there, our theoretical prediction can be verified by the hysteresis measurement.
On the other hand, in the low magnetic field region, a negative Fermi-liquid contribution leads to the FFLO superconductivity. This feature is significantly different from the results of other models and due to the unique electronic structure of monolayer FeSe with incipient bands.

In the nodeless $d$-wave pairing state, the geometric term is negative in the same way as in the extended $s$-wave pairing state.
On the other hand, the conventional term becomes negative only in the high magnetic field region, which makes contrast to the extended $s$-wave paring state.
Thus, the FFLO state is stable only in high magnetic field region.
In this case, the negative geometric contribution expands the FFLO region as the quantum-geometry-induced FFLO superconductivity occurs.

We conclude from the results that quantum geometry may play an essential role in superconductors. The relevant phenomena range from the previously studied Meissner effect and BKT transition to the FFLO superconductivity.
Since the superfluid weight is essential for various superconducting phenomena related to the CMMCP, this work may stimulate further studies exploring novel superconducting phenomena.
Interestingly, the geometric contribution shows various behaviors, which depend on the order parameter of superconductivity. Therefore, we expect to see rich phenomena due to quantum geometry, and it can be used to verify unconventional superconducting states.  

\begin{acknowledgments}
We are grateful to J. Ishizuka, S. Kanasugi, and K. Kimura for fruitful discussions. 
We are especially grateful to T. Yamashita for pointing out the possibility of the negative geometric contribution to the superfluid weight.
This work was supported by JSPS KAKENHI (Grants Nos. JP18H01178, JP18H05227, JP19H05825, JP21K18145, JP21K13880, JP22H01181, JP22J22520) and SPIRITS 2020 of Kyoto University.
\end{acknowledgments}

\appendix
\section{Derivation of the superfluid weight\label{sec:sfw_dirivation}}
Here we show the detailed calculation of Sec.~\ref{sec:sfw_mf}.
\subsection{Bogoliubov-de Gennes Hamiltonian with in-plane magnetic field and finite center of mass momenta of Cooper pairs}
First, we derive the Bogoliubov-de Gennes Hamiltonian in the finite center of mass momentum pairing state.
We start from the two-dimensional attractive model with an in-plane magnetic field,
\begin{align}
    \hat{\mathcal{H}} &= \hat{H}+\hat{H}_{\rm zem} +\hat{H}_{\rm int},\\
    \hat{H} &= \sum_{\bm k}\sum_{\sigma}\hat{\bm c}^\dagger_\sigma(\bm k)H_0(\bm k)\hat{\bm c}_{\sigma}(\bm k),\\
    \hat{H}_{\rm zem} &=\sum_{\bm k}\sum_{\sigma\sigma^\prime}(h\sigma_z)_{\sigma\sigma^\prime}\hat{\bm c}^\dagger_{\sigma}(\bm k)\hat{\bm c}_{\sigma^\prime}(\bm k),\\
    \hat{H}_{\rm int} &= \sum_{\bm k\bm k^\prime}\sum_{ls}\hat{c}^\dagger_{l\uparrow}(\bm k+\bm q)\hat{c}_{s\downarrow}^\dagger(-\bm k+\bm q)\notag\\
    &\times V_{ls}(\bm k,\bm k^\prime)
    \hat{c}_{s\downarrow}(-\bm k^\prime+\bm q)\hat{c}_{l\uparrow}(\bm k^\prime+\bm q),
\end{align}
where $\hat{H}$ is the normal state Hamiltonian.
In $\hat{H}_{\rm zem}$, a Zeeman field $(h\sigma_z)_{\sigma\sigma^\prime}$ is induced by the in-plane magnetic field, because we take the spin quantization axis along the magnetic field.
Note that we can freely choose the spin quantization axis due to rotational symmetry in the spin space.
The last term in the Hamiltonian $\hat{H}_{\rm int}$ represents an attractive potential $V_{ls}(\bm k,\bm k^\prime)$ between two electrons with the momentum $\bm k + \bm q$ and $-\bm k + \bm q$.

Applying the BCS mean-field theory to $\hat{H}_{\rm int}$, 
we get the Bogoliubov-de Gennes Hamiltonian of Eq.~\eqref{eq:BdG} in the main text,
\begin{eqnarray}
	\hat{H}_{\BdG}(\bm q) = \sum_{\bm k}\hat{\psi}^\dagger(\bm k,\bm q)H_{\BdG}(\bm k,\bm q)\hat{\psi}(\bm k,\bm q)+{\rm const}.\notag\\
\end{eqnarray}
Here, we explicitly write the constant term as,
\begin{align}
    {\rm const} &= \sum_{\bm k}{\rm tr}\left[H_0(\bm k)-h\bm 1\right]\notag\\
    &-\sum_{\bm k}\sum_{ij}\Delta^\dagger_{ij}(\bm k)\braket{c_{j\downarrow}(-\bm k+\bm q)c_{i\uparrow}(\bm k+\bm q)}.
\end{align}
Thus, the free energy containing the constant term can be written by,
\begin{eqnarray}
    F(\bm q) =  -k_BT\sum_{\bm k}\sum_{a}\ln\left[1+e^{-\beta\left(E_a(\bm k,\bm q)+h\right)}\right]\notag\\
    +{\rm const}.
\end{eqnarray}
For the calculation of the condensation energy, we take into account the constant term.

\subsection{Superfluid weight}
Next, we derive the superfluid weight given by Eq.~\eqref{eq:D_s} in the main text.
The derivative of the free energy with respect to CMMCP is written by,
\begin{eqnarray}
    \partial_{q_\mu}\partial_{q_\nu}F(\bm q) = \sum_{\bm k}\sum_a\left[
    f_h(E_a(\bm k,\bm q))\partial_{q_\mu}\partial_{q_\nu}E_a(\bm k,\bm q)\right.\notag\\
    \left.
    +f_h^\prime(E_a(\bm k,\bm q))\partial_{q_\mu}E_a(\bm k,\bm q)\partial_{q_\nu}E_a(\bm k,\bm q)
    \right].\notag\\
\end{eqnarray}
Using the Hellmann-Feynman`s theorem with respect to $\bm q$,
\begin{eqnarray}
    &&J_{ab}^{\mu}(\bm k,\bm q) =\delta_{a,b}\partial_{q_\mu}E_{a}(\bm k,\bm q)\notag\\
    &&+(E_b(\bm k,\bm q)-E_a(\bm k,\bm q))\braket{\psi_a(\bm k,\bm q)\vert\partial_{q_\mu}\psi_b(\bm k,\bm q)},
\end{eqnarray}
we can rewritte,
\begin{eqnarray}
        &&\partial_{q_\mu}E_{a}(\bm k,\bm q) = J_{aa}^{\mu}(\bm k,\bm q),\\
        &&\partial_{q_\mu}\partial_{q_\nu}E_{a}(\bm k,\bm q) = J_{aa}^{\mu\nu}(\bm k,\bm q)\notag\\
        &&+\sum_{b\neq(a)}\left(
        \dfrac{J_{ab}^{\mu}(\bm k,\bm q)J_{ba}^{\nu}(\bm k,\bm q)}{E_a(\bm k,\bm q)-E_b({\bm k,\bm q})}
        +{\rm c.c}\right),
\end{eqnarray}
where
\begin{align}
        &J_{ab}^{\mu}(\bm k,\bm q)=\bra{\psi_a(\bm k,\bm q)}\partial_{q_\mu}H_{\BdG}(\bm k,\bm q)\ket{\psi_b(\bm k,\bm q)},\notag\\\\
        &J_{aa}^{\mu\nu}(\bm k,\bm q) = \bra{\psi_a(\bm k,\bm q)}\partial_{q_\nu}\partial_{q_\mu}H_{\BdG}(\bm k,\bm q)\ket{\psi_a(\bm k,\bm q)}.
\end{align}
Since we have
\begin{eqnarray}
    &\partial_{q_\mu}H_{\BdG}(\bm k,\bm q) = \partial_{k_\mu}H_{+}(\bm k,\bm q),\\
    &\partial_{q_\nu}\partial_{q_\mu}H_{\BdG}(\bm k,\bm q) = \partial_{k_\nu}\partial_{k_\mu}H_{-}(\bm k,\bm q),
\end{eqnarray}
taking the limit $\bm q\rightarrow0$, we get the superfluid weight,
\begin{eqnarray}
    &&D_{\mu \nu}^{\rm s} = D_{\mu\nu}^{\rm para} + D_{\mu\nu}^{\rm diag},\\
    &&D_{\mu\nu}^{\rm diag} = \sum_{\bm k}\sum_{a}f_h(E_a(\bm k))J_{aa}^{\mu\nu}(\bm k),\\
    &&D_{\mu\nu}^{\rm para} = \sum_{\bm k}\sum_{ab}\dfrac{f_h(E_a(\bm k))-f_h(E_b(\bm k))}{E_a(\bm k)-E_b(\bm k)}J_{ab}^{+\mu}(\bm k)J_{ba}^{+\nu}(\bm k),\notag\\
\end{eqnarray}
with
\begin{eqnarray}
    J_{ab}^{\pm\mu}(\bm k) = \bra{\psi_a(\bm k)}\partial_{k_\mu}H_{\pm}(\bm k)\ket{\psi_b(\bm k)},\\
    J_{ab}^{\mu\nu}(\bm k) = \bra{\psi_a(\bm k)}\partial_{k_\nu}\partial_{k_\mu}H_{-}(\bm k)\ket{\psi_b(\bm k)}.
\end{eqnarray}
By using the Hellmann-Feynman`s theorem with respect to $\bm k$,
\begin{align}
    J_{ab}^{-\mu}(\bm k)+d \bm \Delta_{ab}^{\mu}(\bm k) &=\delta_{ab}\partial_{k_\mu}E_{a}(\bm k) \notag\\
    &+\left(E_b(\bm k)-E_a({\bm k})\right) \braket{\psi_a(\bm k)\vert\partial_{k_\mu}\psi_b(\bm k)},
\end{align}
the diamagnetic term $D^{\rm diag}_{\mu\nu}$ can be rewritten as,
\begin{eqnarray}
    D_{\mu\nu}^{\rm diag} = -\sum_{\bm k}\sum_{ab}\dfrac{f_h(E_a(\bm k))-f_h(E_b(\bm k))}{E_a(\bm k)-E_b(\bm k)}\notag\\
    \times J_{ab}^{-\mu}(\bm k)\left(J_{ba}^{-\nu}(\bm k)+ d \bm \Delta_{ba}^{\nu}(\bm k)\right).
\end{eqnarray}
Thus, we get Eqs.~\eqref{eq:sfw_dia} and \eqref{eq:sfw_para} in the main text.

\subsection{The case of $\Delta(k) = \Delta 1$}
Here, we derive the superfluid weight in the isotropic $s$-wave pairing state,
\begin{eqnarray}
   \bm \Delta(\bm k) = \Delta \bm 1.
\end{eqnarray}
In this case, the gap term vanishes, and the BdG Hamiltonian in the band representation is written as,
\begin{eqnarray}
   \tilde{H}_\BdG(\bm k) 
   &=&\left(
   \begin{array}{cc}
    \bm \epsilon(\bm k)  & \Delta \bm 1 \\
    \Delta \bm 1  & -\bm \epsilon(\bm k)
   \end{array}
   \right) + h\bm 1,
\end{eqnarray}
where
\begin{eqnarray}
   \bm \epsilon(\bm k) = U(\bm k)H_0(\bm k)U^\dagger(\bm k).
\end{eqnarray}
We can easily obtain the eigenvector of the Hamiltonian,
\begin{eqnarray}
   &\ket{\psi_a(\bm k)} = \sum_{n}\left(
    \begin{array}{c}
         \left(\delta_{a,n}u_n(\bm k)-\delta_{a,n+f}v_n(\bm k)\right) \ket{u_n(\bm k)}\\
         \left(\delta_{a,n}v^*_n(\bm k)+\delta_{a,n+f}u_n(\bm k)\right) \ket{u_n(\bm k)}
    \end{array}
   \right)\label{eq:psi_egien_exp}.\notag\\
 \end{eqnarray}
Here, $u_{n}(\bm k)$ and $v_{n}(\bm k)$ are
given by
\begin{eqnarray}
   u_n(\bm k) = \dfrac{1}{\sqrt{2}}\sqrt{1+\dfrac{\epsilon_n(\bm k)}{E^{\rm s}_n(\bm k)}},\\ v_n(\bm k) = \dfrac{\Delta}{\vert\Delta\vert\sqrt{2}}\sqrt{1-\dfrac{\epsilon_n(\bm k)}{E^{\rm s}_n(\bm k)}},
\end{eqnarray}
and $E_a(\bm k) = \sum_{n}\left(\delta_{a,n}-\delta_{a,n+f}\right)E_n^{\rm s}(\bm k)$ with $E_{n}^{\rm s}(\bm k) = \sqrt{\epsilon_n(\bm k)^2+\vert\Delta\vert^2}$.
Thus, we get the coefficient in the superfluid weight formula, Eq.~\eqref{eq:c_sfw}, as
\begin{eqnarray}
    C_{nmpq}^{\uparrow\uparrow\downarrow\downarrow}(\bm k) = \delta_{n,q}\delta_{m,p}\sum_{\bm k}\sum_{ab}\dfrac{f_h(E_{a}(\bm k))-f_h(E_{b}(\bm k))}{E_{a}(\bm k)-E_{b}(\bm k)}\notag\\
    \times(\delta_{a,n}-\delta_{a,n+f})(\delta_{b,m}-\delta_{b,m+f})u_n(\bm k)v_n^*(\bm k)u_m(\bm k)v_m(\bm k)\label{eq:c_iso_s}.\notag\\
\end{eqnarray}
Since Eq.~\eqref{eq:c_iso_s} becomes finite only for $n=q$ and $m=p$, we find that the multi-gap term vanishes.
Inserting Eq.~\eqref{eq:c_iso_s} into Eqs.~\eqref{conv} and \eqref{geom}, we obtain Eqs.~\eqref{eq:Ds_conv_ts} and \eqref{eq:Ds_conv_ts} in the main text.

\section{Derivation of the negative geometric contribution\label{sec:two_degree}}
We derive the superfluid weight in the model of Sec.~\ref{sec:geomtric_sfw_two_band} when the gap function in the band representation is
\begin{eqnarray}
    \tilde{\bm \Delta}(\bm k) = \tilde{d}(\bm k)\rho_z.
\end{eqnarray}
The eigenvector of the BdG Hamiltonian is obtained as,
\begin{align}
    &\ket{\psi_a(\bm k)} \notag\\
    &= \sum_n\left(
        \begin{array}{c}
        \left(\delta_{a,n}u_n(\bm k) - (\rho_z)_{nn}\delta_{a,n+2}v_n(\bm k)\right)\ket{u_n(\bm k)}\\
         \left((\rho_z)_{nn}\delta_{a,n}v_n(\bm k) + \delta_{a,n+2}u_n(\bm k)\right)\ket{u_n(\bm k)} 
        \end{array}
    \right).\label{eq:eigen_dz}
\end{align}
where
\begin{align}
   &u_n(\bm k) = \dfrac{1}{\sqrt{2}}\sqrt{1+\dfrac{\epsilon_n(\bm k)}{E^{\rm s}_n(\bm k)}},\\
   &v_n(\bm k) = \dfrac{\tilde{d}_z(\bm k)}{\vert\tilde{d}_z(\bm k)\vert\sqrt{2}}\sqrt{1-\dfrac{\epsilon_n(\bm k)}{E^{\rm s}_n(\bm k)}},
\end{align}
and 
$E^{\rm s}_n(\bm k) = \sqrt{\epsilon_n(\bm k)^2+ \vert\tilde{d}_z(\bm k)\vert^2}$.
Here, we use the eigenvalue equation,
\begin{eqnarray}
    &\bm f(\bm k)\cdot \bm \rho \ket{u_{1(2)}(\bm k)} = (-)\vert\bm f(\bm k)\vert\ket{u_{1(2)}(\bm k)},\\
    &\epsilon_{1(2)}(\bm k) = \xi(\bm k)\pm\vert\bm f(\bm k)\vert.
\end{eqnarray}
We would like stress that the eigenvector Eq.~\eqref{eq:eigen_dz} contains $(\rho_z)_{nn}$ different from Eq.~\eqref{eq:psi_egien_exp} since the gap function is proportional to $\rho_z$.
As a consequence, the coefficient in the superfluid weight formula Eq.~\eqref{eq:c_sfw} is written as,
\begin{eqnarray}
    &&C_{nmpq}^{\uparrow\uparrow\downarrow\downarrow}(\bm k) = \delta_{n,q}\delta_{m,p}\sum_{\bm k}\sum_{ab}\dfrac{f_h(E_{a}(\bm k))-f_h(E_{b}(\bm k))}{E_{a}(\bm k)-E_{b}(\bm k)}\notag\\
    &&\times(\delta_{a,n}-\delta_{a,n+f})(\delta_{b,m}-\delta_{b,m+f})u_n(\bm k)v_n^*(\bm k)u_m(\bm k)v_m(\bm k)\notag\\
    &&\times(\rho_z)_{nn}(\rho_z)_{mm}.
\end{eqnarray}
Using this, we get the conventional term Eq.~\eqref{eq:conv_dz} and the geometric term Eq.~\eqref{eq:geom_dz} in the main text.
It is emphasized that the formula of the geometric term is different from that for the isotropic $s$-wave pairing state Eq.~\eqref{eq:Ds_geom_ts}, because of the $\rho_z$ component arising from the gap function. This is the origin of the negative geometric contribution discussed in Sec.~\ref{sec:geomtric_sfw_two_band}.

\section{Negative geometric contribution in the Gao's model for inter-sublattice pairing state \label{appendix:negative_geom_gao}}
We show the negative geometric contribution to the superfluid weight in the Gao's model.
As mentioned in the main text, in the Gao's model Eq.~\eqref{eq:hamiltonian_gao}, the orbital space can be diagonalized with ${\bm k}$-independent unitary transformation. 
Thus, we can also diagonalize the sublattice space by the unitary matrix proportional to
$
    (a(\bm k)\tau_z+b(\bm k)\tau_x)
$
with $a(\bm k) = h_T(\bm k)$ and $b(\bm k) = \epsilon_{A1}(\bm k)-h_A(\bm k)-h_{xy}(\bm k)$.
Here, 
\begin{eqnarray}
    \epsilon_{A1}(\bm k)&=&\dfrac{h_A(\bm k)+h_B(\bm k)+2h_{xy}(\bm k)}{2}\notag\\
    &+&\dfrac{\sqrt{(h_A(\bm k)-h_B(\bm k))^2+4h_T(\bm k)^2}}{2},
\end{eqnarray}
is an energy eigenvalue of the Gao's model.
As a result, the unitary matrix which diagonalizes the Gao's model is obtained as,
\begin{align}
    &U_{\rm Gao}^\dagger(\bm k) = 
    \notag\\&
    \dfrac{1}{\sqrt{2(a(\bm k)^2+b(\bm k)^2)}}
    (a(\bm k)\tau_z+b(\bm k)\tau_x)\otimes(\rho_z+\rho_x).
\end{align}
After the above unitary transformation, the band representation of the gap function in the inter-sublattice pairing state is obtained as, 
\begin{eqnarray}
    &&\tilde{\bm \Delta}(\bm k) 
    = \Delta(\bm k)
    \notag\\&&
    \times\left(\dfrac{b(\bm k)^2-a(\bm k)^2}{a(\bm k)^2+b(\bm k)^2}\tau_x+\dfrac{2b(\bm k)a(\bm k)}{a(\bm k)^2+b(\bm k)^2}\tau_z\right)\otimes\rho_0.~\label{eq:gap_inter_sub}
\end{eqnarray}
The gap function is orbital-independent as it is proportional to $\rho_0$. On the other hand, the $\tau_z$ component in the sublattice space is expected to give a negative geometric term as we discussed in Sec.~\ref{sec:geomtric_sfw_two_band}.
Analogy with the discussion in Sec.~\ref{sec:geomtric_sfw_two_band} becomes clearer by focusing on the contribution from the lines on $\vert k_x\vert =\vert k_y\vert$,
where, $a(\bm k)^2 = b(\bm k)^2$ and $\tilde{\bm \Delta}(\bm k) = \tilde{\Delta}(\bm k) \tau_z\otimes\rho_0$ with $\tilde{\Delta}(\bm k)=\Delta(\bm k)\dfrac{2b(\bm k)a(\bm k)}{a(\bm k)^2+b(\bm k)^2}$.
In this case, $D^{\rm geom2}=0$, and
contribution to $D^{\rm geom1}$ is obtained as,
\begin{widetext}
\begin{eqnarray}
\dfrac{1}{2}
&&
\sum_{(n_\tau,n_\rho)\neq(m_\tau,m_\rho)}\sum_{\sigma\sigma^\prime}(\tau_z)_{n_\tau n_\tau}(\tau_z)_{m_\tau m_\tau}
    \dfrac{f_h(S_\sigma E^{\rm s}_{(n_\tau,n_\rho)}(\bm k))-f_h(S_{\sigma^\prime} E^{\rm s}_{(m_\tau,m_\rho)}(\bm k))}{S_\sigma E^{\rm s}_{(n_\tau,n_\rho)}(\bm k)-S_{\sigma^\prime} E^{\rm s}_{(m_\tau,m_\rho)}(\bm k)}\notag\\
    &&\times\left(
    \dfrac{S_\sigma S_{\sigma^\prime}\vert\tilde{\Delta}(\bm k)\vert^2}{E^{\rm s}_{(n_\tau,n_\rho)}(\bm k)E^{\rm s}_{(m_\tau,m_\rho)}(\bm k)}
    \right)
    \left\{\epsilon_{(n_\tau,n_\rho)}(\bm k)-\epsilon_{(m_\tau, m_\rho)}(\bm k)\right\}^2 g_{(n_\tau,n_\rho)(m_\tau,m_\rho)}^{\mu\nu}(\bm k)\label{eq:gao_geom_x=y},
\end{eqnarray}
\end{widetext}
similarly to Eq.~\eqref{eq:geom_dz}.
Because the Berry connection is finite only for $n_{\rho}=m_{\rho}$, we have only to consider the contribution in this condition. Then, the constraint $(n_\tau,n_\rho)\neq(m_\tau,m_\rho)$ leads to $n_{\tau} \neq m_{\tau}$, and therefore,
$(\tau_z)_{n_\tau n_\tau}(\tau_z)_{m_\tau m_\tau}=-1$.
Thus, the geometric term arising from the lines $\vert k_x\vert =\vert k_y\vert$ have to be negative.

\section{Superfluid weight in the Gao's model for isotropic $s$-wave pairing state at $h=0$\label{sec:geom_gao}}

\begin{figure}[htbp]
    \centering
    \includegraphics[width=0.5\textwidth]{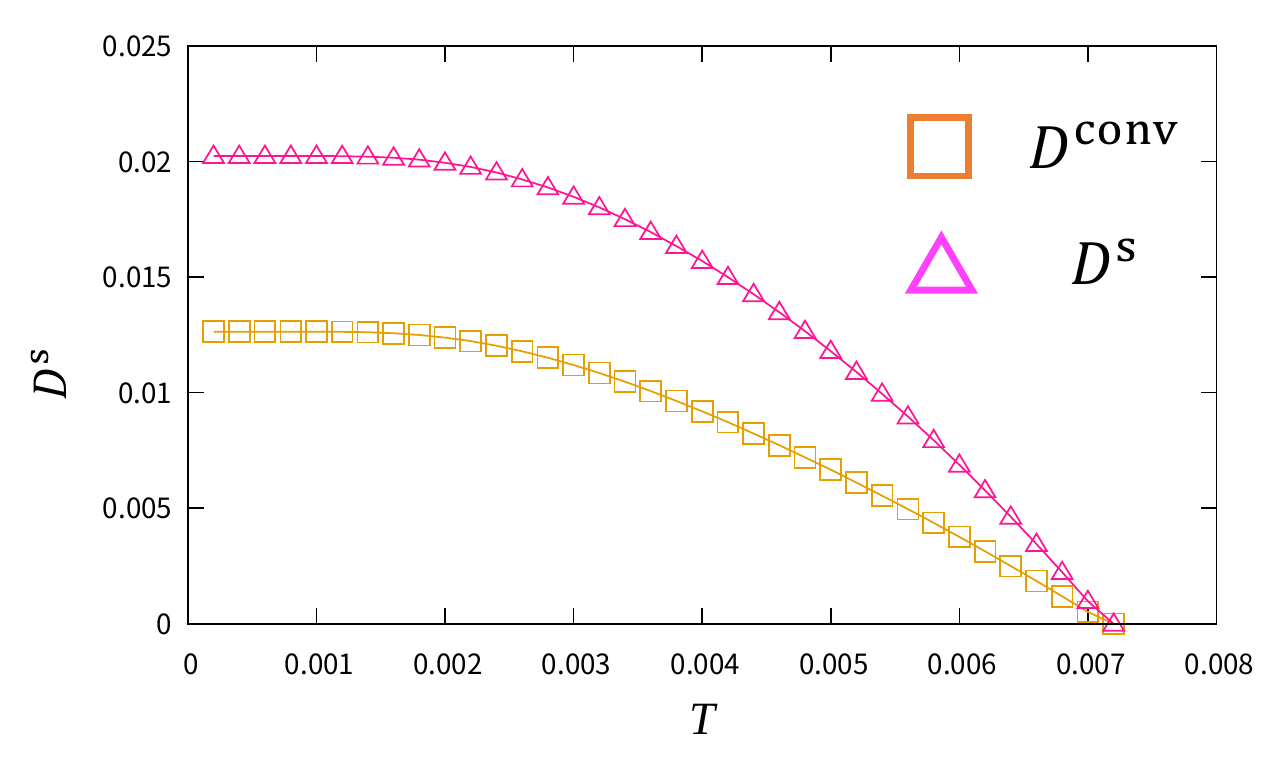}
    \caption{The superfluid weight of the Gao's model in the isotropic $s$-wave pairing state at zero magnetic field.
    We set mass enhancement factor $z=1/5$.
    The orange and pink lines show the conventional term and the total superfluid weight, respectively.
    Thus, the difference between the pink and orange lines shows the geometric contribution to the superfluid weight.}
    \label{fig:sfw_gao_u-077}
\end{figure}

In the previous study~\cite{kitamura2021superconductivity}, the superfluid weight of realistic 10-orbital model for monolayer FeSe was calculated, which reveals monolayer FeSe to have sizeable geometric contribution to the superfluid weight.
To study the quantum geometric effect on FFLO superconductivity in monolayer FeSe, we adopt the Gao's model as an effective model. Thus, it should be verified that sizable geometric contribution comparable to the realistic 10-orbital model is reproduced in the Gao's model.

Here, taking into account the mass enhancement, we show that the Gao's model qualitatively reproduces the geometric contribution obtained in the realistic 10-orbital model.
Since the calculations in Ref.~\onlinecite{kitamura2021superconductivity} were carried out at zero magnetic field, 
we show the superfluid weight of Gao's model at zero magnetic field in Fig.~\ref{fig:sfw_gao_u-077}.
The isotropic $s$-wave pairing state is assumed, and the mass enhancement is set as $z=1/5$.
The orange and pink lines show the conventional term $D^{\rm conv}$ and the total superfluid weight $D^{\rm s}$, respectively; $D^{\rm s}-D^{\rm conv}$ shows the geometric contribution.

Consistent with the realistic 10-orbital model of monolayer FeSe~\cite{kitamura2021superconductivity}, we find a sizeable geometric contribution in the Gao's model (for example, compare Fig.~\ref{fig:sfw_gao_u-077} with Fig.~2 in Ref.~\onlinecite{kitamura2021superconductivity}).
Thus, we adopt the Gao's model with mass enhancement factor $z=1/5$ as an effective model, and it is considered valid for studying the quantum geometric effect on FFLO superconductivity.
We also assume a large mass enhancement factor $z=1/10$ as an extreme case in Sec.~\ref{sec:ex_s_pd}, to demonstrate quantum-geometry-induced FFLO superconductivity in the extended $s$-wave pairing state.




\section{Geometric contribution to superfluid weight in bulk iron-based superconductors with glide-mirror symmetry\label{sec:geom_bulk_iron}}

In this Appendix, we discuss the sign of the geometric term in bulk iron-based superconductors.
The Fermi surfaces of iron-based superconductors are mainly constructed by the $d_{xz}$- and $d_{yz}$-orbitals of iron atoms. 
For comparison with the Gao's model for monolayer FeSe, we consider the following two-orbital two-sublattice model,
\begin{align}
    &H_0(\bm k) = \tau_0\otimes H_{\rm sub}(\bm k) + \tau_x\otimes H_{\rm T}(\bm k)\label{eq:hamilotonian_iron},\\
    &H_{\rm sub}(\bm k) = h_0(\bm k)\rho_0 + h_{xy}(\bm k)\rho_x,\\
    &H_{\rm T}(\bm k) = \dfrac{h_{Tx}(\bm k)+h_{Ty}(\bm k)}{2} \rho_{0} +\dfrac{h_{Tx}(\bm k)-h_{Ty}(\bm k)}{2} \rho_{z},\notag\\
\end{align}
which satisfies the symmetry of canonical iron-based superconductors.
For example, the Raghu`s model~\cite{raghu2008minimal} for the bulk iron-based superconductors has the same form as Eq.~\eqref{eq:hamilotonian_iron}.
Here, 
we adopt the two-dimensional model, since most iron-based superconductors are quasi-two-dimensional systems.
Owing to the four-fold rotational symmetry, the relationship, $h_{Tx}(k_x, k_y) = h_{Ty}(-k_y,k_x)$, must be satisfied which implies $h_{Tx}(\bm k) \neq h_{Ty}(\bm k)$.
We do not specify the details of hopping parameters. 

There are two main differences between the Gao's model Eq.~\eqref{eq:hamiltonian_gao} and Eq.~\eqref{eq:hamilotonian_iron}.
One is that the intra-sublattice hopping terms are sublattice-independent in Eq.~\eqref{eq:hamilotonian_iron}, while they are different between the two sublattices in the Gao's model Eq.~\eqref{eq:hamiltonian_gao} due to glide-mirror symmetry breaking.
The other is that although the inter-sublattice hoppings are different between the $d_{xz}$ and $d_{yz}$ orbitals, i.e. $h_{Tx}(\bm k) \neq h_{Ty}(\bm k)$ in the canonical iron-based superconductors, the Gao's model ignores the difference.

In contrast to the Gao's model, the sublattice space of the model  Eq.~\eqref{eq:hamilotonian_iron} can be diagonalized by the $\bm k$-independent unitary matrix $\dfrac{\tau_x+\tau_z}{\sqrt{2}}$, and we get
\begin{widetext}
\begin{eqnarray}
&&\left[\dfrac{\tau_x+\tau_z}{\sqrt{2}}\otimes\rho_0\right]H_0(\bm k)\left[\dfrac{\tau_x+\tau_z}{\sqrt{2}}\otimes\rho_0\right]\notag\\
&&=	\left(
			\begin{array}{cc}
				(h_0+\dfrac{h_{Tx}+h_{Ty}}{2})\rho_0+h_{xy}\rho_x+\dfrac{h_{Tx}-h_{Ty}}{2}\rho_z&0\\
				0&(h-\dfrac{h_{Tx}+h_{Ty}}{2})\rho_0+h_{xy}\rho_x-\dfrac{h_{Tx}-h_{Ty}}{2}\rho_z
			\end{array}
	\right).
\end{eqnarray}
\end{widetext}
Here, we suppressed the $\bm k$-dependence.
This Hamiltonian is diagonalized by,
\begin{eqnarray}
&\dfrac{\tau_0+\tau_z}{2}\otimes U_{\rho+}^\dagger+\dfrac{\tau_0-\tau_z}{2}\otimes U_{\rho-}^\dagger,\\
&U_{\rho_+}^\dagger = \dfrac{1}{\sqrt{2}}\left(\begin{array}{cc}
    u & -v \\
    v & u
\end{array}\right),\
U_{\rho_-}^\dagger = \dfrac{1}{\sqrt{2}}\left(\begin{array}{cc}
    v & -u \\
    u & v
\end{array}\right),
\end{eqnarray}
where
\begin{eqnarray}
&u = \sqrt{1+\dfrac{h_{T_x}-h_{T_y}}{2\epsilon_{\rm orb}}},\
v = \sqrt{1-\dfrac{h_{T_x}-h_{T_y}}{2\epsilon_{\rm orb}}},\\
&\epsilon_{\rm orb} = \sqrt{\left(\dfrac{h_{T_x}-h_{T_y}}{2}\right)^2+h_{xy}^2}.
\end{eqnarray}
Therefore, the unitary matrix diagonalizing Eq.~\eqref{eq:hamilotonian_iron} is obtained as,
\begin{eqnarray}
    \dfrac{1}{\sqrt{2}}\left(\begin{array}{cc}
    1 & 0\\
    1 & 0
    \end{array}\right)\otimes U^\dagger_{\rho+}
    +\dfrac{1}{\sqrt{2}}\left(\begin{array}{cc}
    0 & 1\\
    0 & -1
    \end{array}\right)\otimes U^\dagger_{\rho-}\label{eq:iron_unitary}.
\end{eqnarray}
Thus, the Bloch wave function can be written by the tensor product,
\begin{eqnarray}
    \ket{u_{n_\tau n_\rho}(\bm k)} = \ket{\tau_{n_\tau} }\otimes\ket{\rho_{n_\rho}(\bm k)},
\end{eqnarray}
in which only the orbital space depends on the wave vector, $\bm k$.

Then, we consider the inter-sublattice pairing state represented by,
\begin{eqnarray}
    \bm \Delta(\bm k) = \Delta(\bm k)\tau_x\otimes\rho_0.
\end{eqnarray}
Using Eq.~\eqref{eq:iron_unitary}, we get the band representation of the gap function as,
\begin{eqnarray}
    \tilde{\bm \Delta}(\bm k) = \Delta(\bm k)\tau_z\otimes\rho_0.
\end{eqnarray}
Thus, the situation is the same as the case of Eq.~\eqref{eq:gap_inter_sub} at $|k_x| = |k_y|$, and the geometric term has the same form as Eq.~\eqref{eq:gao_geom_x=y}.
However, now the Berry connection is finite only when $n_{\tau}=m_{\tau}$, since the unitary matrix of the sublattice space does not depend on $\bm k$.
Therefore, $(\tau_z)_{n_\tau n_\tau}(\tau_z)_{m_\tau m_\tau}=1$ is satisfied, and the geometric term is supposed to be positive in contrast to the Gao's model.
Thus, we conclude that the negative geometric contribution to the superfluid weight, demonstrated in the main text, is owing to the glide-mirror symmetry breaking characteristic of monolayer FeSe on a substrate.

\section{FFLO state in the low magnetic field region in the presence of finite intra-sublattice pairing \label{sec:fflo_low_magnetir}}

\begin{figure}[htbp]
    \centering
    \includegraphics[width=0.5\textwidth]{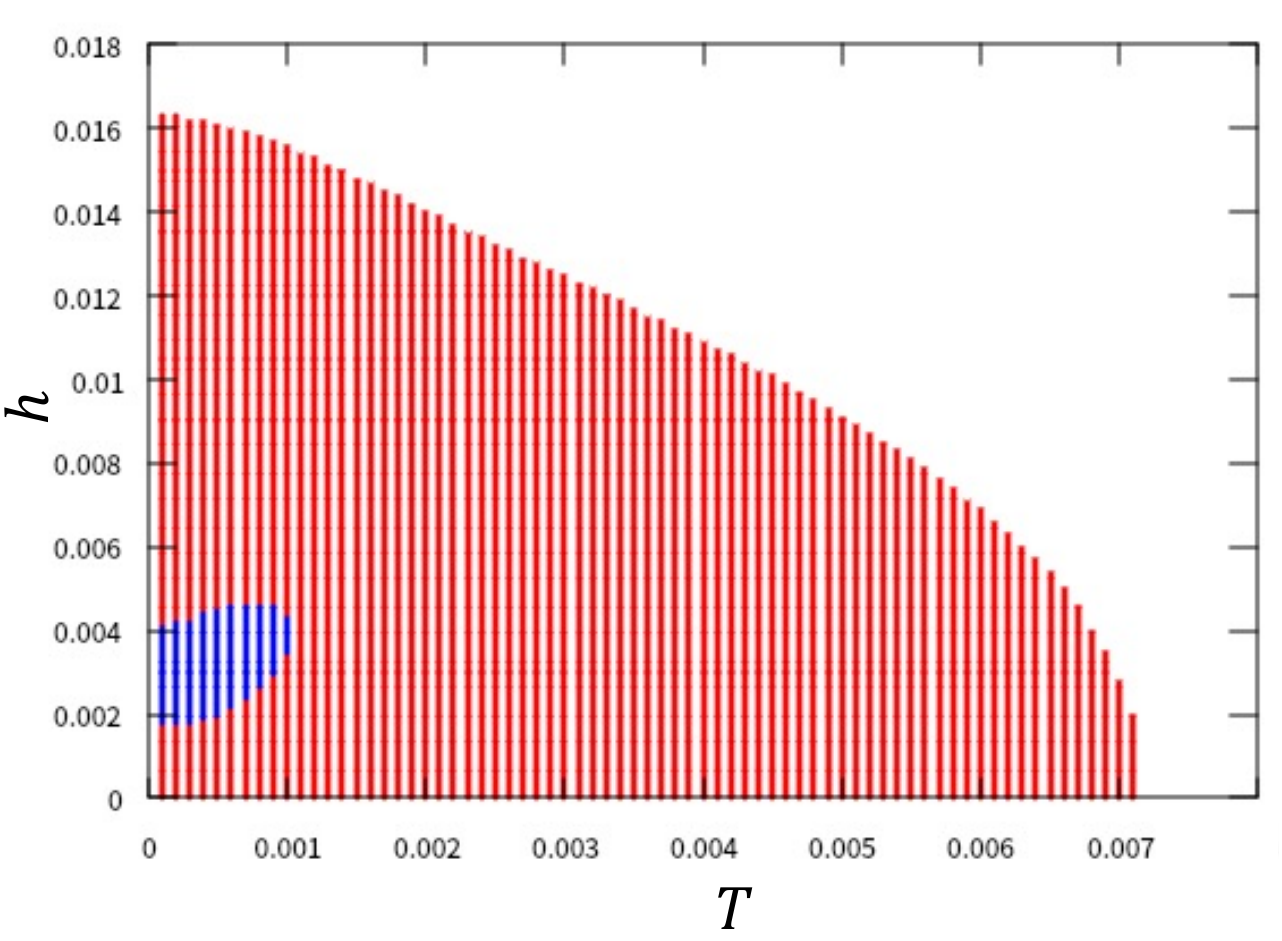}
    \caption{The superconducting phase diagram for the gap function, Eq.~\eqref{eq:s+exs}.
    We set the mass enhancement factor as $z=1/5$.}
    \label{fig:pd_gao_s_exs}
\end{figure}

We show that the low-field FFLO state shown in Sec.~\ref{sec:ex_s_pd} is stable when a small intra-sublattice pairing exists.
Here, we consider the gap function,
\begin{eqnarray}
    \bm \Delta(\bm k) = \Delta_0\tau_0\otimes\rho_0+\Delta(\bm k)\tau_x\otimes\rho_0\label{eq:s+exs},
\end{eqnarray}
where $\Delta(\bm k)\propto\cos k_x/2\cos k_y/2$.
To get this gap function, we set the attractive interaction as $V_{ls}(\bm k,\bm k^\prime) = V_0\delta_{ls}+V_1\cos k_x/2\cos k_y/2\cos k_x^\prime/2\cos k_y^\prime/2\left(\tau_x\otimes\rho_0\right)_{ls}$ with $10V_0 =V_1$.
In this case, the superconducting state is gapped due to finite intra-sublattice pairing.

In Fig.~\ref{fig:pd_gao_s_exs}, we show the superconducting phase diagram for Eq.~\eqref{eq:s+exs}.
From the figure, we confirm that the FFLO state in the low magnetic field region is stable even when the intra-sublattice pairing is finite.
Thus, the low-field FFLO state is expected to be ubiquitous when the incipient bands give sizable condensation energy of superconductivity.

\bibliography{main}

\begin{thebibliography}{97}%
\makeatletter
\providecommand \@ifxundefined [1]{%
 \@ifx{#1\undefined}
}%
\providecommand \@ifnum [1]{%
 \ifnum #1\expandafter \@firstoftwo
 \else \expandafter \@secondoftwo
 \fi
}%
\providecommand \@ifx [1]{%
 \ifx #1\expandafter \@firstoftwo
 \else \expandafter \@secondoftwo
 \fi
}%
\providecommand \natexlab [1]{#1}%
\providecommand \enquote  [1]{``#1''}%
\providecommand \bibnamefont  [1]{#1}%
\providecommand \bibfnamefont [1]{#1}%
\providecommand \citenamefont [1]{#1}%
\providecommand \href@noop [0]{\@secondoftwo}%
\providecommand \href [0]{\begingroup \@sanitize@url \@href}%
\providecommand \@href[1]{\@@startlink{#1}\@@href}%
\providecommand \@@href[1]{\endgroup#1\@@endlink}%
\providecommand \@sanitize@url [0]{\catcode `\\12\catcode `\$12\catcode
  `\&12\catcode `\#12\catcode `\^12\catcode `\_12\catcode `\%12\relax}%
\providecommand \@@startlink[1]{}%
\providecommand \@@endlink[0]{}%
\providecommand \url  [0]{\begingroup\@sanitize@url \@url }%
\providecommand \@url [1]{\endgroup\@href {#1}{\urlprefix }}%
\providecommand \urlprefix  [0]{URL }%
\providecommand \Eprint [0]{\href }%
\providecommand \doibase [0]{http://dx.doi.org/}%
\providecommand \selectlanguage [0]{\@gobble}%
\providecommand \bibinfo  [0]{\@secondoftwo}%
\providecommand \bibfield  [0]{\@secondoftwo}%
\providecommand \translation [1]{[#1]}%
\providecommand \BibitemOpen [0]{}%
\providecommand \bibitemStop [0]{}%
\providecommand \bibitemNoStop [0]{.\EOS\space}%
\providecommand \EOS [0]{\spacefactor3000\relax}%
\providecommand \BibitemShut  [1]{\csname bibitem#1\endcsname}%
\let\auto@bib@innerbib\@empty
\bibitem [{\citenamefont {Hasan}\ and\ \citenamefont
  {Kane}(2010)}]{hasan2010topological}%
  \BibitemOpen
  \bibfield  {author} {\bibinfo {author} {\bibfnamefont {M.~Z.}\ \bibnamefont
  {Hasan}}\ and\ \bibinfo {author} {\bibfnamefont {C.~L.}\ \bibnamefont
  {Kane}},\ }\href
  {https://journals.aps.org/rmp/abstract/10.1103/RevModPhys.82.3045} {\bibfield
   {journal} {\bibinfo  {journal} {Rev. Mod. Phys.}\ }\textbf {\bibinfo
  {volume} {82}},\ \bibinfo {pages} {3045} (\bibinfo {year}
  {2010})}\BibitemShut {NoStop}%
\bibitem [{\citenamefont {Qi}\ and\ \citenamefont
  {Zhang}(2011)}]{qi2011topological}%
  \BibitemOpen
  \bibfield  {author} {\bibinfo {author} {\bibfnamefont {X.-L.}\ \bibnamefont
  {Qi}}\ and\ \bibinfo {author} {\bibfnamefont {S.-C.}\ \bibnamefont {Zhang}},\
  }\href {https://journals.aps.org/rmp/abstract/10.1103/RevModPhys.83.1057}
  {\bibfield  {journal} {\bibinfo  {journal} {Rev. Mod. Phys.}\ }\textbf
  {\bibinfo {volume} {83}},\ \bibinfo {pages} {1057} (\bibinfo {year}
  {2011})}\BibitemShut {NoStop}%
\bibitem [{\citenamefont {Berry}(1984)}]{berry1984quantal}%
  \BibitemOpen
  \bibfield  {author} {\bibinfo {author} {\bibfnamefont {M.~V.}\ \bibnamefont
  {Berry}},\ }\href {\doibase 10.1098/rspa.1984.0023} {\bibfield  {journal}
  {\bibinfo  {journal} {Proceedings of the Royal Society of London. A.
  Mathematical and Physical Sciences}\ }\textbf {\bibinfo {volume} {392}},\
  \bibinfo {pages} {45} (\bibinfo {year} {1984})},\ \Eprint
  {http://arxiv.org/abs/https://royalsocietypublishing.org/doi/pdf/10.1098/rspa.1984.0023}
  {https://royalsocietypublishing.org/doi/pdf/10.1098/rspa.1984.0023}
  \BibitemShut {NoStop}%
\bibitem [{\citenamefont {Provost}\ and\ \citenamefont
  {Vallee}(1980)}]{provost1980riemannian}%
  \BibitemOpen
  \bibfield  {author} {\bibinfo {author} {\bibfnamefont {J.~P.}\ \bibnamefont
  {Provost}}\ and\ \bibinfo {author} {\bibfnamefont {G.}~\bibnamefont
  {Vallee}},\ }\href {\doibase 10.1007/BF02193559} {\bibfield  {journal}
  {\bibinfo  {journal} {Communications in Mathematical Physics}\ }\textbf
  {\bibinfo {volume} {76}},\ \bibinfo {pages} {289} (\bibinfo {year}
  {1980})}\BibitemShut {NoStop}%
\bibitem [{\citenamefont {{ R. Resta}}(2011)}]{resta2011the}%
  \BibitemOpen
  \bibfield  {author} {\bibinfo {author} {\bibnamefont {{ R. Resta}}},\ }\href
  {\doibase 10.1140/epjb/e2010-10874-4} {\bibfield  {journal} {\bibinfo
  {journal} {Eur. Phys. J. B}\ }\textbf {\bibinfo {volume} {79}},\ \bibinfo
  {pages} {121} (\bibinfo {year} {2011})}\BibitemShut {NoStop}%
\bibitem [{\citenamefont {Thouless}\ \emph {et~al.}(1982)\citenamefont
  {Thouless}, \citenamefont {Kohmoto}, \citenamefont {Nightingale},\ and\
  \citenamefont {den Nijs}}]{thouless1982quantized}%
  \BibitemOpen
  \bibfield  {author} {\bibinfo {author} {\bibfnamefont {D.~J.}\ \bibnamefont
  {Thouless}}, \bibinfo {author} {\bibfnamefont {M.}~\bibnamefont {Kohmoto}},
  \bibinfo {author} {\bibfnamefont {M.~P.}\ \bibnamefont {Nightingale}}, \ and\
  \bibinfo {author} {\bibfnamefont {M.}~\bibnamefont {den Nijs}},\ }\href
  {\doibase 10.1103/PhysRevLett.49.405} {\bibfield  {journal} {\bibinfo
  {journal} {Phys. Rev. Lett.}\ }\textbf {\bibinfo {volume} {49}},\ \bibinfo
  {pages} {405} (\bibinfo {year} {1982})}\BibitemShut {NoStop}%
\bibitem [{\citenamefont {Xiao}\ \emph {et~al.}(2010)\citenamefont {Xiao},
  \citenamefont {Chang},\ and\ \citenamefont {Niu}}]{xiao2010berry}%
  \BibitemOpen
  \bibfield  {author} {\bibinfo {author} {\bibfnamefont {D.}~\bibnamefont
  {Xiao}}, \bibinfo {author} {\bibfnamefont {M.-C.}\ \bibnamefont {Chang}}, \
  and\ \bibinfo {author} {\bibfnamefont {Q.}~\bibnamefont {Niu}},\ }\href
  {\doibase 10.1103/RevModPhys.82.1959} {\bibfield  {journal} {\bibinfo
  {journal} {Rev. Mod. Phys.}\ }\textbf {\bibinfo {volume} {82}},\ \bibinfo
  {pages} {1959} (\bibinfo {year} {2010})}\BibitemShut {NoStop}%
\bibitem [{\citenamefont {Nagaosa}\ \emph {et~al.}(2010)\citenamefont
  {Nagaosa}, \citenamefont {Sinova}, \citenamefont {Onoda}, \citenamefont
  {MacDonald},\ and\ \citenamefont {Ong}}]{Nagaosa2010}%
  \BibitemOpen
  \bibfield  {author} {\bibinfo {author} {\bibfnamefont {N.}~\bibnamefont
  {Nagaosa}}, \bibinfo {author} {\bibfnamefont {J.}~\bibnamefont {Sinova}},
  \bibinfo {author} {\bibfnamefont {S.}~\bibnamefont {Onoda}}, \bibinfo
  {author} {\bibfnamefont {A.~H.}\ \bibnamefont {MacDonald}}, \ and\ \bibinfo
  {author} {\bibfnamefont {N.~P.}\ \bibnamefont {Ong}},\ }\href {\doibase
  10.1103/RevModPhys.82.1539} {\bibfield  {journal} {\bibinfo  {journal} {Rev.
  Mod. Phys.}\ }\textbf {\bibinfo {volume} {82}},\ \bibinfo {pages} {1539}
  (\bibinfo {year} {2010})}\BibitemShut {NoStop}%
\bibitem [{\citenamefont {Pi\'echon}\ \emph {et~al.}(2016)\citenamefont
  {Pi\'echon}, \citenamefont {Raoux}, \citenamefont {Fuchs},\ and\
  \citenamefont {Montambaux}}]{piechon2016geometric}%
  \BibitemOpen
  \bibfield  {author} {\bibinfo {author} {\bibfnamefont {F.}~\bibnamefont
  {Pi\'echon}}, \bibinfo {author} {\bibfnamefont {A.}~\bibnamefont {Raoux}},
  \bibinfo {author} {\bibfnamefont {J.-N.}\ \bibnamefont {Fuchs}}, \ and\
  \bibinfo {author} {\bibfnamefont {G.}~\bibnamefont {Montambaux}},\ }\href
  {\doibase 10.1103/PhysRevB.94.134423} {\bibfield  {journal} {\bibinfo
  {journal} {Phys. Rev. B}\ }\textbf {\bibinfo {volume} {94}},\ \bibinfo
  {pages} {134423} (\bibinfo {year} {2016})}\BibitemShut {NoStop}%
\bibitem [{\citenamefont {Marzari}\ and\ \citenamefont
  {Vanderbilt}(1997)}]{marzari1997maximally}%
  \BibitemOpen
  \bibfield  {author} {\bibinfo {author} {\bibfnamefont {N.}~\bibnamefont
  {Marzari}}\ and\ \bibinfo {author} {\bibfnamefont {D.}~\bibnamefont
  {Vanderbilt}},\ }\href {\doibase 10.1103/PhysRevB.56.12847} {\bibfield
  {journal} {\bibinfo  {journal} {Phys. Rev. B}\ }\textbf {\bibinfo {volume}
  {56}},\ \bibinfo {pages} {12847} (\bibinfo {year} {1997})}\BibitemShut
  {NoStop}%
\bibitem [{\citenamefont {Neupert}\ \emph {et~al.}(2013)\citenamefont
  {Neupert}, \citenamefont {Chamon},\ and\ \citenamefont
  {Mudry}}]{Neupert2013}%
  \BibitemOpen
  \bibfield  {author} {\bibinfo {author} {\bibfnamefont {T.}~\bibnamefont
  {Neupert}}, \bibinfo {author} {\bibfnamefont {C.}~\bibnamefont {Chamon}}, \
  and\ \bibinfo {author} {\bibfnamefont {C.}~\bibnamefont {Mudry}},\ }\href
  {\doibase 10.1103/PhysRevB.87.245103} {\bibfield  {journal} {\bibinfo
  {journal} {Phys. Rev. B}\ }\textbf {\bibinfo {volume} {87}},\ \bibinfo
  {pages} {245103} (\bibinfo {year} {2013})}\BibitemShut {NoStop}%
\bibitem [{\citenamefont {Srivastava}\ and\ \citenamefont
  {Imamo\ifmmode~\breve{g}\else \u{g}\fi{}lu}(2015)}]{Srivastava2015}%
  \BibitemOpen
  \bibfield  {author} {\bibinfo {author} {\bibfnamefont {A.}~\bibnamefont
  {Srivastava}}\ and\ \bibinfo {author} {\bibfnamefont {A.~m.~c.}\ \bibnamefont
  {Imamo\ifmmode~\breve{g}\else \u{g}\fi{}lu}},\ }\href {\doibase
  10.1103/PhysRevLett.115.166802} {\bibfield  {journal} {\bibinfo  {journal}
  {Phys. Rev. Lett.}\ }\textbf {\bibinfo {volume} {115}},\ \bibinfo {pages}
  {166802} (\bibinfo {year} {2015})}\BibitemShut {NoStop}%
\bibitem [{\citenamefont {Gao}\ \emph {et~al.}(2014)\citenamefont {Gao},
  \citenamefont {Yang},\ and\ \citenamefont {Niu}}]{gao2014field}%
  \BibitemOpen
  \bibfield  {author} {\bibinfo {author} {\bibfnamefont {Y.}~\bibnamefont
  {Gao}}, \bibinfo {author} {\bibfnamefont {S.~A.}\ \bibnamefont {Yang}}, \
  and\ \bibinfo {author} {\bibfnamefont {Q.}~\bibnamefont {Niu}},\ }\href
  {\doibase 10.1103/PhysRevLett.112.166601} {\bibfield  {journal} {\bibinfo
  {journal} {Phys. Rev. Lett.}\ }\textbf {\bibinfo {volume} {112}},\ \bibinfo
  {pages} {166601} (\bibinfo {year} {2014})}\BibitemShut {NoStop}%
\bibitem [{\citenamefont {Gao}\ and\ \citenamefont
  {Xiao}(2019)}]{gao2019nonreciprocal}%
  \BibitemOpen
  \bibfield  {author} {\bibinfo {author} {\bibfnamefont {Y.}~\bibnamefont
  {Gao}}\ and\ \bibinfo {author} {\bibfnamefont {D.}~\bibnamefont {Xiao}},\
  }\href {\doibase 10.1103/PhysRevLett.122.227402} {\bibfield  {journal}
  {\bibinfo  {journal} {Phys. Rev. Lett.}\ }\textbf {\bibinfo {volume} {122}},\
  \bibinfo {pages} {227402} (\bibinfo {year} {2019})}\BibitemShut {NoStop}%
\bibitem [{\citenamefont {Lapa}\ and\ \citenamefont
  {Hughes}(2019)}]{lapa2019semiclassical}%
  \BibitemOpen
  \bibfield  {author} {\bibinfo {author} {\bibfnamefont {M.~F.}\ \bibnamefont
  {Lapa}}\ and\ \bibinfo {author} {\bibfnamefont {T.~L.}\ \bibnamefont
  {Hughes}},\ }\href {\doibase 10.1103/PhysRevB.99.121111} {\bibfield
  {journal} {\bibinfo  {journal} {Phys. Rev. B}\ }\textbf {\bibinfo {volume}
  {99}},\ \bibinfo {pages} {121111} (\bibinfo {year} {2019})}\BibitemShut
  {NoStop}%
\bibitem [{\citenamefont {Daido}\ \emph {et~al.}(2020)\citenamefont {Daido},
  \citenamefont {Shitade},\ and\ \citenamefont
  {Yanase}}]{daido2020thermodynamic}%
  \BibitemOpen
  \bibfield  {author} {\bibinfo {author} {\bibfnamefont {A.}~\bibnamefont
  {Daido}}, \bibinfo {author} {\bibfnamefont {A.}~\bibnamefont {Shitade}}, \
  and\ \bibinfo {author} {\bibfnamefont {Y.}~\bibnamefont {Yanase}},\ }\href
  {\doibase 10.1103/PhysRevB.102.235149} {\bibfield  {journal} {\bibinfo
  {journal} {Phys. Rev. B}\ }\textbf {\bibinfo {volume} {102}},\ \bibinfo
  {pages} {235149} (\bibinfo {year} {2020})}\BibitemShut {NoStop}%
\bibitem [{\citenamefont {Kitamura}\ \emph {et~al.}(2021)\citenamefont
  {Kitamura}, \citenamefont {Ishizuka}, \citenamefont {Daido},\ and\
  \citenamefont {Yanase}}]{kitamura2021thermodynamic}%
  \BibitemOpen
  \bibfield  {author} {\bibinfo {author} {\bibfnamefont {T.}~\bibnamefont
  {Kitamura}}, \bibinfo {author} {\bibfnamefont {J.}~\bibnamefont {Ishizuka}},
  \bibinfo {author} {\bibfnamefont {A.}~\bibnamefont {Daido}}, \ and\ \bibinfo
  {author} {\bibfnamefont {Y.}~\bibnamefont {Yanase}},\ }\href {\doibase
  10.1103/PhysRevB.103.245114} {\bibfield  {journal} {\bibinfo  {journal}
  {Phys. Rev. B}\ }\textbf {\bibinfo {volume} {103}},\ \bibinfo {pages}
  {245114} (\bibinfo {year} {2021})}\BibitemShut {NoStop}%
\bibitem [{\citenamefont {Mitscherling}\ and\ \citenamefont
  {Holder}(2022)}]{mitscherling2022bound}%
  \BibitemOpen
  \bibfield  {author} {\bibinfo {author} {\bibfnamefont {J.}~\bibnamefont
  {Mitscherling}}\ and\ \bibinfo {author} {\bibfnamefont {T.}~\bibnamefont
  {Holder}},\ }\href
  {https://journals.aps.org/prb/abstract/10.1103/PhysRevB.105.085154}
  {\bibfield  {journal} {\bibinfo  {journal} {Phys. Rev. B Condens. Matter}\
  }\textbf {\bibinfo {volume} {105}},\ \bibinfo {pages} {085154} (\bibinfo
  {year} {2022})}\BibitemShut {NoStop}%
\bibitem [{\citenamefont {Ahn}\ \emph {et~al.}(2021)\citenamefont {Ahn},
  \citenamefont {Guo}, \citenamefont {Nagaosa},\ and\ \citenamefont
  {Vishwanath}}]{ahn2021reimannian}%
  \BibitemOpen
  \bibfield  {author} {\bibinfo {author} {\bibfnamefont {J.}~\bibnamefont
  {Ahn}}, \bibinfo {author} {\bibfnamefont {G.-Y.}\ \bibnamefont {Guo}},
  \bibinfo {author} {\bibfnamefont {N.}~\bibnamefont {Nagaosa}}, \ and\
  \bibinfo {author} {\bibfnamefont {A.}~\bibnamefont {Vishwanath}},\ }\href
  {https://www.nature.com/articles/s41567-021-01465-z} {\bibfield  {journal}
  {\bibinfo  {journal} {Nat. Phys.}\ }\textbf {\bibinfo {volume} {18}},\
  \bibinfo {pages} {290} (\bibinfo {year} {2021})}\BibitemShut {NoStop}%
\bibitem [{\citenamefont {Rhim}\ \emph {et~al.}(2020)\citenamefont {Rhim},
  \citenamefont {Kim},\ and\ \citenamefont {Yang}}]{rhim2020quantum}%
  \BibitemOpen
  \bibfield  {author} {\bibinfo {author} {\bibfnamefont {J.-W.}\ \bibnamefont
  {Rhim}}, \bibinfo {author} {\bibfnamefont {K.}~\bibnamefont {Kim}}, \ and\
  \bibinfo {author} {\bibfnamefont {B.-J.}\ \bibnamefont {Yang}},\ }\href
  {https://www.nature.com/articles/s41586-020-2540-1} {\bibfield  {journal}
  {\bibinfo  {journal} {Nature}\ }\textbf {\bibinfo {volume} {584}},\ \bibinfo
  {pages} {59} (\bibinfo {year} {2020})}\BibitemShut {NoStop}%
\bibitem [{\citenamefont {Wang}\ \emph {et~al.}(2021)\citenamefont {Wang},
  \citenamefont {Cano}, \citenamefont {Millis}, \citenamefont {Liu},\ and\
  \citenamefont {Yang}}]{wang2021exact}%
  \BibitemOpen
  \bibfield  {author} {\bibinfo {author} {\bibfnamefont {J.}~\bibnamefont
  {Wang}}, \bibinfo {author} {\bibfnamefont {J.}~\bibnamefont {Cano}}, \bibinfo
  {author} {\bibfnamefont {A.~J.}\ \bibnamefont {Millis}}, \bibinfo {author}
  {\bibfnamefont {Z.}~\bibnamefont {Liu}}, \ and\ \bibinfo {author}
  {\bibfnamefont {B.}~\bibnamefont {Yang}},\ }\href
  {https://journals.aps.org/prl/abstract/10.1103/PhysRevLett.127.246403}
  {\bibfield  {journal} {\bibinfo  {journal} {Phys. Rev. Lett.}\ }\textbf
  {\bibinfo {volume} {127}},\ \bibinfo {pages} {246403} (\bibinfo {year}
  {2021})}\BibitemShut {NoStop}%
\bibitem [{\citenamefont {Hwang}\ \emph {et~al.}(2021)\citenamefont {Hwang},
  \citenamefont {Rhim},\ and\ \citenamefont {Yang}}]{hwang2021geometric}%
  \BibitemOpen
  \bibfield  {author} {\bibinfo {author} {\bibfnamefont {Y.}~\bibnamefont
  {Hwang}}, \bibinfo {author} {\bibfnamefont {J.-W.}\ \bibnamefont {Rhim}}, \
  and\ \bibinfo {author} {\bibfnamefont {B.-J.}\ \bibnamefont {Yang}},\ }\href
  {https://www.nature.com/articles/s41467-021-26765-z} {\bibfield  {journal}
  {\bibinfo  {journal} {Nat. Commun.}\ }\textbf {\bibinfo {volume} {12}},\
  \bibinfo {pages} {6433} (\bibinfo {year} {2021})}\BibitemShut {NoStop}%
\bibitem [{\citenamefont {Mera}\ and\ \citenamefont
  {Mitscherling}(2022)}]{mera2022nontrivial}%
  \BibitemOpen
  \bibfield  {author} {\bibinfo {author} {\bibfnamefont {B.}~\bibnamefont
  {Mera}}\ and\ \bibinfo {author} {\bibfnamefont {J.}~\bibnamefont
  {Mitscherling}},\ }\href {https://arxiv.org/abs/2205.07900} {\  (\bibinfo
  {year} {2022})},\ \Eprint {http://arxiv.org/abs/2205.07900} {arXiv:2205.07900
  [cond-mat.mes-hall]} \BibitemShut {NoStop}%
\bibitem [{\citenamefont {Julku}\ \emph
  {et~al.}(2021{\natexlab{a}})\citenamefont {Julku}, \citenamefont {Bruun},\
  and\ \citenamefont {T\"orm\"a}}]{julku2021excitations}%
  \BibitemOpen
  \bibfield  {author} {\bibinfo {author} {\bibfnamefont {A.}~\bibnamefont
  {Julku}}, \bibinfo {author} {\bibfnamefont {G.~M.}\ \bibnamefont {Bruun}}, \
  and\ \bibinfo {author} {\bibfnamefont {P.}~\bibnamefont {T\"orm\"a}},\ }\href
  {\doibase 10.1103/PhysRevB.104.144507} {\bibfield  {journal} {\bibinfo
  {journal} {Phys. Rev. B}\ }\textbf {\bibinfo {volume} {104}},\ \bibinfo
  {pages} {144507} (\bibinfo {year} {2021}{\natexlab{a}})}\BibitemShut
  {NoStop}%
\bibitem [{\citenamefont {Julku}\ \emph
  {et~al.}(2021{\natexlab{b}})\citenamefont {Julku}, \citenamefont {Bruun},\
  and\ \citenamefont {T\"orm\"a}}]{julku2021quantum}%
  \BibitemOpen
  \bibfield  {author} {\bibinfo {author} {\bibfnamefont {A.}~\bibnamefont
  {Julku}}, \bibinfo {author} {\bibfnamefont {G.~M.}\ \bibnamefont {Bruun}}, \
  and\ \bibinfo {author} {\bibfnamefont {P.}~\bibnamefont {T\"orm\"a}},\ }\href
  {\doibase 10.1103/PhysRevLett.127.170404} {\bibfield  {journal} {\bibinfo
  {journal} {Phys. Rev. Lett.}\ }\textbf {\bibinfo {volume} {127}},\ \bibinfo
  {pages} {170404} (\bibinfo {year} {2021}{\natexlab{b}})}\BibitemShut
  {NoStop}%
\bibitem [{\citenamefont {Topp}\ \emph {et~al.}(2021)\citenamefont {Topp},
  \citenamefont {Eckhardt}, \citenamefont {Kennes}, \citenamefont {Sentef},\
  and\ \citenamefont {T\"orm\"a}}]{topp2021light}%
  \BibitemOpen
  \bibfield  {author} {\bibinfo {author} {\bibfnamefont {G.~E.}\ \bibnamefont
  {Topp}}, \bibinfo {author} {\bibfnamefont {C.~J.}\ \bibnamefont {Eckhardt}},
  \bibinfo {author} {\bibfnamefont {D.~M.}\ \bibnamefont {Kennes}}, \bibinfo
  {author} {\bibfnamefont {M.~A.}\ \bibnamefont {Sentef}}, \ and\ \bibinfo
  {author} {\bibfnamefont {P.}~\bibnamefont {T\"orm\"a}},\ }\href {\doibase
  10.1103/PhysRevB.104.064306} {\bibfield  {journal} {\bibinfo  {journal}
  {Phys. Rev. B}\ }\textbf {\bibinfo {volume} {104}},\ \bibinfo {pages}
  {064306} (\bibinfo {year} {2021})}\BibitemShut {NoStop}%
\bibitem [{\citenamefont {Solnyshkov}\ \emph {et~al.}(2021)\citenamefont
  {Solnyshkov}, \citenamefont {Leblanc}, \citenamefont {Bessonart},
  \citenamefont {Nalitov}, \citenamefont {Ren}, \citenamefont {Liao},
  \citenamefont {Li},\ and\ \citenamefont {Malpuech}}]{solnyshkov2021quantum}%
  \BibitemOpen
  \bibfield  {author} {\bibinfo {author} {\bibfnamefont {D.~D.}\ \bibnamefont
  {Solnyshkov}}, \bibinfo {author} {\bibfnamefont {C.}~\bibnamefont {Leblanc}},
  \bibinfo {author} {\bibfnamefont {L.}~\bibnamefont {Bessonart}}, \bibinfo
  {author} {\bibfnamefont {A.}~\bibnamefont {Nalitov}}, \bibinfo {author}
  {\bibfnamefont {J.}~\bibnamefont {Ren}}, \bibinfo {author} {\bibfnamefont
  {Q.}~\bibnamefont {Liao}}, \bibinfo {author} {\bibfnamefont {F.}~\bibnamefont
  {Li}}, \ and\ \bibinfo {author} {\bibfnamefont {G.}~\bibnamefont
  {Malpuech}},\ }\href {\doibase 10.1103/PhysRevB.103.125302} {\bibfield
  {journal} {\bibinfo  {journal} {Phys. Rev. B}\ }\textbf {\bibinfo {volume}
  {103}},\ \bibinfo {pages} {125302} (\bibinfo {year} {2021})}\BibitemShut
  {NoStop}%
\bibitem [{\citenamefont {Liao}\ \emph {et~al.}(2021)\citenamefont {Liao},
  \citenamefont {Leblanc}, \citenamefont {Ren}, \citenamefont {Li},
  \citenamefont {Li}, \citenamefont {Solnyshkov}, \citenamefont {Malpuech},
  \citenamefont {Yao},\ and\ \citenamefont {Fu}}]{liao2021experimental}%
  \BibitemOpen
  \bibfield  {author} {\bibinfo {author} {\bibfnamefont {Q.}~\bibnamefont
  {Liao}}, \bibinfo {author} {\bibfnamefont {C.}~\bibnamefont {Leblanc}},
  \bibinfo {author} {\bibfnamefont {J.}~\bibnamefont {Ren}}, \bibinfo {author}
  {\bibfnamefont {F.}~\bibnamefont {Li}}, \bibinfo {author} {\bibfnamefont
  {Y.}~\bibnamefont {Li}}, \bibinfo {author} {\bibfnamefont {D.}~\bibnamefont
  {Solnyshkov}}, \bibinfo {author} {\bibfnamefont {G.}~\bibnamefont
  {Malpuech}}, \bibinfo {author} {\bibfnamefont {J.}~\bibnamefont {Yao}}, \
  and\ \bibinfo {author} {\bibfnamefont {H.}~\bibnamefont {Fu}},\ }\href
  {\doibase 10.1103/PhysRevLett.127.107402} {\bibfield  {journal} {\bibinfo
  {journal} {Phys. Rev. Lett.}\ }\textbf {\bibinfo {volume} {127}},\ \bibinfo
  {pages} {107402} (\bibinfo {year} {2021})}\BibitemShut {NoStop}%
\bibitem [{\citenamefont {Ahn}\ \emph {et~al.}(2020)\citenamefont {Ahn},
  \citenamefont {Guo},\ and\ \citenamefont {Nagaosa}}]{ahn2020low-frequency}%
  \BibitemOpen
  \bibfield  {author} {\bibinfo {author} {\bibfnamefont {J.}~\bibnamefont
  {Ahn}}, \bibinfo {author} {\bibfnamefont {G.-Y.}\ \bibnamefont {Guo}}, \ and\
  \bibinfo {author} {\bibfnamefont {N.}~\bibnamefont {Nagaosa}},\ }\href
  {\doibase 10.1103/PhysRevX.10.041041} {\bibfield  {journal} {\bibinfo
  {journal} {Phys. Rev. X}\ }\textbf {\bibinfo {volume} {10}},\ \bibinfo
  {pages} {041041} (\bibinfo {year} {2020})}\BibitemShut {NoStop}%
\bibitem [{\citenamefont {Watanabe}\ and\ \citenamefont
  {Yanase}(2021)}]{watanabe2021chiral}%
  \BibitemOpen
  \bibfield  {author} {\bibinfo {author} {\bibfnamefont {H.}~\bibnamefont
  {Watanabe}}\ and\ \bibinfo {author} {\bibfnamefont {Y.}~\bibnamefont
  {Yanase}},\ }\href {\doibase 10.1103/PhysRevX.11.011001} {\bibfield
  {journal} {\bibinfo  {journal} {Phys. Rev. X}\ }\textbf {\bibinfo {volume}
  {11}},\ \bibinfo {pages} {011001} (\bibinfo {year} {2021})}\BibitemShut
  {NoStop}%
\bibitem [{\citenamefont {Ahn}\ and\ \citenamefont
  {Nagaosa}(2021)}]{ahn2021superconductivity-induced}%
  \BibitemOpen
  \bibfield  {author} {\bibinfo {author} {\bibfnamefont {J.}~\bibnamefont
  {Ahn}}\ and\ \bibinfo {author} {\bibfnamefont {N.}~\bibnamefont {Nagaosa}},\
  }\href {\doibase 10.1103/PhysRevB.104.L100501} {\bibfield  {journal}
  {\bibinfo  {journal} {Phys. Rev. B}\ }\textbf {\bibinfo {volume} {104}},\
  \bibinfo {pages} {L100501} (\bibinfo {year} {2021})}\BibitemShut {NoStop}%
\bibitem [{\citenamefont {Ozawa}\ and\ \citenamefont
  {Mera}(2021)}]{ozawa2021relations}%
  \BibitemOpen
  \bibfield  {author} {\bibinfo {author} {\bibfnamefont {T.}~\bibnamefont
  {Ozawa}}\ and\ \bibinfo {author} {\bibfnamefont {B.}~\bibnamefont {Mera}},\
  }\href {\doibase 10.1103/PhysRevB.104.045103} {\bibfield  {journal} {\bibinfo
   {journal} {Phys. Rev. B}\ }\textbf {\bibinfo {volume} {104}},\ \bibinfo
  {pages} {045103} (\bibinfo {year} {2021})}\BibitemShut {NoStop}%
\bibitem [{\citenamefont {Mera}\ and\ \citenamefont
  {Ozawa}(2021{\natexlab{a}})}]{mera2021kahler}%
  \BibitemOpen
  \bibfield  {author} {\bibinfo {author} {\bibfnamefont {B.}~\bibnamefont
  {Mera}}\ and\ \bibinfo {author} {\bibfnamefont {T.}~\bibnamefont {Ozawa}},\
  }\href {\doibase 10.1103/PhysRevB.104.045104} {\bibfield  {journal} {\bibinfo
   {journal} {Phys. Rev. B}\ }\textbf {\bibinfo {volume} {104}},\ \bibinfo
  {pages} {045104} (\bibinfo {year} {2021}{\natexlab{a}})}\BibitemShut
  {NoStop}%
\bibitem [{\citenamefont {Mera}\ and\ \citenamefont
  {Ozawa}(2021{\natexlab{b}})}]{mera2021engineering}%
  \BibitemOpen
  \bibfield  {author} {\bibinfo {author} {\bibfnamefont {B.}~\bibnamefont
  {Mera}}\ and\ \bibinfo {author} {\bibfnamefont {T.}~\bibnamefont {Ozawa}},\
  }\href {\doibase 10.1103/PhysRevB.104.115160} {\bibfield  {journal} {\bibinfo
   {journal} {Phys. Rev. B}\ }\textbf {\bibinfo {volume} {104}},\ \bibinfo
  {pages} {115160} (\bibinfo {year} {2021}{\natexlab{b}})}\BibitemShut
  {NoStop}%
\bibitem [{\citenamefont {Mera}\ \emph {et~al.}(2022)\citenamefont {Mera},
  \citenamefont {Zhang},\ and\ \citenamefont {Goldman}}]{mera2021relating}%
  \BibitemOpen
  \bibfield  {author} {\bibinfo {author} {\bibfnamefont {B.}~\bibnamefont
  {Mera}}, \bibinfo {author} {\bibfnamefont {A.}~\bibnamefont {Zhang}}, \ and\
  \bibinfo {author} {\bibfnamefont {N.}~\bibnamefont {Goldman}},\ }\href
  {\doibase 10.21468/SciPostPhys.12.1.018} {\bibfield  {journal} {\bibinfo
  {journal} {SciPost Phys.}\ }\textbf {\bibinfo {volume} {12}},\ \bibinfo
  {pages} {18} (\bibinfo {year} {2022})}\BibitemShut {NoStop}%
\bibitem [{\citenamefont {Peotta}\ and\ \citenamefont
  {T{\"o}rm{\"a}}(2015)}]{peotta2015superfluidity}%
  \BibitemOpen
  \bibfield  {author} {\bibinfo {author} {\bibfnamefont {S.}~\bibnamefont
  {Peotta}}\ and\ \bibinfo {author} {\bibfnamefont {P.}~\bibnamefont
  {T{\"o}rm{\"a}}},\ }\href {\doibase 10.1038/ncomms9944} {\bibfield  {journal}
  {\bibinfo  {journal} {Nature Communications}\ }\textbf {\bibinfo {volume}
  {6}},\ \bibinfo {pages} {8944} (\bibinfo {year} {2015})}\BibitemShut
  {NoStop}%
\bibitem [{\citenamefont {Liang}\ \emph {et~al.}(2017)\citenamefont {Liang},
  \citenamefont {Vanhala}, \citenamefont {Peotta}, \citenamefont {Siro},
  \citenamefont {Harju},\ and\ \citenamefont {T\"orm\"a}}]{liang2017band}%
  \BibitemOpen
  \bibfield  {author} {\bibinfo {author} {\bibfnamefont {L.}~\bibnamefont
  {Liang}}, \bibinfo {author} {\bibfnamefont {T.~I.}\ \bibnamefont {Vanhala}},
  \bibinfo {author} {\bibfnamefont {S.}~\bibnamefont {Peotta}}, \bibinfo
  {author} {\bibfnamefont {T.}~\bibnamefont {Siro}}, \bibinfo {author}
  {\bibfnamefont {A.}~\bibnamefont {Harju}}, \ and\ \bibinfo {author}
  {\bibfnamefont {P.}~\bibnamefont {T\"orm\"a}},\ }\href {\doibase
  10.1103/PhysRevB.95.024515} {\bibfield  {journal} {\bibinfo  {journal} {Phys.
  Rev. B}\ }\textbf {\bibinfo {volume} {95}},\ \bibinfo {pages} {024515}
  (\bibinfo {year} {2017})}\BibitemShut {NoStop}%
\bibitem [{\citenamefont {Tinkham}(2004)}]{tinkham2004introduction}%
  \BibitemOpen
  \bibfield  {author} {\bibinfo {author} {\bibfnamefont {M.}~\bibnamefont
  {Tinkham}},\ }\href {http://www.worldcat.org/isbn/0486435032} {\emph
  {\bibinfo {title} {Introduction to Superconductivity}}},\ \bibinfo {edition}
  {2nd}\ ed.\ (\bibinfo  {publisher} {Dover Publications, New York},\ \bibinfo
  {year} {2004})\BibitemShut {NoStop}%
\bibitem [{\citenamefont {Jujo}(2001)}]{jujo2001fermi}%
  \BibitemOpen
  \bibfield  {author} {\bibinfo {author} {\bibfnamefont {T.}~\bibnamefont
  {Jujo}},\ }\href {\doibase 10.1143/JPSJ.70.1349} {\bibfield  {journal}
  {\bibinfo  {journal} {Journal of the Physical Society of Japan}\ }\textbf
  {\bibinfo {volume} {70}},\ \bibinfo {pages} {1349} (\bibinfo {year}
  {2001})},\ \Eprint
  {http://arxiv.org/abs/https://doi.org/10.1143/JPSJ.70.1349}
  {https://doi.org/10.1143/JPSJ.70.1349} \BibitemShut {NoStop}%
\bibitem [{\citenamefont {Taie}\ \emph {et~al.}(2015)\citenamefont {Taie},
  \citenamefont {Ozawa}, \citenamefont {Ichinose}, \citenamefont {Nishio},
  \citenamefont {Nakajima},\ and\ \citenamefont
  {Takahashi}}]{taie2015coherent}%
  \BibitemOpen
  \bibfield  {author} {\bibinfo {author} {\bibfnamefont {S.}~\bibnamefont
  {Taie}}, \bibinfo {author} {\bibfnamefont {H.}~\bibnamefont {Ozawa}},
  \bibinfo {author} {\bibfnamefont {T.}~\bibnamefont {Ichinose}}, \bibinfo
  {author} {\bibfnamefont {T.}~\bibnamefont {Nishio}}, \bibinfo {author}
  {\bibfnamefont {S.}~\bibnamefont {Nakajima}}, \ and\ \bibinfo {author}
  {\bibfnamefont {Y.}~\bibnamefont {Takahashi}},\ }\href {\doibase
  10.1126/sciadv.1500854} {\bibfield  {journal} {\bibinfo  {journal} {Science
  Advances}\ }\textbf {\bibinfo {volume} {1}} (\bibinfo {year} {2015}),\
  10.1126/sciadv.1500854},\ \Eprint
  {http://arxiv.org/abs/https://advances.sciencemag.org/content/1/10/e1500854.full.pdf}
  {https://advances.sciencemag.org/content/1/10/e1500854.full.pdf} \BibitemShut
  {NoStop}%
\bibitem [{\citenamefont {Ozawa}\ \emph {et~al.}(2017)\citenamefont {Ozawa},
  \citenamefont {Taie}, \citenamefont {Ichinose},\ and\ \citenamefont
  {Takahashi}}]{ozawa2017interaction-driven}%
  \BibitemOpen
  \bibfield  {author} {\bibinfo {author} {\bibfnamefont {H.}~\bibnamefont
  {Ozawa}}, \bibinfo {author} {\bibfnamefont {S.}~\bibnamefont {Taie}},
  \bibinfo {author} {\bibfnamefont {T.}~\bibnamefont {Ichinose}}, \ and\
  \bibinfo {author} {\bibfnamefont {Y.}~\bibnamefont {Takahashi}},\ }\href
  {\doibase 10.1103/PhysRevLett.118.175301} {\bibfield  {journal} {\bibinfo
  {journal} {Phys. Rev. Lett.}\ }\textbf {\bibinfo {volume} {118}},\ \bibinfo
  {pages} {175301} (\bibinfo {year} {2017})}\BibitemShut {NoStop}%
\bibitem [{\citenamefont {Julku}\ \emph {et~al.}(2016)\citenamefont {Julku},
  \citenamefont {Peotta}, \citenamefont {Vanhala}, \citenamefont {Kim},\ and\
  \citenamefont {T\"orm\"a}}]{julku2016geometric}%
  \BibitemOpen
  \bibfield  {author} {\bibinfo {author} {\bibfnamefont {A.}~\bibnamefont
  {Julku}}, \bibinfo {author} {\bibfnamefont {S.}~\bibnamefont {Peotta}},
  \bibinfo {author} {\bibfnamefont {T.~I.}\ \bibnamefont {Vanhala}}, \bibinfo
  {author} {\bibfnamefont {D.-H.}\ \bibnamefont {Kim}}, \ and\ \bibinfo
  {author} {\bibfnamefont {P.}~\bibnamefont {T\"orm\"a}},\ }\href {\doibase
  10.1103/PhysRevLett.117.045303} {\bibfield  {journal} {\bibinfo  {journal}
  {Phys. Rev. Lett.}\ }\textbf {\bibinfo {volume} {117}},\ \bibinfo {pages}
  {045303} (\bibinfo {year} {2016})}\BibitemShut {NoStop}%
\bibitem [{\citenamefont {He}\ \emph {et~al.}(2021)\citenamefont {He},
  \citenamefont {Ding},\ and\ \citenamefont {Zhu}}]{he2021geometry}%
  \BibitemOpen
  \bibfield  {author} {\bibinfo {author} {\bibfnamefont {P.}~\bibnamefont
  {He}}, \bibinfo {author} {\bibfnamefont {H.-T.}\ \bibnamefont {Ding}}, \ and\
  \bibinfo {author} {\bibfnamefont {S.-L.}\ \bibnamefont {Zhu}},\ }\href
  {\doibase 10.1103/PhysRevA.103.043329} {\bibfield  {journal} {\bibinfo
  {journal} {Phys. Rev. A}\ }\textbf {\bibinfo {volume} {103}},\ \bibinfo
  {pages} {043329} (\bibinfo {year} {2021})}\BibitemShut {NoStop}%
\bibitem [{\citenamefont {Huhtinen}\ \emph {et~al.}(2022)\citenamefont
  {Huhtinen}, \citenamefont {Herzog-Arbeitman}, \citenamefont {Chew},
  \citenamefont {Bernevig},\ and\ \citenamefont
  {T{\"o}rm{\"a}}}]{huhtinen2022revisiting}%
  \BibitemOpen
  \bibfield  {author} {\bibinfo {author} {\bibfnamefont {K.-E.}\ \bibnamefont
  {Huhtinen}}, \bibinfo {author} {\bibfnamefont {J.}~\bibnamefont
  {Herzog-Arbeitman}}, \bibinfo {author} {\bibfnamefont {A.}~\bibnamefont
  {Chew}}, \bibinfo {author} {\bibfnamefont {B.~A.}\ \bibnamefont {Bernevig}},
  \ and\ \bibinfo {author} {\bibfnamefont {P.}~\bibnamefont {T{\"o}rm{\"a}}},\
  }\href {https://arxiv.org/abs/2203.11133} {\  (\bibinfo {year} {2022})},\
  \Eprint {http://arxiv.org/abs/2203.11133} {arXiv:2203.11133
  [cond-mat.supr-con]} \BibitemShut {NoStop}%
\bibitem [{\citenamefont {Cao}\ \emph {et~al.}(2018)\citenamefont {Cao},
  \citenamefont {Fatemi}, \citenamefont {Fang}, \citenamefont {Watanabe},
  \citenamefont {Taniguchi}, \citenamefont {Kaxiras},\ and\ \citenamefont
  {Jarillo-Herrero}}]{cao2018unconventional}%
  \BibitemOpen
  \bibfield  {author} {\bibinfo {author} {\bibfnamefont {Y.}~\bibnamefont
  {Cao}}, \bibinfo {author} {\bibfnamefont {V.}~\bibnamefont {Fatemi}},
  \bibinfo {author} {\bibfnamefont {S.}~\bibnamefont {Fang}}, \bibinfo {author}
  {\bibfnamefont {K.}~\bibnamefont {Watanabe}}, \bibinfo {author}
  {\bibfnamefont {T.}~\bibnamefont {Taniguchi}}, \bibinfo {author}
  {\bibfnamefont {E.}~\bibnamefont {Kaxiras}}, \ and\ \bibinfo {author}
  {\bibfnamefont {P.}~\bibnamefont {Jarillo-Herrero}},\ }\href {\doibase
  10.1038/nature26160} {\bibfield  {journal} {\bibinfo  {journal} {Nature}\
  }\textbf {\bibinfo {volume} {556}},\ \bibinfo {pages} {43} (\bibinfo {year}
  {2018})}\BibitemShut {NoStop}%
\bibitem [{\citenamefont {Hu}\ \emph {et~al.}(2019)\citenamefont {Hu},
  \citenamefont {Hyart}, \citenamefont {Pikulin},\ and\ \citenamefont
  {Rossi}}]{hu2019geometric}%
  \BibitemOpen
  \bibfield  {author} {\bibinfo {author} {\bibfnamefont {X.}~\bibnamefont
  {Hu}}, \bibinfo {author} {\bibfnamefont {T.}~\bibnamefont {Hyart}}, \bibinfo
  {author} {\bibfnamefont {D.~I.}\ \bibnamefont {Pikulin}}, \ and\ \bibinfo
  {author} {\bibfnamefont {E.}~\bibnamefont {Rossi}},\ }\href {\doibase
  10.1103/PhysRevLett.123.237002} {\bibfield  {journal} {\bibinfo  {journal}
  {Phys. Rev. Lett.}\ }\textbf {\bibinfo {volume} {123}},\ \bibinfo {pages}
  {237002} (\bibinfo {year} {2019})}\BibitemShut {NoStop}%
\bibitem [{\citenamefont {Julku}\ \emph {et~al.}(2020)\citenamefont {Julku},
  \citenamefont {Peltonen}, \citenamefont {Liang}, \citenamefont {Heikkil\"a},\
  and\ \citenamefont {T\"orm\"a}}]{julku2020superfluid}%
  \BibitemOpen
  \bibfield  {author} {\bibinfo {author} {\bibfnamefont {A.}~\bibnamefont
  {Julku}}, \bibinfo {author} {\bibfnamefont {T.~J.}\ \bibnamefont {Peltonen}},
  \bibinfo {author} {\bibfnamefont {L.}~\bibnamefont {Liang}}, \bibinfo
  {author} {\bibfnamefont {T.~T.}\ \bibnamefont {Heikkil\"a}}, \ and\ \bibinfo
  {author} {\bibfnamefont {P.}~\bibnamefont {T\"orm\"a}},\ }\href {\doibase
  10.1103/PhysRevB.101.060505} {\bibfield  {journal} {\bibinfo  {journal}
  {Phys. Rev. B}\ }\textbf {\bibinfo {volume} {101}},\ \bibinfo {pages}
  {060505} (\bibinfo {year} {2020})}\BibitemShut {NoStop}%
\bibitem [{\citenamefont {Xie}\ \emph {et~al.}(2020)\citenamefont {Xie},
  \citenamefont {Song}, \citenamefont {Lian},\ and\ \citenamefont
  {Bernevig}}]{xie2020topology-bonded}%
  \BibitemOpen
  \bibfield  {author} {\bibinfo {author} {\bibfnamefont {F.}~\bibnamefont
  {Xie}}, \bibinfo {author} {\bibfnamefont {Z.}~\bibnamefont {Song}}, \bibinfo
  {author} {\bibfnamefont {B.}~\bibnamefont {Lian}}, \ and\ \bibinfo {author}
  {\bibfnamefont {B.~A.}\ \bibnamefont {Bernevig}},\ }\href {\doibase
  10.1103/PhysRevLett.124.167002} {\bibfield  {journal} {\bibinfo  {journal}
  {Phys. Rev. Lett.}\ }\textbf {\bibinfo {volume} {124}},\ \bibinfo {pages}
  {167002} (\bibinfo {year} {2020})}\BibitemShut {NoStop}%
\bibitem [{\citenamefont {Peri}\ \emph {et~al.}(2021)\citenamefont {Peri},
  \citenamefont {Song}, \citenamefont {Bernevig},\ and\ \citenamefont
  {Huber}}]{peri2021fragile}%
  \BibitemOpen
  \bibfield  {author} {\bibinfo {author} {\bibfnamefont {V.}~\bibnamefont
  {Peri}}, \bibinfo {author} {\bibfnamefont {Z.-D.}\ \bibnamefont {Song}},
  \bibinfo {author} {\bibfnamefont {B.~A.}\ \bibnamefont {Bernevig}}, \ and\
  \bibinfo {author} {\bibfnamefont {S.~D.}\ \bibnamefont {Huber}},\ }\href
  {\doibase 10.1103/PhysRevLett.126.027002} {\bibfield  {journal} {\bibinfo
  {journal} {Phys. Rev. Lett.}\ }\textbf {\bibinfo {volume} {126}},\ \bibinfo
  {pages} {027002} (\bibinfo {year} {2021})}\BibitemShut {NoStop}%
\bibitem [{\citenamefont {Rossi}(2021)}]{rossi2021quantum}%
  \BibitemOpen
  \bibfield  {author} {\bibinfo {author} {\bibfnamefont {E.}~\bibnamefont
  {Rossi}},\ }\href {\doibase https://doi.org/10.1016/j.cossms.2021.100952}
  {\bibfield  {journal} {\bibinfo  {journal} {Current Opinion in Solid State
  and Materials Science}\ }\textbf {\bibinfo {volume} {25}},\ \bibinfo {pages}
  {100952} (\bibinfo {year} {2021})}\BibitemShut {NoStop}%
\bibitem [{\citenamefont {T{\"o}rm{\"a}}\ \emph {et~al.}(2022)\citenamefont
  {T{\"o}rm{\"a}}, \citenamefont {Peotta},\ and\ \citenamefont
  {Bernevig}}]{torma2022superconductivity}%
  \BibitemOpen
  \bibfield  {author} {\bibinfo {author} {\bibfnamefont {P.}~\bibnamefont
  {T{\"o}rm{\"a}}}, \bibinfo {author} {\bibfnamefont {S.}~\bibnamefont
  {Peotta}}, \ and\ \bibinfo {author} {\bibfnamefont {B.~A.}\ \bibnamefont
  {Bernevig}},\ }\href {https://www.nature.com/articles/s42254-022-00466-y}
  {\bibfield  {journal} {\bibinfo  {journal} {Nature Reviews Physics}\ ,\
  \bibinfo {pages} {1}} (\bibinfo {year} {2022})}\BibitemShut {NoStop}%
\bibitem [{\citenamefont {Hu}\ \emph {et~al.}(2022)\citenamefont {Hu},
  \citenamefont {Hyart}, \citenamefont {Pikulin},\ and\ \citenamefont
  {Rossi}}]{hu2022quantum}%
  \BibitemOpen
  \bibfield  {author} {\bibinfo {author} {\bibfnamefont {X.}~\bibnamefont
  {Hu}}, \bibinfo {author} {\bibfnamefont {T.}~\bibnamefont {Hyart}}, \bibinfo
  {author} {\bibfnamefont {D.~I.}\ \bibnamefont {Pikulin}}, \ and\ \bibinfo
  {author} {\bibfnamefont {E.}~\bibnamefont {Rossi}},\ }\href {\doibase
  10.1103/PhysRevB.105.L140506} {\bibfield  {journal} {\bibinfo  {journal}
  {Phys. Rev. B}\ }\textbf {\bibinfo {volume} {105}},\ \bibinfo {pages}
  {L140506} (\bibinfo {year} {2022})}\BibitemShut {NoStop}%
\bibitem [{\citenamefont {Herzog-Arbeitman}\ \emph {et~al.}(2022)\citenamefont
  {Herzog-Arbeitman}, \citenamefont {Peri}, \citenamefont {Schindler},
  \citenamefont {Huber},\ and\ \citenamefont
  {Bernevig}}]{herzogarbeitman2021superfluid}%
  \BibitemOpen
  \bibfield  {author} {\bibinfo {author} {\bibfnamefont {J.}~\bibnamefont
  {Herzog-Arbeitman}}, \bibinfo {author} {\bibfnamefont {V.}~\bibnamefont
  {Peri}}, \bibinfo {author} {\bibfnamefont {F.}~\bibnamefont {Schindler}},
  \bibinfo {author} {\bibfnamefont {S.~D.}\ \bibnamefont {Huber}}, \ and\
  \bibinfo {author} {\bibfnamefont {B.~A.}\ \bibnamefont {Bernevig}},\ }\href
  {\doibase 10.1103/PhysRevLett.128.087002} {\bibfield  {journal} {\bibinfo
  {journal} {Phys. Rev. Lett.}\ }\textbf {\bibinfo {volume} {128}},\ \bibinfo
  {pages} {087002} (\bibinfo {year} {2022})}\BibitemShut {NoStop}%
\bibitem [{\citenamefont {Tian}\ \emph {et~al.}(2021)\citenamefont {Tian},
  \citenamefont {Che}, \citenamefont {Xu}, \citenamefont {Cheung},
  \citenamefont {Watanabe}, \citenamefont {Taniguchi}, \citenamefont
  {Randeria}, \citenamefont {Zhang}, \citenamefont {Lau},\ and\ \citenamefont
  {Bockrath}}]{tian2021evidence}%
  \BibitemOpen
  \bibfield  {author} {\bibinfo {author} {\bibfnamefont {H.}~\bibnamefont
  {Tian}}, \bibinfo {author} {\bibfnamefont {S.}~\bibnamefont {Che}}, \bibinfo
  {author} {\bibfnamefont {T.}~\bibnamefont {Xu}}, \bibinfo {author}
  {\bibfnamefont {P.}~\bibnamefont {Cheung}}, \bibinfo {author} {\bibfnamefont
  {K.}~\bibnamefont {Watanabe}}, \bibinfo {author} {\bibfnamefont
  {T.}~\bibnamefont {Taniguchi}}, \bibinfo {author} {\bibfnamefont
  {M.}~\bibnamefont {Randeria}}, \bibinfo {author} {\bibfnamefont
  {F.}~\bibnamefont {Zhang}}, \bibinfo {author} {\bibfnamefont {C.~N.}\
  \bibnamefont {Lau}}, \ and\ \bibinfo {author} {\bibfnamefont {M.~W.}\
  \bibnamefont {Bockrath}},\ }\href@noop {} {\  (\bibinfo {year} {2021})},\
  \Eprint {http://arxiv.org/abs/2112.13401} {arXiv:2112.13401
  [cond-mat.supr-con]} \BibitemShut {NoStop}%
\bibitem [{\citenamefont {Kitamura}\ \emph {et~al.}(2022)\citenamefont
  {Kitamura}, \citenamefont {Yamashita}, \citenamefont {Ishizuka},
  \citenamefont {Daido},\ and\ \citenamefont
  {Yanase}}]{kitamura2021superconductivity}%
  \BibitemOpen
  \bibfield  {author} {\bibinfo {author} {\bibfnamefont {T.}~\bibnamefont
  {Kitamura}}, \bibinfo {author} {\bibfnamefont {T.}~\bibnamefont {Yamashita}},
  \bibinfo {author} {\bibfnamefont {J.}~\bibnamefont {Ishizuka}}, \bibinfo
  {author} {\bibfnamefont {A.}~\bibnamefont {Daido}}, \ and\ \bibinfo {author}
  {\bibfnamefont {Y.}~\bibnamefont {Yanase}},\ }\href {\doibase
  10.1103/PhysRevResearch.4.023232} {\bibfield  {journal} {\bibinfo  {journal}
  {Phys. Rev. Research}\ }\textbf {\bibinfo {volume} {4}},\ \bibinfo {pages}
  {023232} (\bibinfo {year} {2022})}\BibitemShut {NoStop}%
\bibitem [{\citenamefont {Wang}\ \emph {et~al.}(2012)\citenamefont {Wang},
  \citenamefont {Li}, \citenamefont {Zhang}, \citenamefont {Zhang},
  \citenamefont {Zhang}, \citenamefont {Li}, \citenamefont {Ding},
  \citenamefont {Ou}, \citenamefont {Deng}, \citenamefont {Chang},
  \citenamefont {Wen}, \citenamefont {Song}, \citenamefont {He}, \citenamefont
  {Jia}, \citenamefont {Ji}, \citenamefont {Wang}, \citenamefont {Wang},
  \citenamefont {Chen}, \citenamefont {Ma},\ and\ \citenamefont
  {Xue}}]{wang2012interface-induced}%
  \BibitemOpen
  \bibfield  {author} {\bibinfo {author} {\bibfnamefont {Q.-Y.}\ \bibnamefont
  {Wang}}, \bibinfo {author} {\bibfnamefont {Z.}~\bibnamefont {Li}}, \bibinfo
  {author} {\bibfnamefont {W.-H.}\ \bibnamefont {Zhang}}, \bibinfo {author}
  {\bibfnamefont {Z.-C.}\ \bibnamefont {Zhang}}, \bibinfo {author}
  {\bibfnamefont {J.-S.}\ \bibnamefont {Zhang}}, \bibinfo {author}
  {\bibfnamefont {W.}~\bibnamefont {Li}}, \bibinfo {author} {\bibfnamefont
  {H.}~\bibnamefont {Ding}}, \bibinfo {author} {\bibfnamefont {Y.-B.}\
  \bibnamefont {Ou}}, \bibinfo {author} {\bibfnamefont {P.}~\bibnamefont
  {Deng}}, \bibinfo {author} {\bibfnamefont {K.}~\bibnamefont {Chang}},
  \bibinfo {author} {\bibfnamefont {J.}~\bibnamefont {Wen}}, \bibinfo {author}
  {\bibfnamefont {C.-L.}\ \bibnamefont {Song}}, \bibinfo {author}
  {\bibfnamefont {K.}~\bibnamefont {He}}, \bibinfo {author} {\bibfnamefont
  {J.-F.}\ \bibnamefont {Jia}}, \bibinfo {author} {\bibfnamefont {S.-H.}\
  \bibnamefont {Ji}}, \bibinfo {author} {\bibfnamefont {Y.-Y.}\ \bibnamefont
  {Wang}}, \bibinfo {author} {\bibfnamefont {L.-L.}\ \bibnamefont {Wang}},
  \bibinfo {author} {\bibfnamefont {X.}~\bibnamefont {Chen}}, \bibinfo {author}
  {\bibfnamefont {X.-C.}\ \bibnamefont {Ma}}, \ and\ \bibinfo {author}
  {\bibfnamefont {Q.-K.}\ \bibnamefont {Xue}},\ }\href {\doibase
  10.1088/0256-307x/29/3/037402} {\bibfield  {journal} {\bibinfo  {journal}
  {Chinese Physics Letters}\ }\textbf {\bibinfo {volume} {29}},\ \bibinfo
  {pages} {037402} (\bibinfo {year} {2012})}\BibitemShut {NoStop}%
\bibitem [{\citenamefont {He}\ \emph {et~al.}(2013)\citenamefont {He},
  \citenamefont {He}, \citenamefont {Zhang}, \citenamefont {Zhao},
  \citenamefont {Liu}, \citenamefont {Liu}, \citenamefont {Mou}, \citenamefont
  {Ou}, \citenamefont {Wang}, \citenamefont {Li}, \citenamefont {Wang},
  \citenamefont {Peng}, \citenamefont {Liu}, \citenamefont {Chen},
  \citenamefont {Yu}, \citenamefont {Liu}, \citenamefont {Dong}, \citenamefont
  {Zhang}, \citenamefont {Chen}, \citenamefont {Xu}, \citenamefont {Chen},
  \citenamefont {Ma}, \citenamefont {Xue},\ and\ \citenamefont
  {Zhou}}]{he2013phase}%
  \BibitemOpen
  \bibfield  {author} {\bibinfo {author} {\bibfnamefont {S.}~\bibnamefont
  {He}}, \bibinfo {author} {\bibfnamefont {J.}~\bibnamefont {He}}, \bibinfo
  {author} {\bibfnamefont {W.}~\bibnamefont {Zhang}}, \bibinfo {author}
  {\bibfnamefont {L.}~\bibnamefont {Zhao}}, \bibinfo {author} {\bibfnamefont
  {D.}~\bibnamefont {Liu}}, \bibinfo {author} {\bibfnamefont {X.}~\bibnamefont
  {Liu}}, \bibinfo {author} {\bibfnamefont {D.}~\bibnamefont {Mou}}, \bibinfo
  {author} {\bibfnamefont {Y.-B.}\ \bibnamefont {Ou}}, \bibinfo {author}
  {\bibfnamefont {Q.-Y.}\ \bibnamefont {Wang}}, \bibinfo {author}
  {\bibfnamefont {Z.}~\bibnamefont {Li}}, \bibinfo {author} {\bibfnamefont
  {L.}~\bibnamefont {Wang}}, \bibinfo {author} {\bibfnamefont {Y.}~\bibnamefont
  {Peng}}, \bibinfo {author} {\bibfnamefont {Y.}~\bibnamefont {Liu}}, \bibinfo
  {author} {\bibfnamefont {C.}~\bibnamefont {Chen}}, \bibinfo {author}
  {\bibfnamefont {L.}~\bibnamefont {Yu}}, \bibinfo {author} {\bibfnamefont
  {G.}~\bibnamefont {Liu}}, \bibinfo {author} {\bibfnamefont {X.}~\bibnamefont
  {Dong}}, \bibinfo {author} {\bibfnamefont {J.}~\bibnamefont {Zhang}},
  \bibinfo {author} {\bibfnamefont {C.}~\bibnamefont {Chen}}, \bibinfo {author}
  {\bibfnamefont {Z.}~\bibnamefont {Xu}}, \bibinfo {author} {\bibfnamefont
  {X.}~\bibnamefont {Chen}}, \bibinfo {author} {\bibfnamefont {X.}~\bibnamefont
  {Ma}}, \bibinfo {author} {\bibfnamefont {Q.}~\bibnamefont {Xue}}, \ and\
  \bibinfo {author} {\bibfnamefont {X.~J.}\ \bibnamefont {Zhou}},\ }\href
  {\doibase 10.1038/nmat3648} {\bibfield  {journal} {\bibinfo  {journal}
  {Nature Materials}\ }\textbf {\bibinfo {volume} {12}},\ \bibinfo {pages}
  {605} (\bibinfo {year} {2013})}\BibitemShut {NoStop}%
\bibitem [{\citenamefont {Xu}\ \emph {et~al.}(2021)\citenamefont {Xu},
  \citenamefont {Rong}, \citenamefont {Wang}, \citenamefont {Wu}, \citenamefont
  {Hu}, \citenamefont {Cai}, \citenamefont {Gao}, \citenamefont {Yan},
  \citenamefont {Li}, \citenamefont {Yin}, \citenamefont {Chen}, \citenamefont
  {Huang}, \citenamefont {Zhu}, \citenamefont {Huang}, \citenamefont {Liu},
  \citenamefont {Xu}, \citenamefont {Zhao},\ and\ \citenamefont
  {Zhou}}]{xu2020spectroscopic}%
  \BibitemOpen
  \bibfield  {author} {\bibinfo {author} {\bibfnamefont {Y.}~\bibnamefont
  {Xu}}, \bibinfo {author} {\bibfnamefont {H.}~\bibnamefont {Rong}}, \bibinfo
  {author} {\bibfnamefont {Q.}~\bibnamefont {Wang}}, \bibinfo {author}
  {\bibfnamefont {D.}~\bibnamefont {Wu}}, \bibinfo {author} {\bibfnamefont
  {Y.}~\bibnamefont {Hu}}, \bibinfo {author} {\bibfnamefont {Y.}~\bibnamefont
  {Cai}}, \bibinfo {author} {\bibfnamefont {Q.}~\bibnamefont {Gao}}, \bibinfo
  {author} {\bibfnamefont {H.}~\bibnamefont {Yan}}, \bibinfo {author}
  {\bibfnamefont {C.}~\bibnamefont {Li}}, \bibinfo {author} {\bibfnamefont
  {C.}~\bibnamefont {Yin}}, \bibinfo {author} {\bibfnamefont {H.}~\bibnamefont
  {Chen}}, \bibinfo {author} {\bibfnamefont {J.}~\bibnamefont {Huang}},
  \bibinfo {author} {\bibfnamefont {Z.}~\bibnamefont {Zhu}}, \bibinfo {author}
  {\bibfnamefont {Y.}~\bibnamefont {Huang}}, \bibinfo {author} {\bibfnamefont
  {G.}~\bibnamefont {Liu}}, \bibinfo {author} {\bibfnamefont {Z.}~\bibnamefont
  {Xu}}, \bibinfo {author} {\bibfnamefont {L.}~\bibnamefont {Zhao}}, \ and\
  \bibinfo {author} {\bibfnamefont {X.~J.}\ \bibnamefont {Zhou}},\ }\href
  {\doibase 10.1038/s41467-021-23106-y} {\bibfield  {journal} {\bibinfo
  {journal} {Nature Communications}\ }\textbf {\bibinfo {volume} {12}},\
  \bibinfo {pages} {2840} (\bibinfo {year} {2021})}\BibitemShut {NoStop}%
\bibitem [{\citenamefont {Nozi{\`e}res}\ and\ \citenamefont
  {Schmitt-Rink}(1985)}]{nozieres1985bose}%
  \BibitemOpen
  \bibfield  {author} {\bibinfo {author} {\bibfnamefont {P.}~\bibnamefont
  {Nozi{\`e}res}}\ and\ \bibinfo {author} {\bibfnamefont {S.}~\bibnamefont
  {Schmitt-Rink}},\ }\href {\doibase 10.1007/BF00683774} {\bibfield  {journal}
  {\bibinfo  {journal} {Journal of Low Temperature Physics}\ }\textbf {\bibinfo
  {volume} {59}},\ \bibinfo {pages} {195} (\bibinfo {year} {1985})}\BibitemShut
  {NoStop}%
\bibitem [{\citenamefont {Kasahara}\ \emph {et~al.}(2014)\citenamefont
  {Kasahara}, \citenamefont {Watashige}, \citenamefont {Hanaguri},
  \citenamefont {Kohsaka}, \citenamefont {Yamashita}, \citenamefont
  {Shimoyama}, \citenamefont {Mizukami}, \citenamefont {Endo}, \citenamefont
  {Ikeda}, \citenamefont {Aoyama}, \citenamefont {Terashima}, \citenamefont
  {Uji}, \citenamefont {Wolf}, \citenamefont {von L{\"o}hneysen}, \citenamefont
  {Shibauchi},\ and\ \citenamefont {Matsuda}}]{kasahara2014field-induced}%
  \BibitemOpen
  \bibfield  {author} {\bibinfo {author} {\bibfnamefont {S.}~\bibnamefont
  {Kasahara}}, \bibinfo {author} {\bibfnamefont {T.}~\bibnamefont {Watashige}},
  \bibinfo {author} {\bibfnamefont {T.}~\bibnamefont {Hanaguri}}, \bibinfo
  {author} {\bibfnamefont {Y.}~\bibnamefont {Kohsaka}}, \bibinfo {author}
  {\bibfnamefont {T.}~\bibnamefont {Yamashita}}, \bibinfo {author}
  {\bibfnamefont {Y.}~\bibnamefont {Shimoyama}}, \bibinfo {author}
  {\bibfnamefont {Y.}~\bibnamefont {Mizukami}}, \bibinfo {author}
  {\bibfnamefont {R.}~\bibnamefont {Endo}}, \bibinfo {author} {\bibfnamefont
  {H.}~\bibnamefont {Ikeda}}, \bibinfo {author} {\bibfnamefont
  {K.}~\bibnamefont {Aoyama}}, \bibinfo {author} {\bibfnamefont
  {T.}~\bibnamefont {Terashima}}, \bibinfo {author} {\bibfnamefont
  {S.}~\bibnamefont {Uji}}, \bibinfo {author} {\bibfnamefont {T.}~\bibnamefont
  {Wolf}}, \bibinfo {author} {\bibfnamefont {H.}~\bibnamefont {von
  L{\"o}hneysen}}, \bibinfo {author} {\bibfnamefont {T.}~\bibnamefont
  {Shibauchi}}, \ and\ \bibinfo {author} {\bibfnamefont {Y.}~\bibnamefont
  {Matsuda}},\ }\href {\doibase 10.1073/pnas.1413477111} {\bibfield  {journal}
  {\bibinfo  {journal} {Proceedings of the National Academy of Sciences}\
  }\textbf {\bibinfo {volume} {111}},\ \bibinfo {pages} {16309} (\bibinfo
  {year} {2014})},\ \Eprint
  {http://arxiv.org/abs/https://www.pnas.org/content/111/46/16309.full.pdf}
  {https://www.pnas.org/content/111/46/16309.full.pdf} \BibitemShut {NoStop}%
\bibitem [{\citenamefont {Kasahara}\ \emph {et~al.}(2016)\citenamefont
  {Kasahara}, \citenamefont {Yamashita}, \citenamefont {Shi}, \citenamefont
  {Kobayashi}, \citenamefont {Shimoyama}, \citenamefont {Watashige},
  \citenamefont {Ishida}, \citenamefont {Terashima}, \citenamefont {Wolf},
  \citenamefont {Hardy}, \citenamefont {Meingast}, \citenamefont
  {L{\"o}hneysen}, \citenamefont {Levchenko}, \citenamefont {Shibauchi},\ and\
  \citenamefont {Matsuda}}]{kasahara2016giant}%
  \BibitemOpen
  \bibfield  {author} {\bibinfo {author} {\bibfnamefont {S.}~\bibnamefont
  {Kasahara}}, \bibinfo {author} {\bibfnamefont {T.}~\bibnamefont {Yamashita}},
  \bibinfo {author} {\bibfnamefont {A.}~\bibnamefont {Shi}}, \bibinfo {author}
  {\bibfnamefont {R.}~\bibnamefont {Kobayashi}}, \bibinfo {author}
  {\bibfnamefont {Y.}~\bibnamefont {Shimoyama}}, \bibinfo {author}
  {\bibfnamefont {T.}~\bibnamefont {Watashige}}, \bibinfo {author}
  {\bibfnamefont {K.}~\bibnamefont {Ishida}}, \bibinfo {author} {\bibfnamefont
  {T.}~\bibnamefont {Terashima}}, \bibinfo {author} {\bibfnamefont
  {T.}~\bibnamefont {Wolf}}, \bibinfo {author} {\bibfnamefont {F.}~\bibnamefont
  {Hardy}}, \bibinfo {author} {\bibfnamefont {C.}~\bibnamefont {Meingast}},
  \bibinfo {author} {\bibfnamefont {H.~v.}\ \bibnamefont {L{\"o}hneysen}},
  \bibinfo {author} {\bibfnamefont {A.}~\bibnamefont {Levchenko}}, \bibinfo
  {author} {\bibfnamefont {T.}~\bibnamefont {Shibauchi}}, \ and\ \bibinfo
  {author} {\bibfnamefont {Y.}~\bibnamefont {Matsuda}},\ }\href {\doibase
  10.1038/ncomms12843} {\bibfield  {journal} {\bibinfo  {journal} {Nature
  Communications}\ }\textbf {\bibinfo {volume} {7}},\ \bibinfo {pages} {12843}
  (\bibinfo {year} {2016})}\BibitemShut {NoStop}%
\bibitem [{\citenamefont {Hanaguri}\ \emph {et~al.}(2019)\citenamefont
  {Hanaguri}, \citenamefont {Kasahara}, \citenamefont {B\"oker}, \citenamefont
  {Eremin}, \citenamefont {Shibauchi},\ and\ \citenamefont
  {Matsuda}}]{hanaguri2019quantum}%
  \BibitemOpen
  \bibfield  {author} {\bibinfo {author} {\bibfnamefont {T.}~\bibnamefont
  {Hanaguri}}, \bibinfo {author} {\bibfnamefont {S.}~\bibnamefont {Kasahara}},
  \bibinfo {author} {\bibfnamefont {J.}~\bibnamefont {B\"oker}}, \bibinfo
  {author} {\bibfnamefont {I.}~\bibnamefont {Eremin}}, \bibinfo {author}
  {\bibfnamefont {T.}~\bibnamefont {Shibauchi}}, \ and\ \bibinfo {author}
  {\bibfnamefont {Y.}~\bibnamefont {Matsuda}},\ }\href {\doibase
  10.1103/PhysRevLett.122.077001} {\bibfield  {journal} {\bibinfo  {journal}
  {Phys. Rev. Lett.}\ }\textbf {\bibinfo {volume} {122}},\ \bibinfo {pages}
  {077001} (\bibinfo {year} {2019})}\BibitemShut {NoStop}%
\bibitem [{\citenamefont {Kasahara}\ \emph {et~al.}(2020)\citenamefont
  {Kasahara}, \citenamefont {Sato}, \citenamefont {Licciardello}, \citenamefont
  {\ifmmode~\check{C}\else \v{C}\fi{}ulo}, \citenamefont
  {Arsenijevi\ifmmode~\acute{c}\else \'{c}\fi{}}, \citenamefont {Ottenbros},
  \citenamefont {Tominaga}, \citenamefont {B\"oker}, \citenamefont {Eremin},
  \citenamefont {Shibauchi}, \citenamefont {Wosnitza}, \citenamefont {Hussey},\
  and\ \citenamefont {Matsuda}}]{kasahara2020evidence}%
  \BibitemOpen
  \bibfield  {author} {\bibinfo {author} {\bibfnamefont {S.}~\bibnamefont
  {Kasahara}}, \bibinfo {author} {\bibfnamefont {Y.}~\bibnamefont {Sato}},
  \bibinfo {author} {\bibfnamefont {S.}~\bibnamefont {Licciardello}}, \bibinfo
  {author} {\bibfnamefont {M.}~\bibnamefont {\ifmmode~\check{C}\else
  \v{C}\fi{}ulo}}, \bibinfo {author} {\bibfnamefont {S.}~\bibnamefont
  {Arsenijevi\ifmmode~\acute{c}\else \'{c}\fi{}}}, \bibinfo {author}
  {\bibfnamefont {T.}~\bibnamefont {Ottenbros}}, \bibinfo {author}
  {\bibfnamefont {T.}~\bibnamefont {Tominaga}}, \bibinfo {author}
  {\bibfnamefont {J.}~\bibnamefont {B\"oker}}, \bibinfo {author} {\bibfnamefont
  {I.}~\bibnamefont {Eremin}}, \bibinfo {author} {\bibfnamefont
  {T.}~\bibnamefont {Shibauchi}}, \bibinfo {author} {\bibfnamefont
  {J.}~\bibnamefont {Wosnitza}}, \bibinfo {author} {\bibfnamefont {N.~E.}\
  \bibnamefont {Hussey}}, \ and\ \bibinfo {author} {\bibfnamefont
  {Y.}~\bibnamefont {Matsuda}},\ }\href {\doibase
  10.1103/PhysRevLett.124.107001} {\bibfield  {journal} {\bibinfo  {journal}
  {Phys. Rev. Lett.}\ }\textbf {\bibinfo {volume} {124}},\ \bibinfo {pages}
  {107001} (\bibinfo {year} {2020})}\BibitemShut {NoStop}%
\bibitem [{\citenamefont {Wang}\ \emph {et~al.}(2015)\citenamefont {Wang},
  \citenamefont {Zhang}, \citenamefont {Xu}, \citenamefont {Zeng},
  \citenamefont {Miao}, \citenamefont {Xu}, \citenamefont {Qian}, \citenamefont
  {Weng}, \citenamefont {Richard}, \citenamefont {Fedorov}, \citenamefont
  {Ding}, \citenamefont {Dai},\ and\ \citenamefont
  {Fang}}]{wang2015topological}%
  \BibitemOpen
  \bibfield  {author} {\bibinfo {author} {\bibfnamefont {Z.}~\bibnamefont
  {Wang}}, \bibinfo {author} {\bibfnamefont {P.}~\bibnamefont {Zhang}},
  \bibinfo {author} {\bibfnamefont {G.}~\bibnamefont {Xu}}, \bibinfo {author}
  {\bibfnamefont {L.~K.}\ \bibnamefont {Zeng}}, \bibinfo {author}
  {\bibfnamefont {H.}~\bibnamefont {Miao}}, \bibinfo {author} {\bibfnamefont
  {X.}~\bibnamefont {Xu}}, \bibinfo {author} {\bibfnamefont {T.}~\bibnamefont
  {Qian}}, \bibinfo {author} {\bibfnamefont {H.}~\bibnamefont {Weng}}, \bibinfo
  {author} {\bibfnamefont {P.}~\bibnamefont {Richard}}, \bibinfo {author}
  {\bibfnamefont {A.~V.}\ \bibnamefont {Fedorov}}, \bibinfo {author}
  {\bibfnamefont {H.}~\bibnamefont {Ding}}, \bibinfo {author} {\bibfnamefont
  {X.}~\bibnamefont {Dai}}, \ and\ \bibinfo {author} {\bibfnamefont
  {Z.}~\bibnamefont {Fang}},\ }\href {\doibase 10.1103/PhysRevB.92.115119}
  {\bibfield  {journal} {\bibinfo  {journal} {Phys. Rev. B}\ }\textbf {\bibinfo
  {volume} {92}},\ \bibinfo {pages} {115119} (\bibinfo {year}
  {2015})}\BibitemShut {NoStop}%
\bibitem [{\citenamefont {Xu}\ \emph {et~al.}(2016)\citenamefont {Xu},
  \citenamefont {Lian}, \citenamefont {Tang}, \citenamefont {Qi},\ and\
  \citenamefont {Zhang}}]{xu2016topological}%
  \BibitemOpen
  \bibfield  {author} {\bibinfo {author} {\bibfnamefont {G.}~\bibnamefont
  {Xu}}, \bibinfo {author} {\bibfnamefont {B.}~\bibnamefont {Lian}}, \bibinfo
  {author} {\bibfnamefont {P.}~\bibnamefont {Tang}}, \bibinfo {author}
  {\bibfnamefont {X.-L.}\ \bibnamefont {Qi}}, \ and\ \bibinfo {author}
  {\bibfnamefont {S.-C.}\ \bibnamefont {Zhang}},\ }\href {\doibase
  10.1103/PhysRevLett.117.047001} {\bibfield  {journal} {\bibinfo  {journal}
  {Phys. Rev. Lett.}\ }\textbf {\bibinfo {volume} {117}},\ \bibinfo {pages}
  {047001} (\bibinfo {year} {2016})}\BibitemShut {NoStop}%
\bibitem [{\citenamefont {Wang}\ \emph {et~al.}(2018)\citenamefont {Wang},
  \citenamefont {Kong}, \citenamefont {Fan}, \citenamefont {Chen},
  \citenamefont {Zhu}, \citenamefont {Liu}, \citenamefont {Cao}, \citenamefont
  {Sun}, \citenamefont {Du}, \citenamefont {Schneeloch}, \citenamefont {Zhong},
  \citenamefont {Gu}, \citenamefont {Fu}, \citenamefont {Ding},\ and\
  \citenamefont {Gao}}]{wang2018evidence}%
  \BibitemOpen
  \bibfield  {author} {\bibinfo {author} {\bibfnamefont {D.}~\bibnamefont
  {Wang}}, \bibinfo {author} {\bibfnamefont {L.}~\bibnamefont {Kong}}, \bibinfo
  {author} {\bibfnamefont {P.}~\bibnamefont {Fan}}, \bibinfo {author}
  {\bibfnamefont {H.}~\bibnamefont {Chen}}, \bibinfo {author} {\bibfnamefont
  {S.}~\bibnamefont {Zhu}}, \bibinfo {author} {\bibfnamefont {W.}~\bibnamefont
  {Liu}}, \bibinfo {author} {\bibfnamefont {L.}~\bibnamefont {Cao}}, \bibinfo
  {author} {\bibfnamefont {Y.}~\bibnamefont {Sun}}, \bibinfo {author}
  {\bibfnamefont {S.}~\bibnamefont {Du}}, \bibinfo {author} {\bibfnamefont
  {J.}~\bibnamefont {Schneeloch}}, \bibinfo {author} {\bibfnamefont
  {R.}~\bibnamefont {Zhong}}, \bibinfo {author} {\bibfnamefont
  {G.}~\bibnamefont {Gu}}, \bibinfo {author} {\bibfnamefont {L.}~\bibnamefont
  {Fu}}, \bibinfo {author} {\bibfnamefont {H.}~\bibnamefont {Ding}}, \ and\
  \bibinfo {author} {\bibfnamefont {H.-J.}\ \bibnamefont {Gao}},\ }\href
  {\doibase 10.1126/science.aao1797} {\bibfield  {journal} {\bibinfo  {journal}
  {Science}\ }\textbf {\bibinfo {volume} {362}},\ \bibinfo {pages} {333}
  (\bibinfo {year} {2018})},\ \Eprint
  {http://arxiv.org/abs/https://science.sciencemag.org/content/362/6412/333.full.pdf}
  {https://science.sciencemag.org/content/362/6412/333.full.pdf} \BibitemShut
  {NoStop}%
\bibitem [{\citenamefont {Zhang}\ \emph {et~al.}(2018)\citenamefont {Zhang},
  \citenamefont {Yaji}, \citenamefont {Hashimoto}, \citenamefont {Ota},
  \citenamefont {Kondo}, \citenamefont {Okazaki}, \citenamefont {Wang},
  \citenamefont {Wen}, \citenamefont {Gu}, \citenamefont {Ding},\ and\
  \citenamefont {Shin}}]{zhang2018obsevation}%
  \BibitemOpen
  \bibfield  {author} {\bibinfo {author} {\bibfnamefont {P.}~\bibnamefont
  {Zhang}}, \bibinfo {author} {\bibfnamefont {K.}~\bibnamefont {Yaji}},
  \bibinfo {author} {\bibfnamefont {T.}~\bibnamefont {Hashimoto}}, \bibinfo
  {author} {\bibfnamefont {Y.}~\bibnamefont {Ota}}, \bibinfo {author}
  {\bibfnamefont {T.}~\bibnamefont {Kondo}}, \bibinfo {author} {\bibfnamefont
  {K.}~\bibnamefont {Okazaki}}, \bibinfo {author} {\bibfnamefont
  {Z.}~\bibnamefont {Wang}}, \bibinfo {author} {\bibfnamefont {J.}~\bibnamefont
  {Wen}}, \bibinfo {author} {\bibfnamefont {G.~D.}\ \bibnamefont {Gu}},
  \bibinfo {author} {\bibfnamefont {H.}~\bibnamefont {Ding}}, \ and\ \bibinfo
  {author} {\bibfnamefont {S.}~\bibnamefont {Shin}},\ }\href {\doibase
  10.1126/science.aan4596} {\bibfield  {journal} {\bibinfo  {journal}
  {Science}\ }\textbf {\bibinfo {volume} {360}},\ \bibinfo {pages} {182}
  (\bibinfo {year} {2018})},\ \Eprint
  {http://arxiv.org/abs/https://science.sciencemag.org/content/360/6385/182.full.pdf}
  {https://science.sciencemag.org/content/360/6385/182.full.pdf} \BibitemShut
  {NoStop}%
\bibitem [{\citenamefont {Machida}\ \emph {et~al.}(2019)\citenamefont
  {Machida}, \citenamefont {Sun}, \citenamefont {Pyon}, \citenamefont {Takeda},
  \citenamefont {Kohsaka}, \citenamefont {Hanaguri}, \citenamefont {Sasagawa},\
  and\ \citenamefont {Tamegai}}]{machida2019zero-energy}%
  \BibitemOpen
  \bibfield  {author} {\bibinfo {author} {\bibfnamefont {T.}~\bibnamefont
  {Machida}}, \bibinfo {author} {\bibfnamefont {Y.}~\bibnamefont {Sun}},
  \bibinfo {author} {\bibfnamefont {S.}~\bibnamefont {Pyon}}, \bibinfo {author}
  {\bibfnamefont {S.}~\bibnamefont {Takeda}}, \bibinfo {author} {\bibfnamefont
  {Y.}~\bibnamefont {Kohsaka}}, \bibinfo {author} {\bibfnamefont
  {T.}~\bibnamefont {Hanaguri}}, \bibinfo {author} {\bibfnamefont
  {T.}~\bibnamefont {Sasagawa}}, \ and\ \bibinfo {author} {\bibfnamefont
  {T.}~\bibnamefont {Tamegai}},\ }\href {\doibase 10.1038/s41563-019-0397-1}
  {\bibfield  {journal} {\bibinfo  {journal} {Nature Materials}\ }\textbf
  {\bibinfo {volume} {18}},\ \bibinfo {pages} {811} (\bibinfo {year}
  {2019})}\BibitemShut {NoStop}%
\bibitem [{\citenamefont {Fulde}\ and\ \citenamefont
  {Ferrell}(1964)}]{flude1964superconductivity}%
  \BibitemOpen
  \bibfield  {author} {\bibinfo {author} {\bibfnamefont {P.}~\bibnamefont
  {Fulde}}\ and\ \bibinfo {author} {\bibfnamefont {R.~A.}\ \bibnamefont
  {Ferrell}},\ }\href {\doibase 10.1103/PhysRev.135.A550} {\bibfield  {journal}
  {\bibinfo  {journal} {Phys. Rev.}\ }\textbf {\bibinfo {volume} {135}},\
  \bibinfo {pages} {A550} (\bibinfo {year} {1964})}\BibitemShut {NoStop}%
\bibitem [{\citenamefont {Larkin}\ and\ \citenamefont
  {Ovchinnikov}(1964)}]{larkin1964nonuniform}%
  \BibitemOpen
  \bibfield  {author} {\bibinfo {author} {\bibfnamefont {A.~I.}\ \bibnamefont
  {Larkin}}\ and\ \bibinfo {author} {\bibfnamefont {Y.~N.}\ \bibnamefont
  {Ovchinnikov}},\ }\href@noop {} {\bibfield  {journal} {\bibinfo  {journal}
  {Zh. Eksp. Teor. Fiz.}\ }\textbf {\bibinfo {volume} {47}},\ \bibinfo {pages}
  {1136} (\bibinfo {year} {1964})},\ \bibinfo {note} {[Sov. Phys. JETP 20, 762
  (1965)]}\BibitemShut {NoStop}%
\bibitem [{\citenamefont {Gao}\ \emph {et~al.}(2016)\citenamefont {Gao},
  \citenamefont {Yu}, \citenamefont {Zhou}, \citenamefont {Huang},\ and\
  \citenamefont {Wang}}]{gao20216hidden}%
  \BibitemOpen
  \bibfield  {author} {\bibinfo {author} {\bibfnamefont {Y.}~\bibnamefont
  {Gao}}, \bibinfo {author} {\bibfnamefont {Y.}~\bibnamefont {Yu}}, \bibinfo
  {author} {\bibfnamefont {T.}~\bibnamefont {Zhou}}, \bibinfo {author}
  {\bibfnamefont {H.}~\bibnamefont {Huang}}, \ and\ \bibinfo {author}
  {\bibfnamefont {Q.-H.}\ \bibnamefont {Wang}},\ }\href {\doibase
  10.1103/PhysRevB.94.144512} {\bibfield  {journal} {\bibinfo  {journal} {Phys.
  Rev. B}\ }\textbf {\bibinfo {volume} {94}},\ \bibinfo {pages} {144512}
  (\bibinfo {year} {2016})}\BibitemShut {NoStop}%
\bibitem [{\citenamefont {Huang}\ and\ \citenamefont
  {Hoffman}(2017)}]{huang2017monolayer}%
  \BibitemOpen
  \bibfield  {author} {\bibinfo {author} {\bibfnamefont {D.}~\bibnamefont
  {Huang}}\ and\ \bibinfo {author} {\bibfnamefont {J.~E.}\ \bibnamefont
  {Hoffman}},\ }\href {\doibase 10.1146/annurev-conmatphys-031016-025242}
  {\bibfield  {journal} {\bibinfo  {journal} {Annual Review of Condensed Matter
  Physics}\ }\textbf {\bibinfo {volume} {8}},\ \bibinfo {pages} {311} (\bibinfo
  {year} {2017})},\ \Eprint
  {http://arxiv.org/abs/https://doi.org/10.1146/annurev-conmatphys-031016-025242}
  {https://doi.org/10.1146/annurev-conmatphys-031016-025242} \BibitemShut
  {NoStop}%
\bibitem [{\citenamefont {Yamakawa}\ and\ \citenamefont
  {Kontani}(2017)}]{yamakawa2017superconductivity}%
  \BibitemOpen
  \bibfield  {author} {\bibinfo {author} {\bibfnamefont {Y.}~\bibnamefont
  {Yamakawa}}\ and\ \bibinfo {author} {\bibfnamefont {H.}~\bibnamefont
  {Kontani}},\ }\href {\doibase 10.1103/PhysRevB.96.045130} {\bibfield
  {journal} {\bibinfo  {journal} {Phys. Rev. B}\ }\textbf {\bibinfo {volume}
  {96}},\ \bibinfo {pages} {045130} (\bibinfo {year} {2017})}\BibitemShut
  {NoStop}%
\bibitem [{\citenamefont {Khodas}\ and\ \citenamefont
  {Chubukov}(2012)}]{khodas2012interpocket}%
  \BibitemOpen
  \bibfield  {author} {\bibinfo {author} {\bibfnamefont {M.}~\bibnamefont
  {Khodas}}\ and\ \bibinfo {author} {\bibfnamefont {A.~V.}\ \bibnamefont
  {Chubukov}},\ }\href {\doibase 10.1103/PhysRevLett.108.247003} {\bibfield
  {journal} {\bibinfo  {journal} {Phys. Rev. Lett.}\ }\textbf {\bibinfo
  {volume} {108}},\ \bibinfo {pages} {247003} (\bibinfo {year}
  {2012})}\BibitemShut {NoStop}%
\bibitem [{\citenamefont {Chen}\ \emph {et~al.}(2015)\citenamefont {Chen},
  \citenamefont {Maiti}, \citenamefont {Linscheid},\ and\ \citenamefont
  {Hirschfeld}}]{chen2015electron}%
  \BibitemOpen
  \bibfield  {author} {\bibinfo {author} {\bibfnamefont {X.}~\bibnamefont
  {Chen}}, \bibinfo {author} {\bibfnamefont {S.}~\bibnamefont {Maiti}},
  \bibinfo {author} {\bibfnamefont {A.}~\bibnamefont {Linscheid}}, \ and\
  \bibinfo {author} {\bibfnamefont {P.~J.}\ \bibnamefont {Hirschfeld}},\ }\href
  {\doibase 10.1103/PhysRevB.92.224514} {\bibfield  {journal} {\bibinfo
  {journal} {Phys. Rev. B}\ }\textbf {\bibinfo {volume} {92}},\ \bibinfo
  {pages} {224514} (\bibinfo {year} {2015})}\BibitemShut {NoStop}%
\bibitem [{\citenamefont {Agterberg}\ \emph {et~al.}(2017)\citenamefont
  {Agterberg}, \citenamefont {Shishidou}, \citenamefont {O'Halloran},
  \citenamefont {Brydon},\ and\ \citenamefont
  {Weinert}}]{agterberg2017resilient}%
  \BibitemOpen
  \bibfield  {author} {\bibinfo {author} {\bibfnamefont {D.~F.}\ \bibnamefont
  {Agterberg}}, \bibinfo {author} {\bibfnamefont {T.}~\bibnamefont
  {Shishidou}}, \bibinfo {author} {\bibfnamefont {J.}~\bibnamefont
  {O'Halloran}}, \bibinfo {author} {\bibfnamefont {P.~M.~R.}\ \bibnamefont
  {Brydon}}, \ and\ \bibinfo {author} {\bibfnamefont {M.}~\bibnamefont
  {Weinert}},\ }\href {\doibase 10.1103/PhysRevLett.119.267001} {\bibfield
  {journal} {\bibinfo  {journal} {Phys. Rev. Lett.}\ }\textbf {\bibinfo
  {volume} {119}},\ \bibinfo {pages} {267001} (\bibinfo {year}
  {2017})}\BibitemShut {NoStop}%
\bibitem [{\citenamefont {Ge}\ \emph {et~al.}(2019)\citenamefont {Ge},
  \citenamefont {Yan}, \citenamefont {Zhang}, \citenamefont {Agterberg},
  \citenamefont {Weinert},\ and\ \citenamefont {Li}}]{ge2019evidence}%
  \BibitemOpen
  \bibfield  {author} {\bibinfo {author} {\bibfnamefont {Z.}~\bibnamefont
  {Ge}}, \bibinfo {author} {\bibfnamefont {C.}~\bibnamefont {Yan}}, \bibinfo
  {author} {\bibfnamefont {H.}~\bibnamefont {Zhang}}, \bibinfo {author}
  {\bibfnamefont {D.}~\bibnamefont {Agterberg}}, \bibinfo {author}
  {\bibfnamefont {M.}~\bibnamefont {Weinert}}, \ and\ \bibinfo {author}
  {\bibfnamefont {L.}~\bibnamefont {Li}},\ }\href
  {https://pubs.acs.org/doi/10.1021/acs.nanolett.9b00135} {\bibfield  {journal}
  {\bibinfo  {journal} {Nano letters}\ }\textbf {\bibinfo {volume} {19}},\
  \bibinfo {pages} {2497} (\bibinfo {year} {2019})}\BibitemShut {NoStop}%
\bibitem [{\citenamefont {Schrodi}\ \emph {et~al.}(2020)\citenamefont
  {Schrodi}, \citenamefont {Aperis},\ and\ \citenamefont
  {Oppeneer}}]{schrodi2020multichannel}%
  \BibitemOpen
  \bibfield  {author} {\bibinfo {author} {\bibfnamefont {F.}~\bibnamefont
  {Schrodi}}, \bibinfo {author} {\bibfnamefont {A.}~\bibnamefont {Aperis}}, \
  and\ \bibinfo {author} {\bibfnamefont {P.~M.}\ \bibnamefont {Oppeneer}},\
  }\href {\doibase 10.1103/PhysRevB.102.180501} {\bibfield  {journal} {\bibinfo
   {journal} {Phys. Rev. B}\ }\textbf {\bibinfo {volume} {102}},\ \bibinfo
  {pages} {180501} (\bibinfo {year} {2020})}\BibitemShut {NoStop}%
\bibitem [{\citenamefont {Kang}\ and\ \citenamefont
  {Fernandes}(2016)}]{kang2016superconductivity}%
  \BibitemOpen
  \bibfield  {author} {\bibinfo {author} {\bibfnamefont {J.}~\bibnamefont
  {Kang}}\ and\ \bibinfo {author} {\bibfnamefont {R.~M.}\ \bibnamefont
  {Fernandes}},\ }\href {\doibase 10.1103/PhysRevLett.117.217003} {\bibfield
  {journal} {\bibinfo  {journal} {Phys. Rev. Lett.}\ }\textbf {\bibinfo
  {volume} {117}},\ \bibinfo {pages} {217003} (\bibinfo {year}
  {2016})}\BibitemShut {NoStop}%
\bibitem [{\citenamefont {Zhang}\ \emph {et~al.}(2016)\citenamefont {Zhang},
  \citenamefont {Yi}, \citenamefont {Liu}, \citenamefont {Li}, \citenamefont
  {Lee}, \citenamefont {Moore}, \citenamefont {Hashimoto}, \citenamefont
  {Nakajima}, \citenamefont {Eisaki}, \citenamefont {Mo}, \citenamefont
  {Hussain}, \citenamefont {Devereaux}, \citenamefont {Shen},\ and\
  \citenamefont {Lu}}]{zhang2016distinctive}%
  \BibitemOpen
  \bibfield  {author} {\bibinfo {author} {\bibfnamefont {Y.}~\bibnamefont
  {Zhang}}, \bibinfo {author} {\bibfnamefont {M.}~\bibnamefont {Yi}}, \bibinfo
  {author} {\bibfnamefont {Z.-K.}\ \bibnamefont {Liu}}, \bibinfo {author}
  {\bibfnamefont {W.}~\bibnamefont {Li}}, \bibinfo {author} {\bibfnamefont
  {J.~J.}\ \bibnamefont {Lee}}, \bibinfo {author} {\bibfnamefont {R.~G.}\
  \bibnamefont {Moore}}, \bibinfo {author} {\bibfnamefont {M.}~\bibnamefont
  {Hashimoto}}, \bibinfo {author} {\bibfnamefont {M.}~\bibnamefont {Nakajima}},
  \bibinfo {author} {\bibfnamefont {H.}~\bibnamefont {Eisaki}}, \bibinfo
  {author} {\bibfnamefont {S.-K.}\ \bibnamefont {Mo}}, \bibinfo {author}
  {\bibfnamefont {Z.}~\bibnamefont {Hussain}}, \bibinfo {author} {\bibfnamefont
  {T.~P.}\ \bibnamefont {Devereaux}}, \bibinfo {author} {\bibfnamefont {Z.-X.}\
  \bibnamefont {Shen}}, \ and\ \bibinfo {author} {\bibfnamefont {D.~H.}\
  \bibnamefont {Lu}},\ }\href {\doibase 10.1103/PhysRevB.94.115153} {\bibfield
  {journal} {\bibinfo  {journal} {Phys. Rev. B}\ }\textbf {\bibinfo {volume}
  {94}},\ \bibinfo {pages} {115153} (\bibinfo {year} {2016})}\BibitemShut
  {NoStop}%
\bibitem [{\citenamefont {Bang}(2019)}]{bang2019phonon}%
  \BibitemOpen
  \bibfield  {author} {\bibinfo {author} {\bibfnamefont {Y.}~\bibnamefont
  {Bang}},\ }\href {https://doi.org/10.1038/s41598-019-40536-3} {\bibfield
  {journal} {\bibinfo  {journal} {Sci. Rep.}\ }\textbf {\bibinfo {volume}
  {9}},\ \bibinfo {pages} {3907} (\bibinfo {year} {2019})}\BibitemShut
  {NoStop}%
\bibitem [{\citenamefont {Rademaker}\ \emph {et~al.}(2021)\citenamefont
  {Rademaker}, \citenamefont {Alvarez-Suchini}, \citenamefont {Nakatsukasa},
  \citenamefont {Wang},\ and\ \citenamefont
  {Johnston}}]{rademaker2021enchnanced}%
  \BibitemOpen
  \bibfield  {author} {\bibinfo {author} {\bibfnamefont {L.}~\bibnamefont
  {Rademaker}}, \bibinfo {author} {\bibfnamefont {G.}~\bibnamefont
  {Alvarez-Suchini}}, \bibinfo {author} {\bibfnamefont {K.}~\bibnamefont
  {Nakatsukasa}}, \bibinfo {author} {\bibfnamefont {Y.}~\bibnamefont {Wang}}, \
  and\ \bibinfo {author} {\bibfnamefont {S.}~\bibnamefont {Johnston}},\ }\href
  {https://link.aps.org/doi/10.1103/PhysRevB.103.144504} {\bibfield  {journal}
  {\bibinfo  {journal} {Phys. Rev. B Condens. Matter}\ }\textbf {\bibinfo
  {volume} {103}},\ \bibinfo {pages} {144504} (\bibinfo {year}
  {2021})}\BibitemShut {NoStop}%
\bibitem [{\citenamefont {Fan}\ \emph {et~al.}(2015)\citenamefont {Fan},
  \citenamefont {Zhang}, \citenamefont {Liu}, \citenamefont {Yan},
  \citenamefont {Ren}, \citenamefont {Peng}, \citenamefont {Xu}, \citenamefont
  {Xie}, \citenamefont {Hu}, \citenamefont {Zhang},\ and\ \citenamefont
  {Feng}}]{fan2015plain}%
  \BibitemOpen
  \bibfield  {author} {\bibinfo {author} {\bibfnamefont {Q.}~\bibnamefont
  {Fan}}, \bibinfo {author} {\bibfnamefont {W.~H.}\ \bibnamefont {Zhang}},
  \bibinfo {author} {\bibfnamefont {X.}~\bibnamefont {Liu}}, \bibinfo {author}
  {\bibfnamefont {Y.~J.}\ \bibnamefont {Yan}}, \bibinfo {author} {\bibfnamefont
  {M.~Q.}\ \bibnamefont {Ren}}, \bibinfo {author} {\bibfnamefont
  {R.}~\bibnamefont {Peng}}, \bibinfo {author} {\bibfnamefont {H.~C.}\
  \bibnamefont {Xu}}, \bibinfo {author} {\bibfnamefont {B.~P.}\ \bibnamefont
  {Xie}}, \bibinfo {author} {\bibfnamefont {J.~P.}\ \bibnamefont {Hu}},
  \bibinfo {author} {\bibfnamefont {T.}~\bibnamefont {Zhang}}, \ and\ \bibinfo
  {author} {\bibfnamefont {D.~L.}\ \bibnamefont {Feng}},\ }\href {\doibase
  10.1038/nphys3450} {\bibfield  {journal} {\bibinfo  {journal} {Nature
  Physics}\ }\textbf {\bibinfo {volume} {11}},\ \bibinfo {pages} {946}
  (\bibinfo {year} {2015})}\BibitemShut {NoStop}%
\bibitem [{\citenamefont {Miyata}\ \emph {et~al.}(2015)\citenamefont {Miyata},
  \citenamefont {Nakayama}, \citenamefont {Sugawara}, \citenamefont {Sato},\
  and\ \citenamefont {Takahashi}}]{miyata2015high-temperature}%
  \BibitemOpen
  \bibfield  {author} {\bibinfo {author} {\bibfnamefont {Y.}~\bibnamefont
  {Miyata}}, \bibinfo {author} {\bibfnamefont {K.}~\bibnamefont {Nakayama}},
  \bibinfo {author} {\bibfnamefont {K.}~\bibnamefont {Sugawara}}, \bibinfo
  {author} {\bibfnamefont {T.}~\bibnamefont {Sato}}, \ and\ \bibinfo {author}
  {\bibfnamefont {T.}~\bibnamefont {Takahashi}},\ }\href {\doibase
  10.1038/nmat4302} {\bibfield  {journal} {\bibinfo  {journal} {Nature
  Materials}\ }\textbf {\bibinfo {volume} {14}},\ \bibinfo {pages} {775}
  (\bibinfo {year} {2015})}\BibitemShut {NoStop}%
\bibitem [{\citenamefont {Hanzawa}\ \emph {et~al.}(2016)\citenamefont
  {Hanzawa}, \citenamefont {Sato}, \citenamefont {Hiramatsu}, \citenamefont
  {Kamiya},\ and\ \citenamefont {Hosono}}]{hanzawa2016electric}%
  \BibitemOpen
  \bibfield  {author} {\bibinfo {author} {\bibfnamefont {K.}~\bibnamefont
  {Hanzawa}}, \bibinfo {author} {\bibfnamefont {H.}~\bibnamefont {Sato}},
  \bibinfo {author} {\bibfnamefont {H.}~\bibnamefont {Hiramatsu}}, \bibinfo
  {author} {\bibfnamefont {T.}~\bibnamefont {Kamiya}}, \ and\ \bibinfo {author}
  {\bibfnamefont {H.}~\bibnamefont {Hosono}},\ }\href {\doibase
  10.1073/pnas.1520810113} {\bibfield  {journal} {\bibinfo  {journal}
  {Proceedings of the National Academy of Sciences}\ }\textbf {\bibinfo
  {volume} {113}},\ \bibinfo {pages} {3986} (\bibinfo {year} {2016})},\ \Eprint
  {http://arxiv.org/abs/https://www.pnas.org/content/113/15/3986.full.pdf}
  {https://www.pnas.org/content/113/15/3986.full.pdf} \BibitemShut {NoStop}%
\bibitem [{\citenamefont {Shiogai}\ \emph {et~al.}(2016)\citenamefont
  {Shiogai}, \citenamefont {Ito}, \citenamefont {Mitsuhashi}, \citenamefont
  {Nojima},\ and\ \citenamefont {Tsukazaki}}]{shiogai2016electric}%
  \BibitemOpen
  \bibfield  {author} {\bibinfo {author} {\bibfnamefont {J.}~\bibnamefont
  {Shiogai}}, \bibinfo {author} {\bibfnamefont {Y.}~\bibnamefont {Ito}},
  \bibinfo {author} {\bibfnamefont {T.}~\bibnamefont {Mitsuhashi}}, \bibinfo
  {author} {\bibfnamefont {T.}~\bibnamefont {Nojima}}, \ and\ \bibinfo {author}
  {\bibfnamefont {A.}~\bibnamefont {Tsukazaki}},\ }\href {\doibase
  10.1038/nphys3530} {\bibfield  {journal} {\bibinfo  {journal} {Nature
  Physics}\ }\textbf {\bibinfo {volume} {12}},\ \bibinfo {pages} {42} (\bibinfo
  {year} {2016})}\BibitemShut {NoStop}%
\bibitem [{\citenamefont {Maletz}\ \emph {et~al.}(2014)\citenamefont {Maletz},
  \citenamefont {Zabolotnyy}, \citenamefont {Evtushinsky}, \citenamefont
  {Thirupathaiah}, \citenamefont {Wolter}, \citenamefont {Harnagea},
  \citenamefont {Yaresko}, \citenamefont {Vasiliev}, \citenamefont {Chareev},
  \citenamefont {B\"ohmer}, \citenamefont {Hardy}, \citenamefont {Wolf},
  \citenamefont {Meingast}, \citenamefont {Rienks}, \citenamefont {B\"uchner},\
  and\ \citenamefont {Borisenko}}]{maletz2014unusual}%
  \BibitemOpen
  \bibfield  {author} {\bibinfo {author} {\bibfnamefont {J.}~\bibnamefont
  {Maletz}}, \bibinfo {author} {\bibfnamefont {V.~B.}\ \bibnamefont
  {Zabolotnyy}}, \bibinfo {author} {\bibfnamefont {D.~V.}\ \bibnamefont
  {Evtushinsky}}, \bibinfo {author} {\bibfnamefont {S.}~\bibnamefont
  {Thirupathaiah}}, \bibinfo {author} {\bibfnamefont {A.~U.~B.}\ \bibnamefont
  {Wolter}}, \bibinfo {author} {\bibfnamefont {L.}~\bibnamefont {Harnagea}},
  \bibinfo {author} {\bibfnamefont {A.~N.}\ \bibnamefont {Yaresko}}, \bibinfo
  {author} {\bibfnamefont {A.~N.}\ \bibnamefont {Vasiliev}}, \bibinfo {author}
  {\bibfnamefont {D.~A.}\ \bibnamefont {Chareev}}, \bibinfo {author}
  {\bibfnamefont {A.~E.}\ \bibnamefont {B\"ohmer}}, \bibinfo {author}
  {\bibfnamefont {F.}~\bibnamefont {Hardy}}, \bibinfo {author} {\bibfnamefont
  {T.}~\bibnamefont {Wolf}}, \bibinfo {author} {\bibfnamefont {C.}~\bibnamefont
  {Meingast}}, \bibinfo {author} {\bibfnamefont {E.~D.~L.}\ \bibnamefont
  {Rienks}}, \bibinfo {author} {\bibfnamefont {B.}~\bibnamefont {B\"uchner}}, \
  and\ \bibinfo {author} {\bibfnamefont {S.~V.}\ \bibnamefont {Borisenko}},\
  }\href {\doibase 10.1103/PhysRevB.89.220506} {\bibfield  {journal} {\bibinfo
  {journal} {Phys. Rev. B}\ }\textbf {\bibinfo {volume} {89}},\ \bibinfo
  {pages} {220506} (\bibinfo {year} {2014})}\BibitemShut {NoStop}%
\bibitem [{\citenamefont {Aichhorn}\ \emph {et~al.}(2010)\citenamefont
  {Aichhorn}, \citenamefont {Biermann}, \citenamefont {Miyake}, \citenamefont
  {Georges},\ and\ \citenamefont {Imada}}]{aichhorn2010theoretical}%
  \BibitemOpen
  \bibfield  {author} {\bibinfo {author} {\bibfnamefont {M.}~\bibnamefont
  {Aichhorn}}, \bibinfo {author} {\bibfnamefont {S.}~\bibnamefont {Biermann}},
  \bibinfo {author} {\bibfnamefont {T.}~\bibnamefont {Miyake}}, \bibinfo
  {author} {\bibfnamefont {A.}~\bibnamefont {Georges}}, \ and\ \bibinfo
  {author} {\bibfnamefont {M.}~\bibnamefont {Imada}},\ }\href {\doibase
  10.1103/PhysRevB.82.064504} {\bibfield  {journal} {\bibinfo  {journal} {Phys.
  Rev. B}\ }\textbf {\bibinfo {volume} {82}},\ \bibinfo {pages} {064504}
  (\bibinfo {year} {2010})}\BibitemShut {NoStop}%
\bibitem [{\citenamefont {Yin}\ \emph {et~al.}(2011)\citenamefont {Yin},
  \citenamefont {Haule},\ and\ \citenamefont {Kotliar}}]{yin2011kinetic}%
  \BibitemOpen
  \bibfield  {author} {\bibinfo {author} {\bibfnamefont {Z.~P.}\ \bibnamefont
  {Yin}}, \bibinfo {author} {\bibfnamefont {K.}~\bibnamefont {Haule}}, \ and\
  \bibinfo {author} {\bibfnamefont {G.}~\bibnamefont {Kotliar}},\ }\href
  {\doibase 10.1038/nmat3120} {\bibfield  {journal} {\bibinfo  {journal}
  {Nature Materials}\ }\textbf {\bibinfo {volume} {10}},\ \bibinfo {pages}
  {932} (\bibinfo {year} {2011})}\BibitemShut {NoStop}%
\bibitem [{\citenamefont {Lee}\ \emph {et~al.}(2014)\citenamefont {Lee},
  \citenamefont {Schmitt}, \citenamefont {Moore}, \citenamefont {Johnston},
  \citenamefont {Cui}, \citenamefont {Li}, \citenamefont {Yi}, \citenamefont
  {Liu}, \citenamefont {Hashimoto}, \citenamefont {Zhang}, \citenamefont {Lu},
  \citenamefont {Devereaux}, \citenamefont {Lee},\ and\ \citenamefont
  {Shen}}]{lee2014interfacial}%
  \BibitemOpen
  \bibfield  {author} {\bibinfo {author} {\bibfnamefont {J.~J.}\ \bibnamefont
  {Lee}}, \bibinfo {author} {\bibfnamefont {F.~T.}\ \bibnamefont {Schmitt}},
  \bibinfo {author} {\bibfnamefont {R.~G.}\ \bibnamefont {Moore}}, \bibinfo
  {author} {\bibfnamefont {S.}~\bibnamefont {Johnston}}, \bibinfo {author}
  {\bibfnamefont {Y.-T.}\ \bibnamefont {Cui}}, \bibinfo {author} {\bibfnamefont
  {W.}~\bibnamefont {Li}}, \bibinfo {author} {\bibfnamefont {M.}~\bibnamefont
  {Yi}}, \bibinfo {author} {\bibfnamefont {Z.~K.}\ \bibnamefont {Liu}},
  \bibinfo {author} {\bibfnamefont {M.}~\bibnamefont {Hashimoto}}, \bibinfo
  {author} {\bibfnamefont {Y.}~\bibnamefont {Zhang}}, \bibinfo {author}
  {\bibfnamefont {D.~H.}\ \bibnamefont {Lu}}, \bibinfo {author} {\bibfnamefont
  {T.~P.}\ \bibnamefont {Devereaux}}, \bibinfo {author} {\bibfnamefont {D.-H.}\
  \bibnamefont {Lee}}, \ and\ \bibinfo {author} {\bibfnamefont {Z.-X.}\
  \bibnamefont {Shen}},\ }\href {\doibase 10.1038/nature13894} {\bibfield
  {journal} {\bibinfo  {journal} {Nature}\ }\textbf {\bibinfo {volume} {515}},\
  \bibinfo {pages} {245} (\bibinfo {year} {2014})}\BibitemShut {NoStop}%
\bibitem [{\citenamefont {Song}\ \emph {et~al.}(2011)\citenamefont {Song},
  \citenamefont {Wang}, \citenamefont {Jiang}, \citenamefont {Li},
  \citenamefont {Wang}, \citenamefont {He}, \citenamefont {Chen}, \citenamefont
  {Ma},\ and\ \citenamefont {Xue}}]{song2011molecular}%
  \BibitemOpen
  \bibfield  {author} {\bibinfo {author} {\bibfnamefont {C.-L.}\ \bibnamefont
  {Song}}, \bibinfo {author} {\bibfnamefont {Y.-L.}\ \bibnamefont {Wang}},
  \bibinfo {author} {\bibfnamefont {Y.-P.}\ \bibnamefont {Jiang}}, \bibinfo
  {author} {\bibfnamefont {Z.}~\bibnamefont {Li}}, \bibinfo {author}
  {\bibfnamefont {L.}~\bibnamefont {Wang}}, \bibinfo {author} {\bibfnamefont
  {K.}~\bibnamefont {He}}, \bibinfo {author} {\bibfnamefont {X.}~\bibnamefont
  {Chen}}, \bibinfo {author} {\bibfnamefont {X.-C.}\ \bibnamefont {Ma}}, \ and\
  \bibinfo {author} {\bibfnamefont {Q.-K.}\ \bibnamefont {Xue}},\ }\href
  {\doibase 10.1103/PhysRevB.84.020503} {\bibfield  {journal} {\bibinfo
  {journal} {Phys. Rev. B}\ }\textbf {\bibinfo {volume} {84}},\ \bibinfo
  {pages} {020503} (\bibinfo {year} {2011})}\BibitemShut {NoStop}%
\bibitem [{\citenamefont {Song}\ \emph {et~al.}(2019)\citenamefont {Song},
  \citenamefont {Yu}, \citenamefont {Lou}, \citenamefont {Xie}, \citenamefont
  {Xu}, \citenamefont {Wen}, \citenamefont {Yao}, \citenamefont {Zhang},
  \citenamefont {Zhu}, \citenamefont {Guo}, \citenamefont {Peng},\ and\
  \citenamefont {Feng}}]{song2019evidence}%
  \BibitemOpen
  \bibfield  {author} {\bibinfo {author} {\bibfnamefont {Q.}~\bibnamefont
  {Song}}, \bibinfo {author} {\bibfnamefont {T.~L.}\ \bibnamefont {Yu}},
  \bibinfo {author} {\bibfnamefont {X.}~\bibnamefont {Lou}}, \bibinfo {author}
  {\bibfnamefont {B.~P.}\ \bibnamefont {Xie}}, \bibinfo {author} {\bibfnamefont
  {H.~C.}\ \bibnamefont {Xu}}, \bibinfo {author} {\bibfnamefont {C.~H.~P.}\
  \bibnamefont {Wen}}, \bibinfo {author} {\bibfnamefont {Q.}~\bibnamefont
  {Yao}}, \bibinfo {author} {\bibfnamefont {S.~Y.}\ \bibnamefont {Zhang}},
  \bibinfo {author} {\bibfnamefont {X.~T.}\ \bibnamefont {Zhu}}, \bibinfo
  {author} {\bibfnamefont {J.~D.}\ \bibnamefont {Guo}}, \bibinfo {author}
  {\bibfnamefont {R.}~\bibnamefont {Peng}}, \ and\ \bibinfo {author}
  {\bibfnamefont {D.~L.}\ \bibnamefont {Feng}},\ }\href {\doibase
  10.1038/s41467-019-08560-z} {\bibfield  {journal} {\bibinfo  {journal}
  {Nature Communications}\ }\textbf {\bibinfo {volume} {10}},\ \bibinfo {pages}
  {758} (\bibinfo {year} {2019})}\BibitemShut {NoStop}%
\bibitem [{\citenamefont {Bang}(2014)}]{bang2014a}%
  \BibitemOpen
  \bibfield  {author} {\bibinfo {author} {\bibfnamefont {Y.}~\bibnamefont
  {Bang}},\ }\href {https://doi.org/10.1088/1367-2630/16/2/023029} {\bibfield
  {journal} {\bibinfo  {journal} {New J. Phys.}\ }\textbf {\bibinfo {volume}
  {16}},\ \bibinfo {pages} {023029} (\bibinfo {year} {2014})}\BibitemShut
  {NoStop}%
\bibitem [{\citenamefont {Linscheid}\ \emph {et~al.}(2016)\citenamefont
  {Linscheid}, \citenamefont {Maiti}, \citenamefont {Wang}, \citenamefont
  {Johnston},\ and\ \citenamefont {Hirschfeld}}]{linscheid2016high}%
  \BibitemOpen
  \bibfield  {author} {\bibinfo {author} {\bibfnamefont {A.}~\bibnamefont
  {Linscheid}}, \bibinfo {author} {\bibfnamefont {S.}~\bibnamefont {Maiti}},
  \bibinfo {author} {\bibfnamefont {Y.}~\bibnamefont {Wang}}, \bibinfo {author}
  {\bibfnamefont {S.}~\bibnamefont {Johnston}}, \ and\ \bibinfo {author}
  {\bibfnamefont {P.~J.}\ \bibnamefont {Hirschfeld}},\ }\href
  {https://link.aps.org/doi/10.1103/PhysRevLett.117.077003} {\bibfield
  {journal} {\bibinfo  {journal} {Phys. Rev. Lett.}\ }\textbf {\bibinfo
  {volume} {117}},\ \bibinfo {pages} {077003} (\bibinfo {year}
  {2016})}\BibitemShut {NoStop}%
\bibitem [{\citenamefont {Maier}\ \emph {et~al.}(2019)\citenamefont {Maier},
  \citenamefont {Mishra}, \citenamefont {Balduzzi},\ and\ \citenamefont
  {Scalapino}}]{maier2019effective}%
  \BibitemOpen
  \bibfield  {author} {\bibinfo {author} {\bibfnamefont {T.~A.}\ \bibnamefont
  {Maier}}, \bibinfo {author} {\bibfnamefont {V.}~\bibnamefont {Mishra}},
  \bibinfo {author} {\bibfnamefont {G.}~\bibnamefont {Balduzzi}}, \ and\
  \bibinfo {author} {\bibfnamefont {D.~J.}\ \bibnamefont {Scalapino}},\
  }\href@noop {} {\bibfield  {journal} {\bibinfo  {journal} {Phys. Rev. B
  Condens. Matter}\ }\textbf {\bibinfo {volume} {99}},\ \bibinfo {pages}
  {140504} (\bibinfo {year} {2019})}\BibitemShut {NoStop}%
\bibitem [{\citenamefont {Mishra}\ \emph {et~al.}(2016)\citenamefont {Mishra},
  \citenamefont {Scalapino},\ and\ \citenamefont {Maier}}]{mishra2016s}%
  \BibitemOpen
  \bibfield  {author} {\bibinfo {author} {\bibfnamefont {V.}~\bibnamefont
  {Mishra}}, \bibinfo {author} {\bibfnamefont {D.~J.}\ \bibnamefont
  {Scalapino}}, \ and\ \bibinfo {author} {\bibfnamefont {T.~A.}\ \bibnamefont
  {Maier}},\ }\href {https://www.nature.com/articles/srep32078} {\bibfield
  {journal} {\bibinfo  {journal} {Sci. Rep.}\ }\textbf {\bibinfo {volume}
  {6}},\ \bibinfo {pages} {32078} (\bibinfo {year} {2016})}\BibitemShut
  {NoStop}%
\bibitem [{\citenamefont {Raghu}\ \emph {et~al.}(2008)\citenamefont {Raghu},
  \citenamefont {Qi}, \citenamefont {Liu}, \citenamefont {Scalapino},\ and\
  \citenamefont {Zhang}}]{raghu2008minimal}%
  \BibitemOpen
  \bibfield  {author} {\bibinfo {author} {\bibfnamefont {S.}~\bibnamefont
  {Raghu}}, \bibinfo {author} {\bibfnamefont {X.-L.}\ \bibnamefont {Qi}},
  \bibinfo {author} {\bibfnamefont {C.-X.}\ \bibnamefont {Liu}}, \bibinfo
  {author} {\bibfnamefont {D.~J.}\ \bibnamefont {Scalapino}}, \ and\ \bibinfo
  {author} {\bibfnamefont {S.-C.}\ \bibnamefont {Zhang}},\ }\href {\doibase
  10.1103/PhysRevB.77.220503} {\bibfield  {journal} {\bibinfo  {journal} {Phys.
  Rev. B}\ }\textbf {\bibinfo {volume} {77}},\ \bibinfo {pages} {220503}
  (\bibinfo {year} {2008})}\BibitemShut {NoStop}%
\end{thebibliography}%

\end{document}